# INTERFACE Force Field for Alumina with Validated Bulk Phases and a pH-Resolved Surface Model Database for Aqueous, Electrolyte, and Molecular Interfaces

Cheng Zhu,[1] Krishan Kanhaiya,[1] Samir Darouich,[2] Sean P. Florez,[1] Karnajit Sen,[2] Patrick Keil,[3] Nawel S. Khelfallah,[4] Eduard Schreiner,[2] Ratan K. Mishra,[2*] Hendrik Heinz[1*]

[1] Department of Chemical and Biological Engineering, University of Colorado Boulder, Boulder, CO 80303-0596, USA

[2] Group Research, BASF SE, Carl-Bosch-Strasse 38, 67056 Ludwigshafen am Rhein, Germany

[3] BASF Coatings GmbH, Glasuritstrasse 1, 48165 Münster, Germany

[4] Chemetall GmbH, Trakhener Strasse 3, 60487 Frankfurt am Main, Germany

* Correspondence to: ratan.mishra@basf.com, hendrik.heinz@colorado.edu

## Abstract

Alumina and aluminum oxyhydroxides underpin numerous chemical-engineering technologies in heterogeneous catalysis, corrosion protection, functional coatings, energy-storage devices, and biomedical components. Yet molecular models that predictively connect phase structure, pH-dependent surface chemistry, electrolyte organization, and adsorption across operating conditions remain limited. Here we introduce the INTERFACE Force Field (IFF) together with a curated, ready-to-use pH-resolved surface model database that provides the most accurate and transferable atomistic description of major alumina phases to date according to experimental benchmarks available. The framework covers α-$Al_2O_3$, γ-$Al_2O_3$, boehmite, diaspore, and gibbsite using a single, physically interpretable parameter set that is directly compatible with CHARMM, AMBER, OPLS-AA, CVFF, and PCFF. Simulations reproduce experimental reference data across structural, thermodynamic, mechanical, and interfacial benchmarks with deviations of only a few percent relative to experiment, exceeding existing force fields and density-functional approaches. We introduce the first transferable treatment of surface ionization and charge regulation across alumina phases over a broad pH range, enabling simulations of realistic solid–electrolyte interfaces without phase-specific reparameterization. Quantitative reliability is demonstrated by reproducing zeta-potential trends and pH-dependent adsorption of an organic corrosion inhibitor at alumina–water interfaces. Predicted adsorption free energies and surface contact times correlate with experiments across an order of magnitude, enabling process-relevant screening and clarifying limitations of classical adsorption models. The approach is 10,000 to 100,000× faster than DFT-machine learned interatomic potentials (MLIPs) and establishes a predictive computational platform for designing alumina-containing catalysts, corrosion-resistant coatings, energy storage and biomaterials under realistic process conditions.

## 1. Introduction

Aluminum (Al) is the most abundant metal in the earth's crust, mostly in the form of bauxite, a mixture of aluminum oxides, hydroxides, and aluminosilicates.[1, 2] The oxides and hydroxides have versatile applications due to excellent mechanical, thermal, electrical, chemical, and biocompatible properties (Figure 1a-e).[3-6] Common phases include α-alumina (α-$Al_2O_3$), γ-alumina (γ-$Al_2O_3$), diaspore (α-AlO(OH)), boehmite (γ-AlO(OH)), and gibbsite ($Al(OH)_3$) (Figure 1f-j). For example, α-alumina is used for engineered ceramics due to wear resistance, for substrates in electronic circuits due to high thermal conductivity, as well as for prosthetics and implants due to biocompatibility.[7-9] The porosity of γ-alumina is suited for catalysts, catalyst supports, and adsorbents.[10-12] Nanostructures of aluminum oxide and hydroxides are also used as coatings on battery electrodes and separators,[13] for engineering drug delivery systems and advanced imaging systems,[14] as well as in reinforcements in polymer and metal matrix composites.[7, 9] High purity α-alumina (sapphire) is used in LEDs and optical lenses. Gibbsite is often included in fire retardants due to water release upon heating,[15] and gibbsite-rich bauxite rock is a key source to produce aluminum metal with wide use due to low mass density, good ductility, and high electrical conductivity.[1, 2] The oxide and hydroxide phases then play a critical role in surface protection and as anti-corrosion coatings of Al and its alloys (Figure 1c).[16] Strength, stiffness, toughness, corrosion resistance, and physical appearance of Al can be improved by alloying elements such as Cu, Si, Mg, Cr, and Ni.[17] Examples of major Al alloys include the 1xxx series to the 7xxx series,[18] as well as the 6xxx series with Si and Mg.[19]

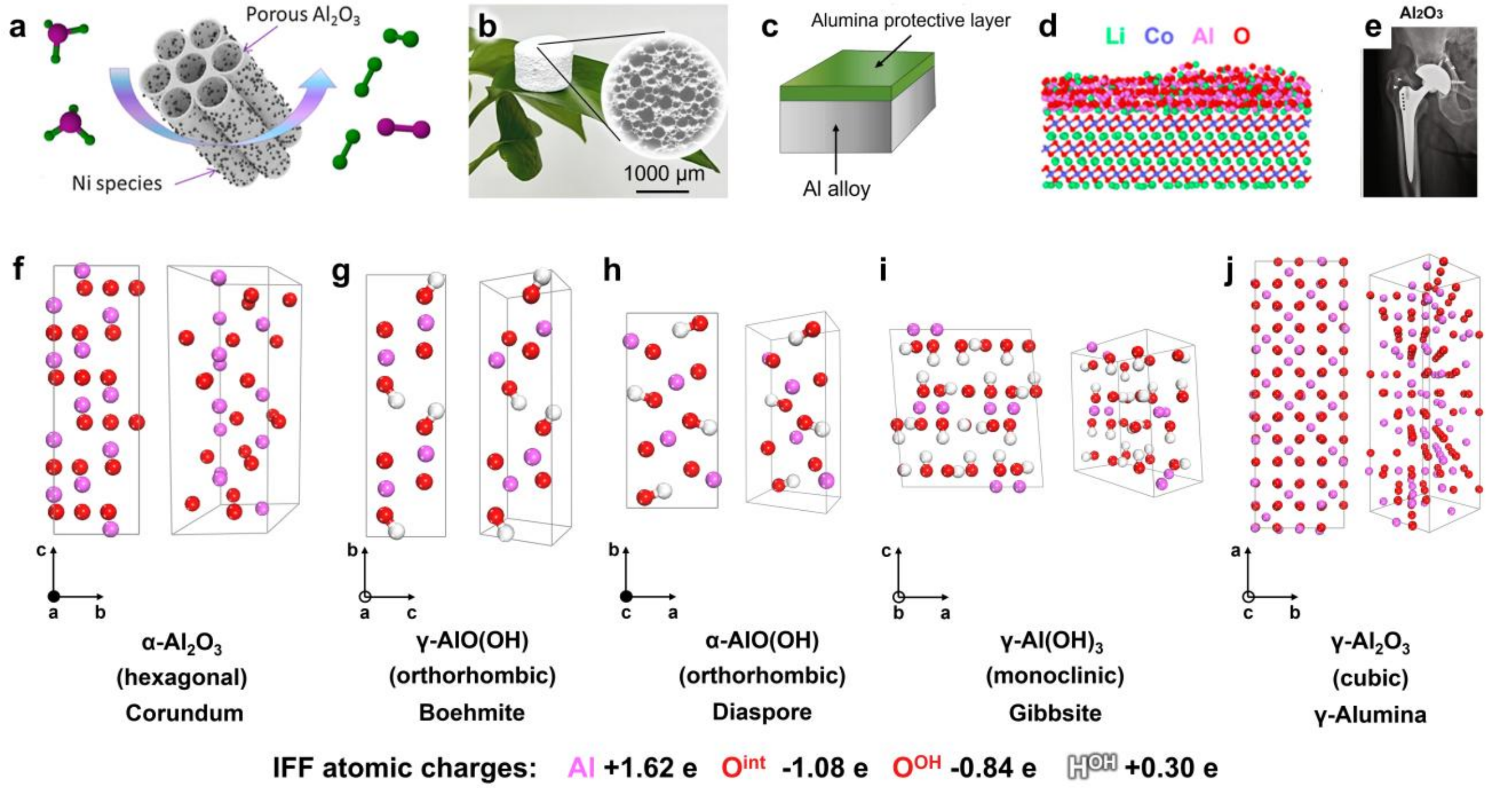

**Figure 1.** Common uses of aluminum oxide and hydroxide phases and their representation in the INTERFACE force field (IFF) for molecular simulations. (a-e) Representative applications include catalyst supports, e.g., ammonia decomposition)[6] (a), lightweight porous ceramics based on protein-foamed precursors[5] (b), coatings for corrosion protection (c), coatings on electrodes and separators in metal ion batteries[13] (d), and biocompatible implants for hip prostheses[4] (e). (f-j) Models of common alumina phases introduced into IFF and the representation of chemical bonding, shown as projections and as perspective views at an oblique angle upon slight rotation of the vertical axis to the left. The phases include α-aluminum oxide (corundum, α-$Al_2O_3$) (f), boehmite (γ-AlO(OH)) (g), diaspore (α-AlO(OH)) (h), gibbsite ($Al(OH)_3$) (i), and γ-aluminum oxide (γ-$Al_2O_3$) (j). γ-$Al_2O_3$ features statistical disorder of Al atoms. The parameters include atomic charges (shown at bottom), Lennard-Jones parameters for van-der-Waals interactions, and bonded parameters for OH groups to reproduce bulk, surface, and reactive properties.[20]

As Al alloys with enhanced corrosion resistance still suffer from pitting and stress-assisted corrosion,[21] various protective coatings such as oxide layers,[22] polymeric and silicone-based coatings,[23] powder coatings,[24] and Zn galvanizing[25] can provide long-term protection. Aluminum oxides ($Al_2O_3$) are attractive protection coatings due to high abundance, low cost, chemical and thermal stability, good strength, wear resistance, hardness, high melting point, and good electrical resistance.[26-28] Transitions between alumina phases as well as hydration and dehydration reactions (Figure 1a-e) can often be tailored by modulating treatment temperatures.[29]

Given the widespread use of $Al_2O_3$ and related compounds, it is important to understand, evaluate and predict structures and properties of pure, porous, and mixed phases, hydration reactions, as well as interactions with metals, oxides, solvents, molecules, polymers, and biopolymers. Fast simulations can complement time-consuming experiments, reach all-atom resolution, and monitor reaction dynamics. However, quantum mechanical and molecular simulations lack accurate, compatible parameters and realistic surface models of $Al_2O_3$-derived phases to address these problems.

The INTERFACE force field (IFF)[30-32] covers a wide range of metals, oxides, minerals, 2D materials, polymers, ionic liquids, gases, building on interpretable representations of chemical bonding and charge distributions in inorganic and organic compounds across the periodic table. Specifically, IFF parameters for oxides and hydroxides, including α-$Al_2O_3$, have demonstrated orders of magnitude improved reliability of bulk and surface properties relative to prior force fields and DFT calculations (Figure S1 and Section S1 in the Supplementary Material).[20] The energy expression is class I, with full class II compatibility, and parameters can be combined with the force fields CHARMM,[33] AMBER,[34] GROMOS,[35] OPLS−AA,[36] PCFF,[37, 38] and COMPASS[39] without further additions.[30, 40] IFF models and parameters are trained and validated in comparison to experimental reference data from the atomic scale to the macroscale, enabling quantitative simulations and predictions of multiphase assembly up to 1000 nm (Figure 2a and Section S1 in the Supplementary Material).[31, 41, 42]

Here we show how the existing IFF parameters for α-$Al_2O_3$[20] can be applied to other alumina phases (Figure 1f-j) and organic interfaces using additional parameters for internal and surface OH groups upon hydration and pH sensitive ionization. We achieve an average accuracy of 2.1% in densities, 4% in surface energies, and 7% in computed bulk moduli (with 12-6 LJ) relative to experiments. Thereby, we address several bottlenecks of prior simulation approaches for $Al_2O_3$ and oxyhydroxides using force fields such as Pedone,[43] CLAYFF,[44, 45] UFF,[46] ReaxFF[47, 48], CVFF,[49] PCFF,[37, 38] COMPASS,[39] COMPASS II, Transferable Gaussian Potentials,[50] and DFT[51] (Figure S1 and Section S1 in the Supplementary Material). The bottlenecks include (1) limited physical interpretability and reliability of the parameters (see details in refs. [30-32]), (2) limited validation of bulk and interfacial properties, often with less than two types of comparisons to experimental measurements,[52-56] (3) missing model representations for the pH-sensitive surface chemistry of alumina phases in contact with aqueous solutions, and (4) high computational expense of DFT methods (~$10^{12}$ times) and DFT/ML methods (~$10^5$ times) for systems with $10^4$ to $10^5$ atoms. In addition, Pedone potentials are difficult to use with hydroxides, surfaces, and organic components. Several force fields such as UFF, ClayFF, ReaxFF have not been tested for multiple alumina-related phases and compatibility with other force fields (CHARMM, AMBER, OPLS-AA, CVFF, and PCFF) that provide access to solvents, organic molecules, and biopolymers. CVFF, PCFF, and COMPASS perform poorly in structural stability and surface properties of alumina phases. Gaussian potentials for alumina[50] have no experimental validation and binding energies of organic molecules deviate significantly (6 to 11 kcal/mol) from the chosen DFT ground truth data. A quantitative performance comparison of IFF for oxides and alumina phases versus other force fields and DFT is given in Figure S1 and Section S1 in the Supplementary Material, building on recent benchmarking for several oxides and hydroxides in ref. [20].

In this contribution, we introduce IFF parameters and a pH-resolved surface model database for alumina phases, aqueous interfaces, and organic hybrid materials. We utilize existing, validated IFF parent parameters for α-$Al_2O_3$[20] and add parameters for OH groups that describe all oxide and hydrated phases, as well as surface ionization as a function of pH in high accuracy (Figure 2). The workflow[20, 30] involved (1) training of the parameters, (2) validation of key structural and interfacial properties in comparison to experimental data, and (3) predictions of properties outside the training range (Figure 2a). Specifically, we validated a series of bulk and surface properties of the five alumina phases, including lattice parameters, densities, (hkl) surface energies, bulk moduli, and selected contact angles. We introduce a suite of all-atom surface models as a function of pH based on experimentally derived surface ionization states (using surface titration and zeta-potential data). These models serve as input, while interfacial structure, ion distributions, and organic assembly behavior are predicted by all-atom IFF MD simulations. As a sample application, we examined the structure and dynamics of alumina-electrolyte interfaces and predict the binding characteristics of the corrosion inhibitor p-hydroxybenzoic acid as a function of pH value in comparison to detailed experimental adsorption isotherms.[57]

The alumina models overcome high uncertainties in prior methods (Figure S1 in the Supplementary Material),[20, 52-55] and introduce detailed surface chemistry.[56] They are necessary to enable trustworthy MD simulations of complex alumina hybrid materials for faster discovery of catalysts, analysis of corrosion processes, composites, and drug delivery systems.

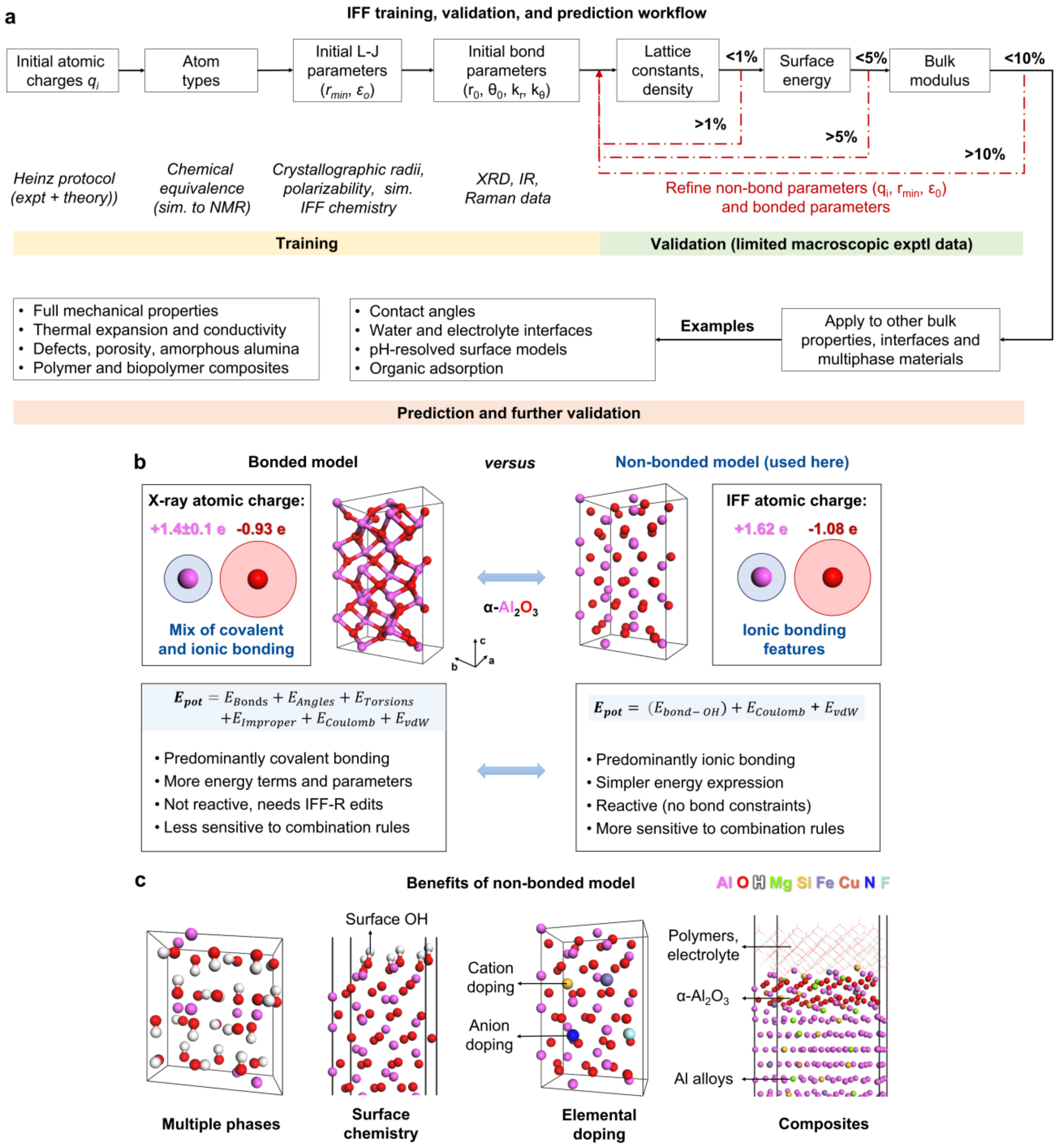

**Figure 2.** Workflow for IFF parameter development and a comparison of benefits of bonded *versus* nonbonded models for alumina phases. (a) The workflow focuses on the representation of chemical bonding and validation of structures and energy differences relative to experimental data. These steps enable predictions of non-fitted properties, including electrolyte and (bio)organic interfaces

(see refs. [30] and [20]). We utilized existing α-$Al_2O_3$ parameters[20] for all phases and added bonded parameters for OH groups to maintain simplicity and transferability. (b) All-atom representation of α-$Al_2O_3$ in bonded versus nonbonded models. Atomic charges from X-ray deformation electron densities in bonded models[58-60] are lower than atomic charges in the nonbonded model where stronger Coulomb interactions account for residual covalent bonding contributions.[20, 58] Key tradeoffs are indicated. Both types of models can represent surface chemistry, interfaces with electrolytes, organic molecules, and (bio)polymers. We introduce nonbonded models here, which require only a fraction of the constants (e.g., no Al-O bond and angle parameters) and retain a bonded term ($E_{bond\text{-}OH}$) for O-H groups. (c) Nonbonded models can represent multiple alumina phases, doping, defects, changing surface chemistry, and reactive interfaces with few or no further parameter assignments. Fully bonded models would require many additional parameters. The representations illustrate the nonbonded and residual bonded interactions (OH) in the IFF nonbonded models.

## 2. Results and Discussion

### 2.1. Rationale for Nonbonded Models

While Al atoms and O atoms in $Al_2O_3$ have formal charges (oxidation states) of +3e for Al and -2e for O, chemical bonding is approximately half covalent by sharing valence electrons in Al-O bonds and half ionic.[58-60] The atomic charges are between +1.3e and +1.7e for all alumina and aluminum hydroxide species, allowing the use of bonded models (if less than half the formal charges) or nonbonded models (if more than half the formal charges) (Figure 2b).[20, 58, 61, 62] We chose nonbonded models, which have several advantages:[20] (1) We eliminate extensive groups of atom types and bonded parameters for bonds, angles, and dihedrals in the oxide and hydroxide frameworks that would be required in bonded models (Figure 2b). (2) We only require few nonbonded parameters (atomic charge and LJ parameters) and bonded parameters for predominantly covalent O-H bonds and (Al)$OH_2^+$ bond angles. (3) We cover all phases of $Al_2O_3$ and oxyhydroxides with one set of compatible IFF parameters for a given energy expression (Figure 2c). (4) We include critical surface chemistry across pH values from 2 to 12 without further angle and torsion parameters.[20, 63] (5) We achieve high accuracy of computed structural, interfacial, and mechanical properties, including doping, defects, and reactive composites (Figure 2c). A minor tradeoff of nonbonded relative to bonded models is higher sensitivity to combination rules (see further discussion of bonded versus nonbonded models in Section S2 in the Supplementary Material).[32]

**2.2. IFF Energy Expression and Parameters**

The IFF parameters utilize the same potential energy expressions as materials oriented and biomolecular force fields,[30] including additive energy terms for harmonic bond stretching (for O-H bonds), harmonic angle bending for $OH_2^+$ groups, Coulomb energy, and a Lennard-Jones potential with 12-6 and 9-6 options (Table 1):

$$E_{pot} = \sum_{ij\ bonded} K_{r,ij}\left(r_{ij} - r_{0,ij}\right)^2 + \sum_{ijk\ bonded} K_{\theta,ijk}\left(\theta_{ijk} - \theta_{0,ijk}\right)^2 + \frac{1}{4\pi\varepsilon_0}\sum_{\substack{ij\ Coulomb \\ (1,2\ and\ 1,3\ excl)}} \frac{q_i q_j}{r_{ij}} +$$

$$\sum_{\substack{ij\ nonbonded \\ (1,2\ and\ 1,3\ excl)}} \varepsilon_{ij}\left[\left(\frac{r_{min,ij}}{r_{ij}}\right)^{12} - 2\left(\frac{r_{min,ij}}{r_{ij}}\right)^{6}\right] \quad (1)$$

with $r_{min,ij} = \frac{r_{min,ii}+r_{min,jj}}{2}$ (CHARMM, AMBER, DREIDING) or

$r_{min,ij} = \sqrt{r_{min,ii} \cdot r_{min,jj}}$ (CVFF, OPLS-AA)

and $\varepsilon_{ij} = \sqrt{\varepsilon_{ii} \cdot \varepsilon_{jj}}$ (all FFs with 12-6 LJ)

$$E_{pot} = \sum_{ij\ bonded} K_{r,ij}\left(r_{ij} - r_{0,ij}\right)^2 + + \sum_{ijk\ bonded} K_{\theta,ijk}\left(\theta_{ijk} - \theta_{0,ijk}\right)^2 + \frac{1}{4\pi\varepsilon_0}\sum_{\substack{ij\ Coulomb \\ (1,2\ and\ 1,3\ excl)}} \frac{q_i q_j}{r_{ij}} +$$

$$\sum_{\substack{ij\ nonbonded \\ (1,2\ and\ 1,3\ excl)}} \varepsilon_{ij}\left[2\left(\frac{r_{min,ij}}{r_{ij}}\right)^{9} - 3\left(\frac{r_{min,ij}}{r_{ij}}\right)^{6}\right] \quad (2)$$

with $r_{min,ij} = \left(\frac{r_{min,ii}^6+r_{min,jj}^6}{2}\right)^{1/6}$ and $\varepsilon_{ij} = \sqrt{\varepsilon_{ii} \cdot \varepsilon_{jj}}\frac{2r_{min,ii}^3 r_{min,jj}^3}{r_{min,ii}^6+r_{min,jj}^6}$ (CFF, PCFF, COMPASS)

The parameters include the harmonic bond stretching constants $K_r$, equilibrium bond lengths $r_0$, the harmonic angle bending constants $K_\theta$, equilibrium bond angles $\theta_0$, atomic charges $q_i$, equilibrium nonbond distances $r_{min,ij}$ (earlier also called $r_0$ or $\sigma_0$), as well as the nonbond potential well depth $\varepsilon_{ij}$. The bonded parameters $K_r, r_0$, $K_\theta$, and $\theta_0$ represent the characteristics of covalent bonds in [Al O-H] groups or in [Al $OH_2^+$] groups in contact with aqueous solution (Figure 3). The energy expressions in Equations (1) and (2) differ in the LJ potential, and force fields using a 12-6 LJ potentials[30] differ further in combination rules for $r_{min,ij}$ (lower part in Equation (1)). The distinction of combination rules in nonbonded models is essential for accuracy since chemical bonding and molecular geometries are exclusively represented by Coulomb and Lennard-Jones interactions (Figure 2b).[20] Therefore, we introduce (1) a set of IFF-CHARMM,

AMBER, DREIDING parameters with a 12-6 LJ potential and an arithmetic mean for $r_{min,ij}$ that has full compatibility with CHARMM,[64] AMBER,[34] and DREIDING,[65] (2) a second set of IFF-CVFF, OPLS parameters with a 12-6 LJ potential and a geometric mean for $r_{min,ij}$ has full compatibility with CVFF[49] and OPLS-AA[36] force fields, and (3) a third set of IFF-PCFF, CFF, COMPASS parameters with a 9-6 LJ potential[30] that has full compatibility with CFF,[37] PCFF,[38] and COMPASS[39] and uses Waldmann-Hagler combination rules[66] (Equation (2)). These sets of IFF LJ parameters were derived using the same workflow (Figure 2a) and achieve best compatibility in each group (see details in Section S3 in the Supplementary Material). If 12-6 LJ parameters are transferred from one group to another, results remain of acceptable quality, although the accuracy is notably reduced, e.g., from 5% to 20% deviation of computed organic adsorption properties relative to experimental reference data.

We use the IFF parent parameters for α-$Al_2O_3$[20] for inner Al and O atoms in all alumina phases (Figure 1f-j) and added parameters for bulk and surface [$Al_2$ O-H] groups, protonated $Al_2OH_2^+$ and deprotonated $Al_2O^-$ surface groups with associated charge distributions (Figure 3). The atomic charges of +1.62e for internal Al and -1.08e for internal O atoms are at the high end of the physical range to represent predominantly ionic bonds and incorporate residual covalent contributions (Figure 3a). The values are consistent with atomic charges from X-ray deformation electron densities (+1.3e to +1.5e),[58-60] trends in the atomization energy of Al compared to other elements, and trends in the first ionization energy of Al compared to other elements (Figure 2b).[28, 58-60, 67] The atomization energy of Al, for example, is lower than for more covalent C and Si atoms, and higher than for more ionic Na, Mg, and Mn atoms, setting lower and upper boundaries for atomic charges (Figure S2 in the Supplementary Material).[58] Similar considerations apply to trends in first ionization energies (Figure S3 in the Supplementary Material). Furthermore, these charges are consistent with the electronegativity difference of 1.83 between O and Al in comparison to similar chemistries (oxides and hydroxides) with known atomic charges.[20, 58] If unjustified charges are used, e.g. +1.2e for Al in COMPASS,[39] consistency across the phases is lost and errors in computed properties increase by an order of magnitude (Figure S1 and Section S1 in the Supplementary Material).

Parameters for $Al_2OH$ surface groups align with changes in chemical bonding during hydration, protonation, and deprotonation reactions, considering contact angles, surface titration, zeta potentials, and pK values (Figure 3b). The models benefit from chemical analogy with silica,[63] clay minerals,[61] hydroxyapatite,[68] and related hydroxide surfaces such as Al-OH, Si-OH, P-OH, Mg-OH, and Ca-OH groups[20] with prior validation in IFF. The atomic charges in electroneutral $Al_2OH$ groups inside hydroxide-containing minerals (-0.84e/+0.30e) were found to be the same as on $Al_2OH$ terminated surfaces (-0.84e/+0.30e) (Figure 3b and Figure S4 in the Supplementary Material). If not in contact with water, surface $Al_2OH$ groups may be better described with slightly lower charges of (-0.83e/+0.29e), however, the difference is almost negligible. Protonated $Al_2OH_2^+$ surface groups below the point of zero charge (typically pH < 8)[69-71] maintain an oxygen charge of -0.84e and carry higher H charges of +0.65e to maintain structural stability of the surface (if bonded terms for Al-O were added, these charges would be lower). Deprotonated $Al_2O^-$ surface groups receive a higher negative charge of -1.10e on the $O^-$ sites, and the nearest two Al atoms reduce their atomic charge from 1.62e to 1.40e (Figure 3b). This single set of atom types and charge distribution, along with LJ parameters and O-H bonded parameters, describes all major phases of aluminum oxides and hydroxides and requires no phase-specific modifications (Fig. 1f-j). Models with pH-resolved aqueous surface chemistry, including the $Al_2OH$, $Al_2OH_2^+$ $Cl^-$, and $Al_2O^-$ $Na^+$ groups, are necessary for realistic simulations of solid-electrolyte and inorganic-organic interfaces.

The parameters in the form of IFF-CHARMM, AMBER; IFF-CVFF, OPLS-AA; and IFF-PCFF, COMPASS are listed in Table 1 (see Section S4 in the Supplementary Material for a detailed interpretation).

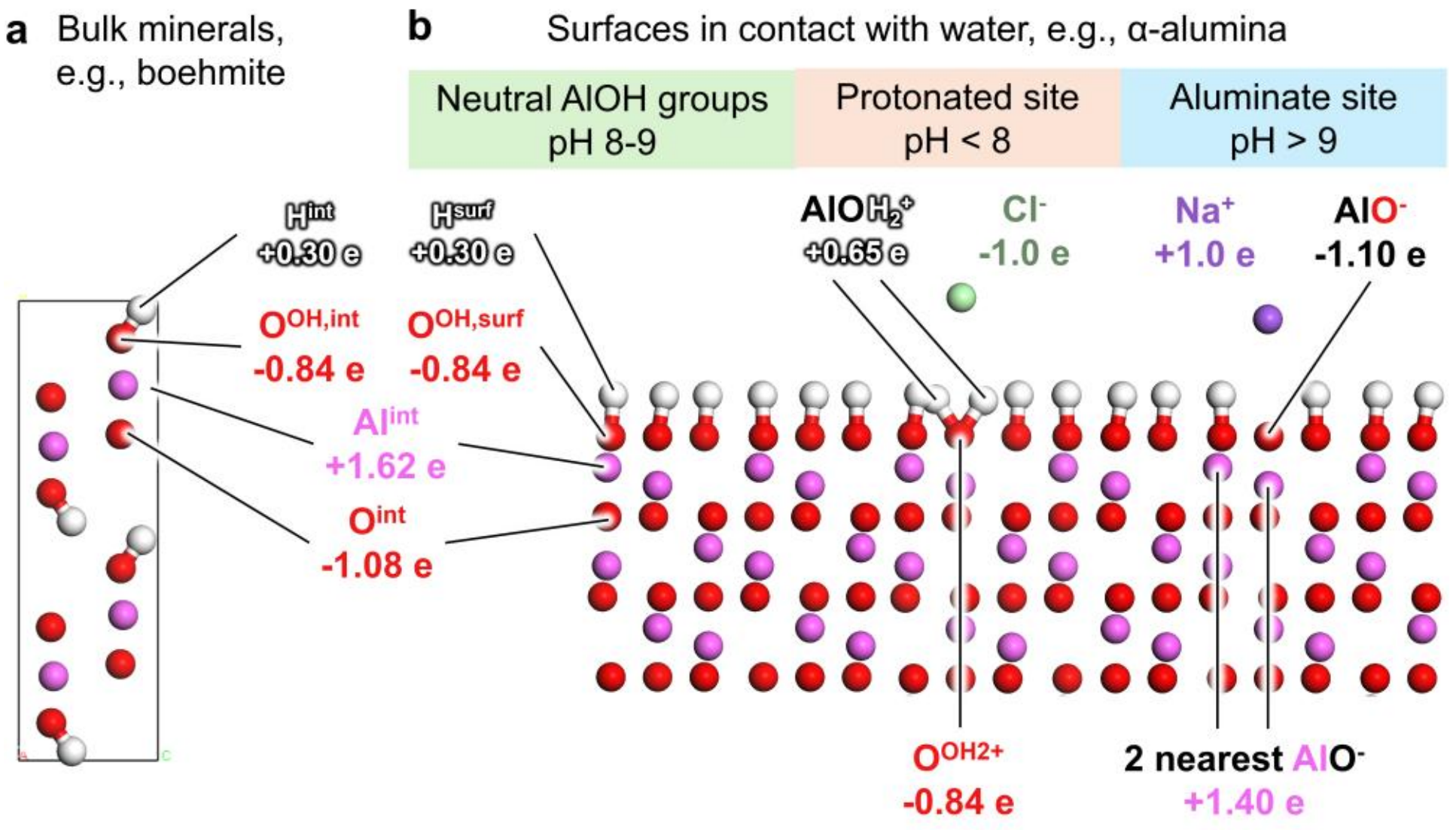

**Figure 3.** Atom types and atomic charges for alumina phases and surfaces. We use one set of atom types and small differences in atomic charges of OH groups at surfaces. (a) Example of a bulk

mineral (boehmite). Internal $Al_2OH$ groups have an atom type of oxygen atoms with -0.84e charge ($O^{OH, int}$) and an atom type of hydrogen atoms with +0.30e charge ($H^{int}$). (b) Surface $Al_2OH$ groups can be native to minerals (e.g., gibbsite) or formed upon hydration (e.g., γ-alumina) and utilize the same atom type and the same charges of -0.84e ($O^{OH, surf}$) and +0.30e ($H^{surf}$). Further protonation or deprotonation depending on pH value leads to adjustments in atomic charges. The charge balance for electroneutral $Al_2(OH_2)^+$ $Cl^-$ (pH < 8) and $Al_2O^-$ $Na^+$ sites (pH > 8) is shown. Relatively high negative O charges at both types of surface sites account for ionic and covalent bonding contributions and maintain structural stability.

**Table 1.** IFF parameters for alumina phases. Nonbonded parameters include atomic charges specific to the environment (Figure 3) and Lennard-Jones (LJ) parameters. Bonded parameters are applied to internal and surface (Al)O-H groups, as well as surface $(Al)OH_2^+$ groups depending on the pH values in aqueous media. Atomic charges and LJ parameters for Al and $O^{int}$ are the same as in ref. [20].

| **Atomic charges** | **Al (e)** | **$O^{int}$ (e)** | **$O^{OH}$ (e)** | **H (e)** |
|---|---|---|---|---|
| Same for all energy functions | **Al**: +1.62<br>In **AlO**$^-$: +1.40<br>(2 Al closest to $O^-$ sites) | -1.08 | Al**O**H: -0.84<br>Al**O**$H_2^+$: -0.84<br>Al**O**$^-$: -1.10 | AlO**H**: 0.30<br>AlO**$H_2^+$**: 0.65 |
| **LJ parameters** | **Al** | **$O^{int}$** | **$O^{OH}$** | **$H^{OH}$** |
| | **$r_{min}$ (Å), $\varepsilon$** | **$r_{min}$ (Å), $\varepsilon$** | **$r_{min}$ (Å), $\varepsilon$** | **$r_{min}$ (Å), $\varepsilon$** |

| | (kcal/mol) | (kcal/mol) | (kcal/mol) | (kcal/mol) |
|---|---|---|---|---|
| IFF-CHARMM, AMBER (12-6) | 1.86, 0.1 | 3.54, 0.09 | 3.47, 0.122 | 1.085, 0.015[a] |
| IFF-CVFF, OPLS (12-6) | 1.72, 0.45 | 3.3, 0.35 | 3.47, 0.122 | 1.085, 0.015[a] |
| IFF-PCFF, COMPASS (9-6) | 1.81, 0.35 | 3.45, 0.2 | 3.47, 0.120 | 1.098, 0.013 |

| Bonded parameters | O–H bond | Angle (in $AlOH_2^+$) |
|---|---|---|
| Same for all energy functions | (Al) O-H: $r_0$ = 0.945 Å<br>$K_r$ = 495 kcal/(mol·Å$^2$) | NA |
| | (Al) O-$H_2^+$: $r_0$ = 1.000 Å<br>$K_r$ = 540.6 kcal/(mol·Å$^2$) | $\theta_0$ = 109.47°<br>$K_\theta$ = 50 kcal/(rad·Å$^2$) |

[a] Simplified H parameters in OH groups of 0.001, 0.001 can also be used with less than 5% change in interfacial properties.

## 2.3. Validation of Lattice Parameters, Surface Energy, and Bulk Modulus

The models allow an excellent representation of the structural and interfacial properties of $Al_2O_3$ and oxyhydroxide compounds (Figure 4 and Tables S1 to S6 in the Supplementary Material). Average deviations from experimental data are 0.8% in lattice parameters (2.1% in density), 4% in surface energies, and 7% in bulk modulus (for 12-6 LJ potentials), plus agreement in contact angles.

For α-alumina, lattice vectors and angles are almost identical to experimental values with average deviations less than 0.15%. The root mean squared displacement (RMSD) of the atoms during simulations was about 0.15 Å, and the density matches experiments within 0.1% relative to XRD data of 3.984 g/cm$^3$ (Figure 4a and Table S1 in the Supplementary Material).[72] The (0001) surface energy of 1.65±0.05 J/m$^2$ is reproduced within error bars,[73] and the $(11\bar{2}0)$ surface energy was computed to be slightly higher at 1.88 J/m$^2$ (Figure 4b, Figure S5 and Table S1 in the Supplementary Material). Experimental measurements suggest a bulk modulus of 254 GPa,[74, 75] which IFF MD simulations reproduce within 4-8% (IFF-CVFF, OPLS and IFF-CHARMM, AMBER), except for IFF-PCFF which is less recommended for mechanical properties due to the 9-6 LJ potential (Figure 4c).[20] Detailed elastic constants for α-$Al_2O_3$ are also reproduced well in MD simulations (Table S6 in the Supplementary Material). The simulated shear modulus of 155 GPa deviates only 6% from the experimental value 165 GPa, and the computed Poisson ratio of 0.26 is close to the experimental measurement of 0.24.[74]

For boehmite, the computed lattice parameters deviate on average by about 1% and RMSD deviations are ~0.5 Å in a 25 Å supercell. The definition of the structures is thus excellent (Figure 5a). The density in MD simulations matches experimental values within 0.5%, except for IFF-CHARMM, AMBER with 3% deviation (Figure 4a and Table S2 in the Supplementary Material). Computed (001) surface energies match likely values of ~0.9 J/m$^2$ within 5% while the cleavage plane of lowest energy is (010), terminated by Al(OH) groups on both sides (Figure 1g), with 0.20 ±0.04 J/m$^2$ cleavage energy (Figure 4b and Table S2 in the Supplementary Material). The experimental bulk modulus is somewhat uncertain to-date, with most likely values of 100 ± 10 GPa. MD simulations with 12-6 LJ potentials yield values of 95 and 100 GPa (70 GPa for PCFF), consistent with excellent lattice parameters and surface energies (Figure 4c and Table S2 in the Supplementary Materials).

Diaspore showed overall ~2% deviation in lattice parameters and up to several percent in individual values (Figure 4a and Table S3 in the Supplementary Material). Computed densities agree within 1-3% of experimental data, and the structural fidelity is high (Figure 5b). The (001) surface energy was computed as ~1.25 $J/m^2$ and is similar to the (100) surface energy of ~1.20 $J/m^2$ (Figure 4b). The experimental bulk modulus of approximately 120 GPa was computed around 110 GPa, except for IFF-PCFF (86 GPa) which routinely leads to underestimations (Figure 4c and Table S3 in the Supplementary Material).

Computations on gibbsite (γ-$Al(OH)_3$) achieved excellent reproduction of the lattice parameters with average deviations <1%, including the monoclinic cell angle, and deviations in density of 2-4% (Figure 4a and Table S4 in the Supplementary Material). γ-$Al(OH)_3$ stands out as a highly hydrous mineral with 2D OH layers that coordinate Al layers in alternation (Figure 1i), are easily displaced and known to form perfect (001) cleavage planes with a computed surface energy of 0.28 ±0.03 $J/m^2$ (Figure 4b). The value aligns with the presence of densely packed OH groups at the interface and is equal to about 4x the surface tension of liquid water (72 $mJ/m^2$). The bulk modulus is expected to be between 30 and 35 GPa according to experimental data at low pressures, and computed values agree within this range, except a lower value for IFF-PCFF (Figure 4c and Table S4 in the Supplementary Material). As a result of the hydrous nature and associated lower cohesion, the $Al(OH)_3$ layers are easily displaced and simulation cells of 69.472 × 60.936 × 58.233 $Å^3$ or larger size were required to maintain stable structures in MD simulations (Figure S6 in the Supplementary Material).

$\gamma$-$Al_2O_3$ is a distinctively disordered cubic phase with a well-defined oxygen lattice and random occupancy of vacancies by Al (32 O sites and 119 fractionally occupied Al sites).[76] Stoichiometric models for simulations require a supercell that consists of at least 3 unit cells, such as 3×1×1 (23.815 × 7.938 × 7.938 $Å^3$). The $\gamma$-$Al_2O_3$ model then includes 64 Al atoms in random positions along with 96 O atoms in well-defined positions (Figure S7 in the Supplementary Material). The lattice parameters are reproduced in excellent agreement with experimental data, showing average deviations of 0.4%, and the density agrees within ~2.5% (Figure 4a and Table S5 in the Supplementary Material). Some rearrangements of Al atoms among the fractionally occupied lattice sites were seen during molecular dynamics, along with high structural stability of the O lattice (Figure S8 in the Supplementary Material). The (001) surface energy is 1.66±0.1 $J/m^2$ according to measurements,[77] identical to $\alpha$-$Al_2O_3$ within the uncertainty, and reproduced within 4% in molecular simulations (Figure 4b). The bulk modulus of 162 GPa is also reproduced by simulations, although with some scatter between the CVFF and CHARMM versions (Figure 4c and Table S5 in the Supplementary Material). Earlier models of $\gamma$-alumina based on DFT calculations by Digne et al. used smaller unit cells with different lattice parameters,[51] which may not capture the Al distribution across vacancies.[76] The density and bulk modulus by Digne et al.[51] agree with experimental data (2-5%) while the (100) surface energy of 1.0 J/m$^2$ was low (1.66 J/m$^2$ in experiments[77]). A detailed analysis of hydration equilibria was described.

Our data across all alumina phases show that higher mass density correlates with higher surface energy and higher bulk modulus (Figure 4a-c). Disordered ($\gamma$-$Al_2O_3$) and hydrated minerals (boehmite, diaspore, gibbsite) have lower densities than $\alpha$-$Al_2O_3$, and surfaces terminated with OH groups (boehmite, diaspore, gibbsite) possess lower surface energy than anhydrous surfaces ($\alpha$-$Al_2O_3$ and $\gamma$-$Al_2O_3$). As previously described,[30] bulk moduli often correlate with surface energies as they are derivatives of the surface energy with respect to displacement of the bounding surfaces.

IFF strongly outperforms prior force fields (Pedone,[20, 43] CLAYFF,[20, 44, 45] UFF,[20, 46] ReaxFF[20, 47, 48], CVFF,[49] PCFF,[37, 38] COMPASS,[39] COMPASS II), providing the simplest, most compatible and transferable parameters (Figure S1 and Section S1 in the Supplementary Material). DFT calculations also show larger uncertainties in several structural, interfacial, and mechanical properties relative to experiment.[20] Specifically, Pedone cannot provide surface and interfacial properties, CLAYFF has deviations in lattice parameters and densities of 5-10% and mismatches of ~30% in surface energy, including boehmite and gibbsite,[20, 44, 45] and UFF has extreme uncertainties on the order of 100%.[20, 46] ReaxFF performance is variable, depending on parameterization, typically with 5-10% deviations in lattice parameters, 50% error in surface energy, and 10-20% in bulk moduli.[20, 47, 48] DFT simulations commonly have 3-4% errors in lattice parameters and density, ~25% errors in surface energy, and

~15% uncertainty in bulk modulus compared to ~2%, 4%, and 7% in IFF simulations (using 12-6 LJ).[20]

In summary, IFF parameters for alumina phases reproduce key features of chemical bonding, structural and energetic properties in outstanding agreement with experimental reference data. The performance is similar for each force field sub-class used (IFF-CHARMM, AMBER; IFF-CVFF, OPLS-AA; IFF-PCFF, COMPASS) and independent of the simulation software (e.g., LAMMPS, NAMD, GROMACS, OpenMM, Forcite and Discover in Materials Studio). For best mechanical properties, the 12-6 LJ versions are preferred over 9-6 LJ versions, which lead to lower moduli unless atomic charges are slightly increased (IFF-PCFF-HQ in ref. [20]).

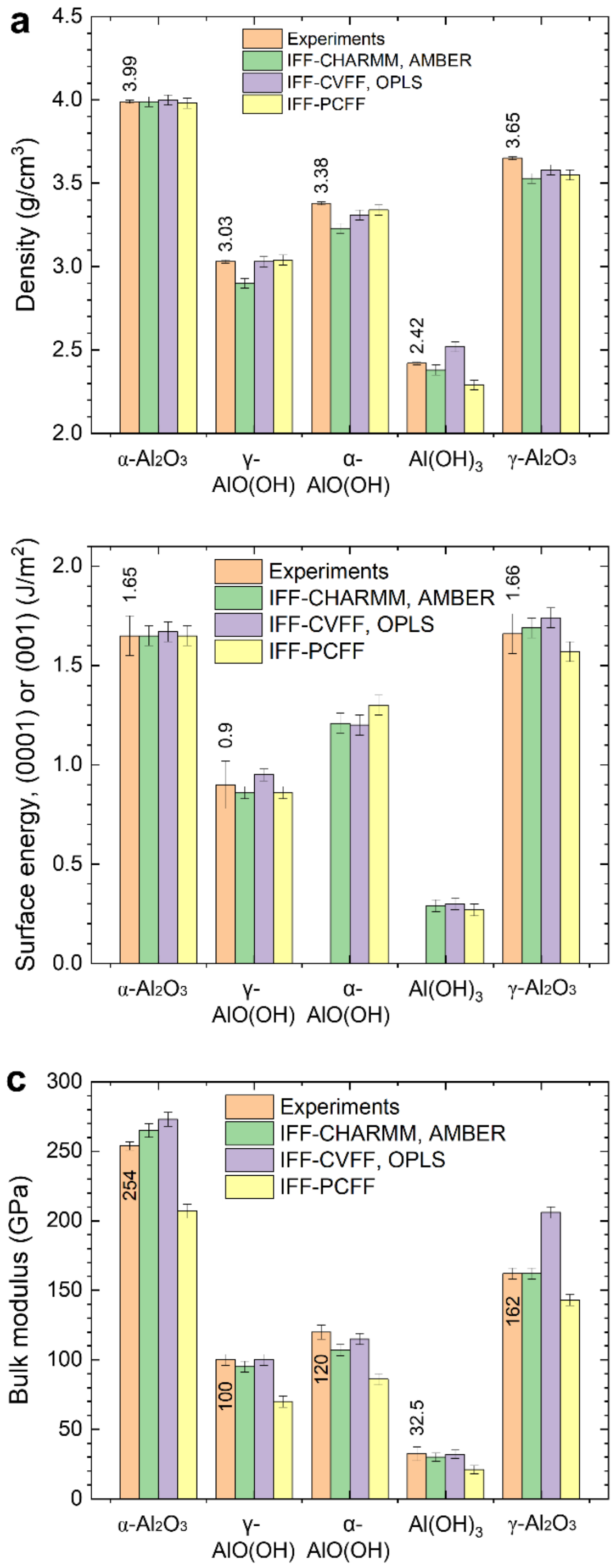

**Figure 4.** Comparison of computed bulk and surface properties in IFF MD with experimental reference data for the 5 alumina phases at 298 K. (a) Densities show 2.1% average deviation. (b) Surface energies of the (0001) plane (corundum) and (001) planes of other phases show 4.0%

average deviation. Experimental data are unavailable for α-AlO(OH) and $Al(OH)_3$. (c) Bulk moduli show 7% average deviation for 12-6 LJ. IFF-PCFF, COMPASS with 9-6 LJ options lead to lower bulk moduli. The overall excellent agreement illustrates the utility of the same nonbonded IFF parameters for multiple phases of aluminum oxides and hydroxides. Details of lattice parameters, surface energies, mechanical properties, and references to experimental data are given in Tables S1 to S6 in the Supplementary Material.

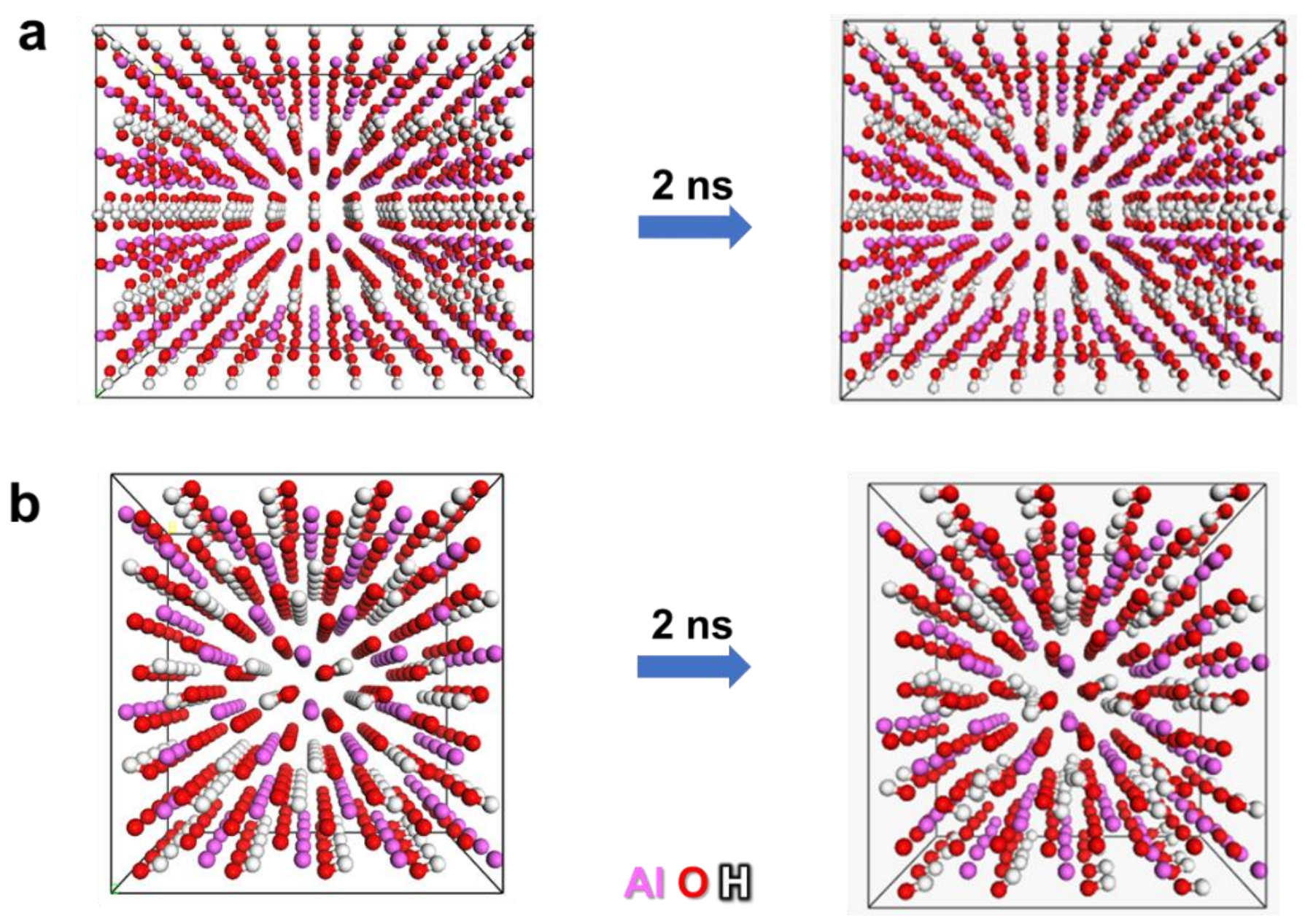

**Figure 5.** Visualization of structural stability during IFF MD simulations (IFF-CVFF) for boehmite and diaspore at 298 K. Starting and equilibrated structures after 2 ns are shown in perspective view. (a) Boehmite (γ-AlO(OH)) with a box size of 28.760 × 24.480 × 14.836 $Å^3$. (b) Diaspore (α-AlO(OH)) structures with a box size of 17.604 × 18.842 ×14.225 $Å^3$. Examples for other phases are included in the Supplementary Material.

### 2.4. Validation of Aqueous Interfacial Properties

α-$Al_2O_3$ undergoes surface hydration in water to form a molecularly thin $Al(OH)_3$ overlayer similar to gibbsite (Figure 6a).[78] The contact angle of this clean hydrated alumina surface is 0° in experiments[79] and we consistently computed 0° in IFF MD simulations (Figure 6a). Contact angles of 0° were obtained for multiple droplet sizes between 500 and 2000 water molecules, as well as for non-hydroxylated surface chemistries and different initial droplet structures (Figures S9 to S12 in the Supplementary Material). Experimental observations of nonzero contact angles on alumina surfaces are also known and have been consistently attributed to surface contaminations.[80] We have previously shown that the same protocol predicts nonzero contact angles for hydrophobic surfaces, e.g., 69±2° for water on $MoS_2$,[81] or 15° to 50° for epoxy precursors on *h*-BN.[82]

Aqueous interfacial properties from MD simulations were further compared with experimental data from crystal truncation rod (CTR) diffraction, which enable accurate determinations of the atomic layer positions near the surface, including contractions relative to the bulk layer spacing (Figure 6b, c).[78] The average layer spacing between the O atoms of the adsorbed water molecules (denoted as $O^{wat}$) from the terminal O layers on the hydrated α-$Al_2O_3$ (0001) surfaces ($O^{OH}$) is 2.30 Å in measurements versus ~2.5 Å in MD simulations (Figure 6b, c).[78] The expansion of interfacial $Al^{sur}$–$O^{OH}$ and $Al^{sur}$–$O^{bulk}$ atomic layer distances is captured well in MD simulations. Specifically, the 1.01 Å and 0.97 Å distances in layers 2-3 and 4-5 are larger than 0.84 Å in inner layers (layers 7-8 and 8-9) in CTR measurements (Figure 6b). MD simulations yield distances of 1.00 Å and 0.95 Å for layers 2-3 and 4-5, as well as 0.87 Å for layers 7-8 and 8-9 in excellent agreement (Figure 6c). Also, $Al^{sur}$–$Al^{sur}$ atomic layer distances are compressed to 0.23 Å compared to $Al^{bulk}$–$Al^{bulk}$ distance of 0.49 Å according to CTR measurements,[78] and consistently show 0.29 Å and 0.49 Å in MD simulations (layers 3-4 and 6-7 in Figure 6b, c). Overall, the α-$Al_2O_3$ (0001) surface models show high accuracy in reproducing the measured atomic layer distances with an average deviation of ~4%, including superficial expansion and compression of atomic layers.

Another perspective of the extended water structure at the interfaces was obtained by radial distribution functions between the surface oxygen atoms and water oxygen atoms in MD simulation (Figure 6d, e). In the radial view, water layers are more disordered and the first peak ranges from 2.4 Å to 3.5 Å. The layers are centered near 3.2, 4.7, 6.8, and weakly at 8.7 Å before approximating the bulk structure of water. The (hkl) wetting properties of γ-$Al_2O_3$, boehmite and diaspore also often involve partial or full hydration reactions leading to similar local gibbsite overlayers. Vice versa, all alumina phases convert to α-$Al_2O_3$ upon heating beyond 1200 °C via dehydration and reconstruction, respectively (Figure S13 in the Supplementary Material).

Gibbsite, in contrast, undergoes only physisorption and is practically insoluble in water. The exothermic hydration (immersion) energy of hexagonal gibbsite crystals of 120 nm width and 8 nm height was measured as 223 ±15 $mJ/m^2$ in experiments.[83] MD simulations of the immersion energy using (400) edge and (001) basal planes in the correct ratio yield 225 ±10 mJ/m$^2$ if assuming slightly smaller charges on the surface OH groups in vacuum (O -0.83e and H +0.29e). Hereby, the flexible SPC water model, TIP3P water model, and PCFF water model lead to the same results with <5% difference (Figure 6f), similar to earlier IFF benchmarks of hydration of silica surfaces.[63] Hydration energies are sensitive to the H charge of the superficial OH groups (see details in Section S5 in the Supplementary Material).

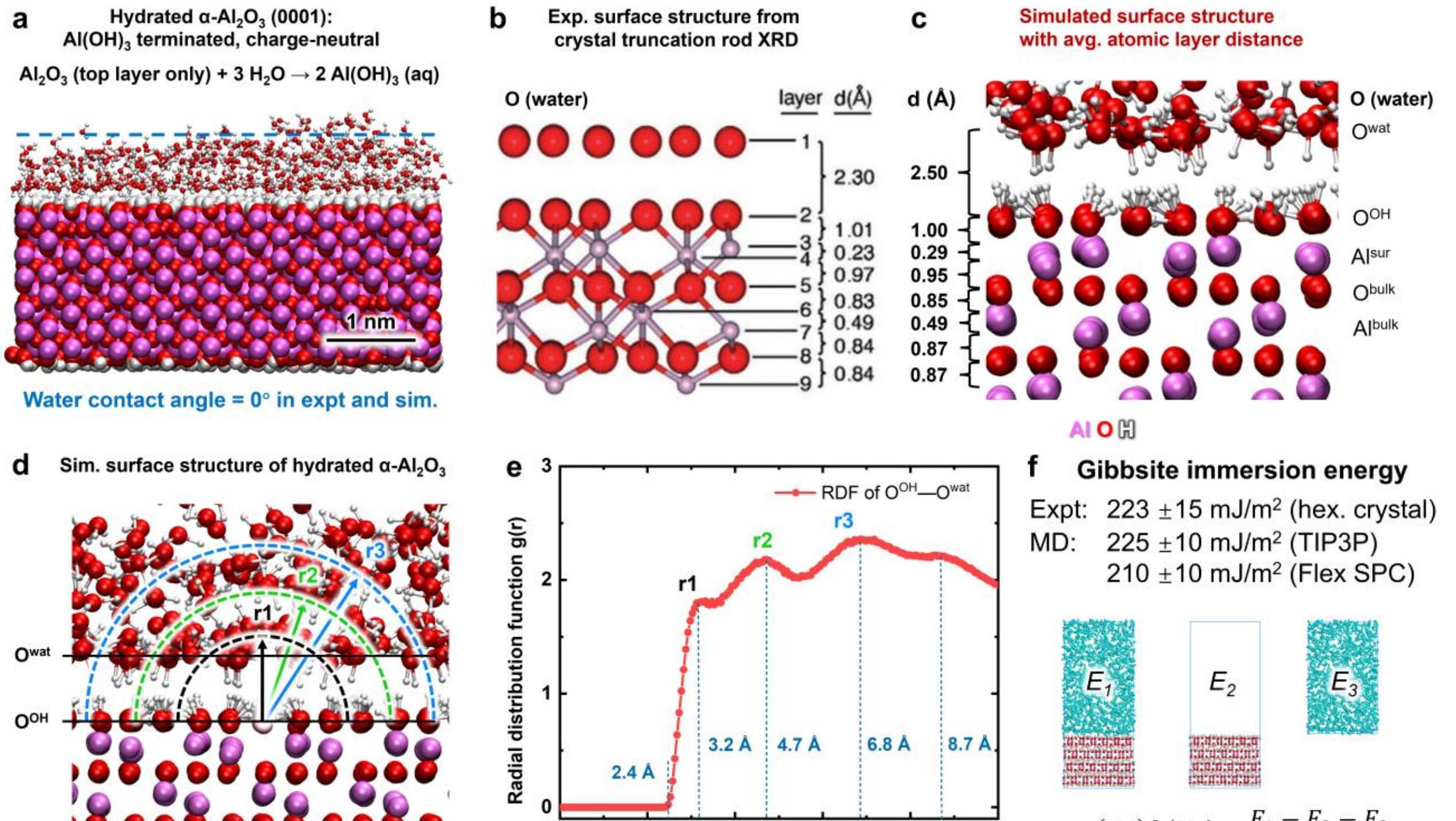

$$\Delta E_{imm}^{(400)\,\&\,(001)} = \frac{E_1 - E_2 - E_3}{2A}$$

**Figure 6.** Interfacial water structure of hydrated α-$Al_2O_3$ (0001) and gibbsite at 298 K. (a) The cleaved surface reacts with water to form a superficial monolayer of gibbsite, $Al(OH)_3$. The simulated water contact angle of 0º agrees with experimental measurements of 0º.[79] (b) Experimentally measured atomic layer distances from crystal truncation rod (CTR) diffraction.[78]

(c) The simulated atomic layer distances as a time average over 5 nanoseconds align well with the experimental data, including surface-specific expansions of $Al^{sur}$–$O^{OH}$ distances and contractions of $Al^{sur}$–$Al^{sur}$ distances. (d) MD Simulations also reveal the location of hydrogen atoms and dynamics of the water interfaces. (e) Radial distribution functions (RDFs) between O atoms on the surface and in bulk water show the formation of distinct layers. IFF-CHARMM parameters were used. (f) IFF MD simulations reproduce the experimental immersion energy of gibbsite nanocrystals in water (given without negative sign).[83]

### 2.5. A Surface Model Database as a Function of pH Value in Aqueous Solution and Electrolyte Dynamics

The surface chemistry is variable depending on the phase of alumina chosen and the pH value in solution.[69, 70] We created a surface model database for common (hkl) cleavage planes of the five alumina phases in vacuum prior to hydration and after exposure to electrolytes (Table 2, Table S7 and Supplementary Files in the Supplementary Material). The cleavage planes are of low surface energy and occupy a large area of crystallite surfaces (Figure S5 and Table S1 in the Supplementary Material).[20] Upon full hydration, all phases and nanostructures of alumina form a hydrated gibbsite-like $Al(OH)_3$ layer on the surface,[78, 84] which is sensitive to the pH value in the electrolyte and requires appropriately ionized surface models (Figures 7 and 8). We characterize these hydrated and ionized surface models for pH values from 2 to 12 based on surface titration measurements for the examples of $\alpha$-$Al_2O_3$ (0001) and $\gamma$-$Al(OH)_3$ (001) surfaces, and analogous models for the other phases can be built using the same protocol, starting with the equilibrium cleave planes in vacuum provided in the Supporting Files. These features have not been systematically incorporated in prior models and earlier simulation studies often led to qualitatively different or unphysical results.

**Table 2.** Guide to IFF surface models for alumina phases and common (hkl) planes after cleavage, hydration, and exposure to electrolytes as a function of pH values. Upon hydration, all surfaces feature a gibbsite-like $Al(OH)_3$ layer with an area density of OH groups between 13.5 per $nm^2$ (gibbsite) and 15.5 per $nm^2$ (corundum) at the point of zero charge of pH ~ 8. In contact with aqueous electrolytes, the surfaces undergo ionization as a function of the pH value across a wide range of positive to negative charge.

| **Alumina phase and (hkl) surface** | **Surface termination in vacuum** | **Surface termination upon hydration at pzc** | **pH value in solution** | **Surface ionization ($e/nm^2$)[a]** | **Percent of ionic OH surface groups** |
|---|---|---|---|---|---|
| $\alpha$-$Al_2O_3$ (0001), (10-10), (11-20) | Al or O | $Al(OH)_3$ | 2 | +1.75 ±0.1 | ~12% $OH_2^+$ |
| $\gamma$-$Al_2O_3$ (001) | Al and O | | 5 | +0.9 ±0.1 | ~6% $OH_2^+$ |
| α-AlOOH (100), (010), (001) | O and OH | | 8.1 ±0.5 | 0 ±0.1 | ~0% (OH only) |
| γ-AlOOH (100), (010) | O and OH | | 10 | -0.75 ±0.1 | ~5% $Al_2O^-$ |
| γ-$Al(OH)_3$ (001) | $Al(OH)_3$ | | 12 | -1.5 ±0.1 | ~10% $Al_2O^-$ |

[a] Based on experimental data from surface titration and zeta potential measurements in refs. [69, 70].

To derive these surface models, we combined knowledge of surface chemistry,[85, 86] acid-base concepts, zeta potential measurements,[57, 71, 87] and quantitative surface titration data.[69, 70] The gibbsite-like hydrated alumina surfaces[78, 84] are prone to protonation and deprotonation reactions (Figure 7a).[85, 86] Zeta potential measurements[71] and surface titration measurements[69, 70] have characterized the surface charge density as a function of pH value, indicating a point of zero charge (pzc) at pH ~ 8.1 ±0.5 for α-alumina and γ-alumina. Earlier studies also report a broader range of the pzc between pH 7 and 9 (see Section S6 in the Supplementary Material). Protonated surfaces with $[Al_2 \cdots OH_2^+]$ groups are present at pH < 8 (Figure 7b) and begin to dissolve into $Al^{3+}(aq)$ species at pH values below 4.5. At pH > 8, we find deprotonated surfaces with $[Al_2 \cdots O^-]$ groups (Figure 7b) that begin to dissolve into $Al(OH)_4^-$ ions at pH values above 9.4.[70] The presence of $Al(OH)_3$ overlayers and $[Al_2 \cdots OH_2^+]$ and $[Al_2 \cdots O^-]$ surface groups, and the absence of potential geminal $Al_2(OH)_2^-$ groups, is supported by XPS[84] and FTIR studies.[88, 89] The coordination environment of the superficial $OH_2^+$, OH, and $O^-$ species involves two nearest Al neighbors through a 3-center coordination bond (Figure 7b, c-e). This geometry is distinctly different, for example, from SiOH groups on silica surfaces with localized Si-OH bonds[63] and reflects the increased ionic character in alumina species. We represent these chemistries and coordination environments in the nonbonded models by adjusting the atomic charge distribution and adding the necessary counter ions on the surface layers for charge neutrality (Figure 7c and Section S4 in the Supplementary Material). The ions can be exchanged for other ions or buffer ions that control the pH value.

The IFF alumina surface model database utilizes experimental surface titration data (Figure 8a, Table 2, Figure S14 and Table S7 in the Supplementary Material)[69, 70] to quantitatively represent the amount of protonated and deprotonated sites. The area density of OH groups on hydrated (0001) α-alumina and on (001) gibbsite surfaces at the point of zero charge is 15.3 per $nm^2$ and 13.6 per $nm^2$, respectively (Figure 8b). Across the pH spectrum from 2 to 12, the surface charge density (σ) ranges from approximately +1.75 $e/nm^2$ to 0 at pH ~8, and then from 0 to -1.5 $e/nm^2$ (Figure 8a-c). The drastic changes in (hkl) surface chemistry of the alumina phases in contact with aqueous electrolytes require specifically designed models at a given pH value for predictive simulations (Figure 8b, c). Positive ionization or negative ionization up to about 1.5 groups per $nm^2$ with ±1.5 elementary charges per $nm^2$ equals to ionization of up to ~10% of all OH groups (Figure 8c and Table 2). The precise surface charge density also depends on the electrolyte composition and concentration (Figure S14 in the Supplementary Material).[69] We assume NaCl solution between 0.0 to 0.5 M concentration, and customization to other electrolytes as needed.

The charge density per area can reach twice that of silica surfaces, which have been extensively studied and shown deterministic effects of the charge density on adsorption, although with a pzc of 3 rather than 8.[63, 90, 91] The wide range of alumina surface charge drastically changes the type of species attracted, for example, molecules with R-$COO^-$ groups vs R-$NH_3^+$ groups, their binding affinity, packing density, and self-assembly. Utilizing adequate surface models is thus critical regardless of the simulation method employed.

Our guidance for surface models can be applied to all alumina phases in contact with electrolytes as they exhibit similar surface chemistry and point of zero charge (Table 2).[71, 84] The workflow to construct customized models of surfaces or nanoparticles with a specific surface charge density is as follows: (1) Build a model of the nanostructure or (hkl) surface with full (or partial) hydration, e.g., charge-neutral $Al(OH)_3$ termination, including correct atom types and charges. (2) According to the desired pH value, decide on the type and area density of ionized surface groups (Table 2, Figure 8c, Figure S14 and Table S7 in the Supplementary Material). (3) Build a model of sufficiently large surface area and identify a statistical distribution of the charged sites (Figure S15 in the Supplementary Material). (4) At the identified sites, add or remove atoms, adjust the atomic charges and modify the atom types as needed (Figure 7c). (5) Double check for charge neutrality and correct, stoichiometric modifications made.

To demonstrate the effect of pH-dependent surface chemistry on the structure and dynamics of the solid-liquid interfaces, we computed the density profiles of water, sodium ions, and chloride ions at pH values of 2, 5, 8, 10, and 12 (Figure 8d, e). The water density profile indicates major peaks at 2.5 Å and 5.8 Å away from superficial O atoms in contact with Al, as well as a minor 3rd peak near 8.8 Å (Figure 8d inset). The water density profile was not affected by the pH value in a major way since ionization rarely surpasses 15% of AlOH groups (Table 2). At low pH values of 2, the major share of superficial chloride ions remains bound to the $AlOH_2^+$ groups and only 13% dissociate, however, at pH values of 5, 45% or ~0.4 $Cl^-$ ions per $nm^2$ dissociate more than 5 Å away from the surface (Figure 8e and Table S8 in the Supplementary Material). A small fraction of $Cl^-$ ions is even found up to 20 Å away from the alumina surface. Thus, a pH value close to the pzc of 8 results in dissociation of a higher fraction of ions from the surface, and larger depth of the solid-electrolyte interface, than pH values further away from the pzc. This trend is related to the lower overall surface charge per unit area and consistent with a steep onset of the positive zeta potential in experiments near the pzc, while the zeta potential curve flattens at pH values further away from the pzc.[57, 87] A similar trend is seen at high pH values for the dissociation of $Na^+$ ions (Figure 8d and Table S8 in the Supplementary Material). Hereby, the negative surface charge at high pH is more concentrated on singular O atoms, as opposed to delocalization of the positive charge over two H atoms and one O atom at low pH, and sodium ions are smaller than chloride ions. Therefore, $Na^+$ ions are more strongly bound and only about 10% dissociate more than 5 Å away from the surface. Consistently, the negative zeta potential at high pH values tends to be smaller than the positive zeta potential for the same amount of positive surface charge at low pH values in the absence of other added electrolytes (-25 to -40 mV at high pH vs. + 30 to +80 mV at

low pH).[57, 87]

The timescale to reach equilibrium of the charged solid-electrolyte interfaces was unusually long at ~20 ns, and ~200 ns were required to obtain a well-resolved statistical analysis of the density profiles (Figure 8d). In comparison, the time to reach the diffuse regime of unconstrained $Na^+$ and $Cl^-$ ions in a 0.1 M NaCl solution was only ~0.5 ns, and the ions would undergo ~2 nm displacement during this time. Accordingly, reversible dissociation of counterions from the alumina surface into solution and re-association with the surfaces is one to two orders of magnitude slower than unconstrained diffusion, associated with the surface charge density and higher barriers for exchange. Lateral mobility of the ions across the surface occurred on a timescale of nanoseconds. The models enable detailed follow-on studies of the unique electrolyte dynamics at alumina-solution interfaces.

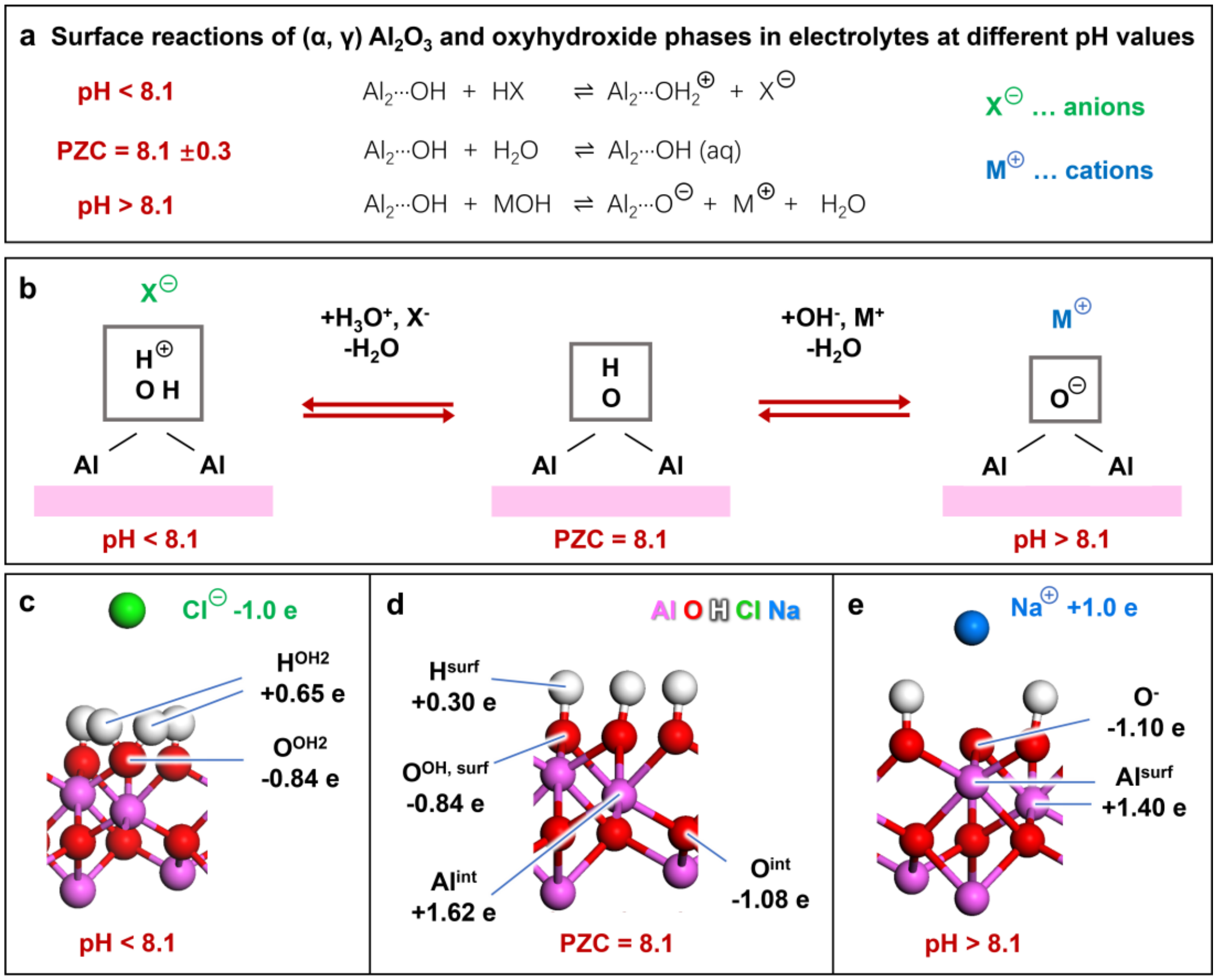

**Figure 7.** Surface chemistry of alumina and oxyhydroxide phases at different pH values. (a) Hydration of minerals leads to formation of superficial gibbsite layers with $Al_2OH$ surface groups. These layers are charge-neutral at the point of zero charge at pH ~8.1. The [$Al_2$ ··· OH] groups undergo protonation reactions at lower pH values and deprotonation reactions at higher pH values.[69, 70] (b) Illustration of the chemical environment of surface terminal groups. Terminal oxygen species have two nearest neighbor Al ions rather than localized Al-OH bonds. (c-e) Representation in IFF atomistic models, for the example of a hydrated α-$Al_2O_3$ (0001) surface, including modifications of atomic charges. Al-O bonds are included to illustrate spatial relationships in the partially covalent bonds. Atomic charges of the neutral OH-terminated surface (d) are modified locally for $OH_2Cl$ defects (c) and $Al_2ONa$ defects as indicated.

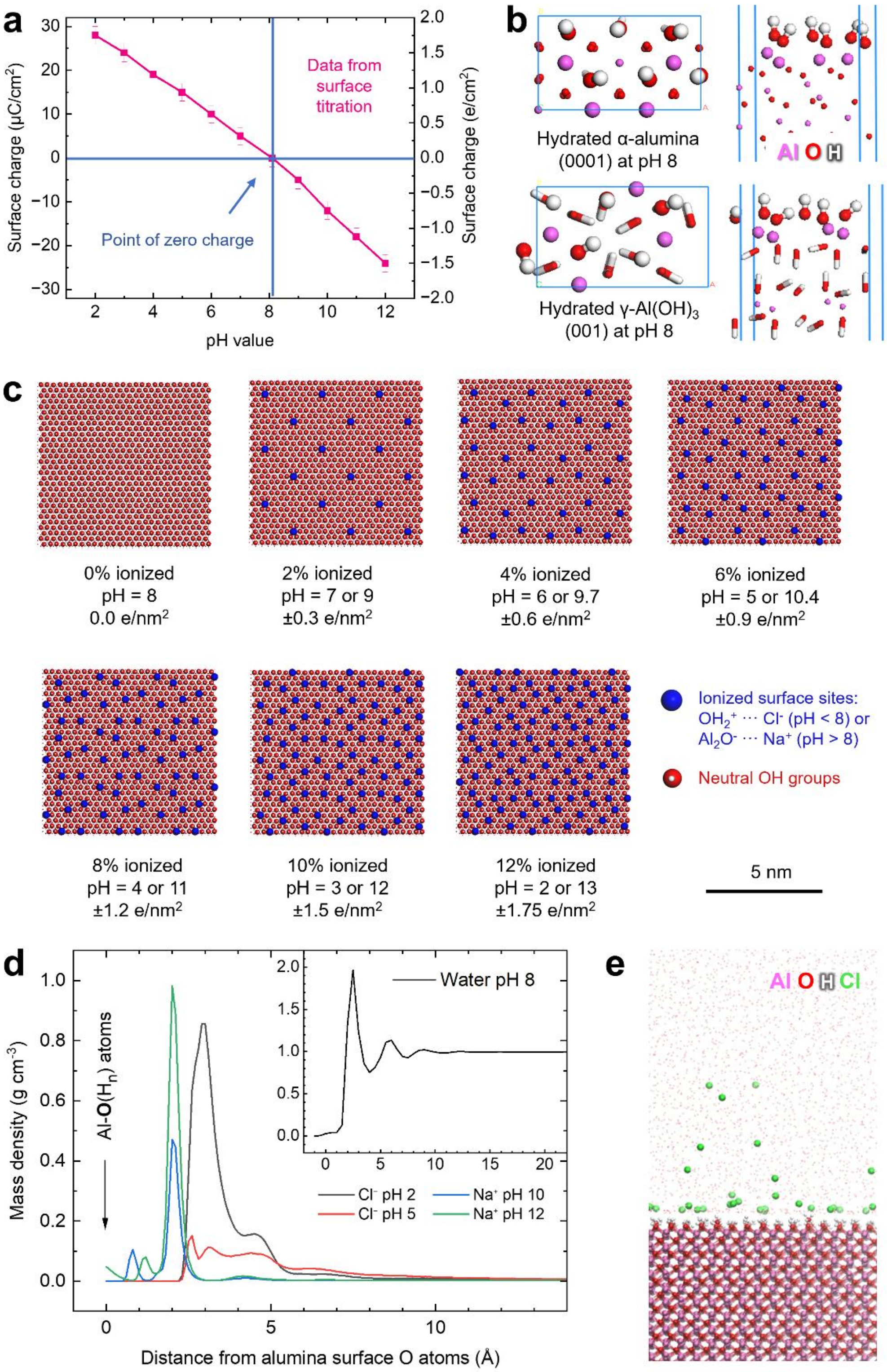

**Figure 8.** Surface charge density of hydrated alumina phases, implementation in atomistic models,

and influence on electrolyte interfaces. (a) The average surface charge density of $\alpha$-$Al_2O_3$ (0001) surfaces in NaCl solution (0 to 0.5 M) as a function of pH value according to surface titration experiments.[69] The slope of the curve is lightly flatter for positively charged surface species at pH < 8.1 than for negatively charged surface species at pH > 8.1. Gibbsite and other hydrated alumina phases follow the same trend. (b) Unit models of electroneutral $\alpha$-$Al_2O_3$ (0001) surfaces and $\gamma$-$Al(OH)_3$ (001) surfaces used to build larger ionized surfaces at specific pH values in (c). On the upper surface, the outer molecular layer of $Al(OH)_3$ is shown in ball-and-stick rendering to distinguish from the rest of the structure. (c) Top view onto $\alpha$-$Al_2O_3$ (0001) surface models for pH values from 2 to 13, showing the range of density of positively or negatively charged groups up to 1.75e per $nm^2$. Incoming molecules or ions near the surface accordingly adapt their interactions. The surface area shown is 24.729 × 23.7945 $Å^2$. (d) Density profiles of electrolytes at distinct pH values illustrate the effects of solution pH on the binding properties at 298 K. Chloride ions at low pH value are bound to $Al$-$OH_2^+$ groups on the surface and diffuse into solution (past 0.5 nm), especially at pH values close to the pzc. Smaller sodium ions at high pH value are bound strongly to $AlO^-$ groups, which have a highly localized negative charge, with a smaller fraction dissociated past 0.5 nm away from the surface. The inset shows the water density profile at pH 8 (no ionization), which remains essentially the same at different pH values. (e) Representative snapshot of the interfacial region at pH = 2 at 298 K with a distribution of positively charged surface sites and chloride counterions that penetrate the solution phase past 1 nm.

## 2.6. Adsorption of Corrosion Inhibitors

The pH-resolved surface models of alumina are therefore critical to predict the adsorption of molecules, polymers, reactants in catalysis, or corrosion inhibitors.[92] As an example, we consider the adsorption of the corrosion inhibitor p-hydroxybenzoic acid on alumina surfaces, for which detailed adsorption isotherms have been measured in experiments between pH 5 and 9 (Figure 9a).[57] The laboratory data show order-of-magnitude differences in the adsorbed amount that are related to changes in alumina surface chemistry (Figure 8). The alumina samples had a pzc of ~7, i.e., slightly below the common value of 8.[57]

For MD simulations, accordingly, we defined zero-charge surface models to be equivalent to pH 7, shifting the pH of the models in the database downwards by one unit (Table 2). The pH value in the simulations was set by modifying the surface chemistry (Figure 8c) for two pH units below and above the pzc. These surface modifications represent the influence of pH value in the models rather than the inclusion of $H_3O^+$ or $OH^-$ ions due to their low concentrations at these pH values (e.g., 5.5 million water molecules would be required for each $H_3O^+$ ion at $10^{-5}$ M at pH 5).[63]

To characterize interfacial dynamics, we quantified the residence time of organic adsorbate at the surface using a distance criterion of 4 Å. The adsorbed amounts of p-hydroxybenzoic acid in MD simulations at pH 5, 7, and 9 quantitatively correlate with experimental adsorption isotherms (Figure 9b-g). Strong adsorption occurred at pH ~ 5 with a residence time of 83% and a binding free energy of -2.6 kcal/mol ($\boldsymbol{\Delta G = -RT\ln K}$ with $\boldsymbol{K = \frac{t_{ads}}{t_{des} \cdot c_{sol}}}$ with $c_{sol}$ = 0.064 M) in very good agreement with -3.6 kcal/mol calculated from the experimentally measured Langmuir adsorption constant ($\boldsymbol{K_S = 0.42 \frac{l}{mmol} = 420\ M^{-1}}$) [57] (Figure 9b, c, see details in Section S7 in the Supplementary Material). The attraction originates from ion paring of the negatively charged carboxylate groups with the positively charged $Al_2(OH_2)^+$ groups on the alumina surface. At the point of zero charge at pH ~ 7 and for increased alkalinity at pH ~ 9, the adsorbed amount decreases sharply in experiments from 1.8 molecules per $nm^2$ to less than 0.2 molecules per $nm^2$ (Figure 9a), and the contact time decreases sharply in MD simulations (Figure 9c-g). Ion paring was no longer feasible at the pzc due to the absence of surface charge, and the presence of a negative surface charge at pH ~9 locally repels negatively charged benzoate groups (the molecules are deprotonated). The contact time of p-hydroxybenzoate with the surface decreased to 13% and 9%, respectively (Figure 9d, f). Temporary contacts involved dipolar interactions and hydrogen bond interactions, which add minor relative binding strength due to surrounding water molecules. The phenolic OH group is partly deprotonated at pH 9 (which we accounted for in the simulation, see Figure 9g). The binding free energies at pH 7 and 9 remain slightly negative at -0.5 and -0.25 kcal/mol, respectively, consistent with minor attraction to the alumina surface in experiments in the correct order (Figure 9a). Further simulations at pH = 3 predicted an increase in binding of p-hydroxybenzoic acid with 98.8% of time bound and ΔG (sim) = -4.2 kcal/mol (Figure 9c). The data was reproducible in independent simulations with uncertainties between ±1% to ±4% in contact times and ±0.2 to ±0.5 kcal/mol in associated binding free energies (Figure 9b-g).

Quantitative, pH-resolved and time resolved 3D molecular-scale insights on alumina surfaces are reported here for the first time to the best of our knowledge, and no prior tools for realistic MD simulations of such systems have been available. The accuracy of IFF MD simulations equals that of silica-organic interfaces reported earlier[90, 91] and detailed experimental measurements for validation are rarely available. In our computational analysis, we found some limitations of the classical Langmuir models of adsorption, which suggest high binding constants across the range of pH values (5, 7, 9) despite steeply decreasing adsorbed amounts (see ref. [57] and Section S7 in the Supplementary Material). These limitations appear to lie in theory rather than in the measurements.

Parameter development and validation for solvents, organic molecules, and biopolymers, as a subset of the periodic table, follow the same concept of the IFF workflow that is suited for compounds across the periodic table.[30] Numerous examples of simulations of inorganic-organic systems in chemical accuracy (<1 kcal/mol, sometimes <0.1 kcal/mol) are known for over two decades,[20, 30, 61, 62, 90, 91, 93-96] including a recent IFF implementation in CHARMM-GUI,[40] and have been previously reviewed.[31] The present study directly evaluates molecular adsorption for p-hydroxybenzoic acid and its pH-dependent protonation states at alumina–water interfaces, with comparison to experimental adsorption data. Applications to other organic molecules, polymers, and biomolecules are enabled by compatibility with established molecular force fields, and their quantitative accuracy at alumina interfaces remains to be assessed through system-specific validation.

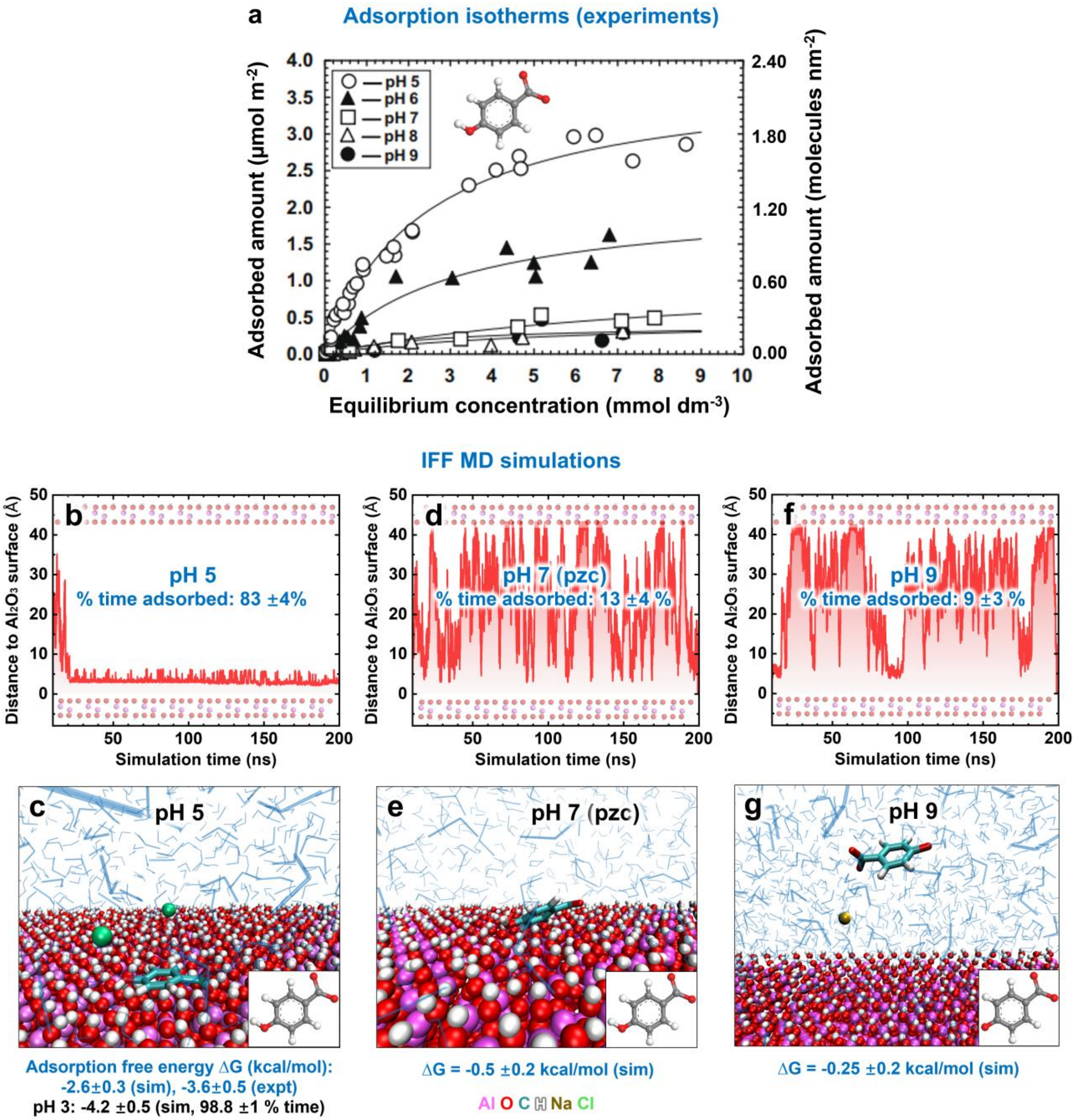

**Figure 9.** Adsorption of p-hydroxybenzoic acid, a corrosion inhibitor, according to experiments[57] and simulations in aqueous solution at different pH values at 298 K. (a) Experimental adsorption

isotherms[57] indicate a strong dependence of the adsorbed amount on the acidity, ranging from near monolayer coverage at pH 5 (~2 molecules per $nm^2$) to weak temporary adsorption (<0.2 molecule per $nm^2$) at pH 8 and 9 (reproduced with permission from ref. [57]). (b-g) MD simulations with surface models of 0.60 $OH_2^+$ groups per $nm^2$ (pH 5), full OH termination (pH 7, pzc), and 0.75 $Al_2O^-$ groups per $nm^2$ (pH 9) show changes in adsorption of single molecules in excellent agreement with experimental adsorption isotherms. Representative traces from selected MD trajectories (b, d, f) show statistical interactions of the molecules with bottom surfaces, top surfaces (equivalent), and preferences towards solution over time. The average time in contact with the surfaces (< 3 Å) was high at pH 5 and decreased to transient adsorption at pH 9. Computed free energies of binding from the equilibrium constants agree with laboratory data at pH 5. The typical pzc of ~8 for hydrated alumina surface models was shifted to the sample-specific pzc of ~7.

### 2.7. Computational Speed Compared to DFT/ML and Foundation Models

IFF MD is between $10^4$ to $10^5$ times faster than DFT machine-learned interatomic potentials (MLIPs) and foundation models for typical system sizes of 10,000s of atoms, and the speedup further increases for larger systems due to lower memory demand. In comparison to DFT-level ab-initio MD for these system sizes, the speedup is on the order of $10^{10}$ to $10^{11}$ times.

Specifically, the wall time for IFF MD simulations of a typical system of 30,000 atoms (e.g., 6,000 alumina units or 10,000 water molecules) is ~300 ns/day on one RTX 3080 or RTX 4090 graphics card.[97, 98] ML-DFT methods such as DeepMD[99]/NequIP[100] for a 15,600-atom protein achieve 0.1 ns/day using a cluster of eight A100 graphics cards, which are equal in price to 48 RTX 4090 cards (6 RTX 4090 ~ 1 A100). Accordingly, IFF-MD is 144,000 times faster than these DFT-ML methods per GPU unit, and requires orders of magnitude less memory.[101] In addition, the ML/DFT potential requires significant upfront training time and convergence tests which are not included in this comparison.

As another example, MD simulations with UMA and MACE models of 572 water molecules (1716 atoms) with an RTX 4090 graphics card yielded ~0.1 ns/day and 0.5 ns/day, respectively.[102] The speedup by using IFF exceeds 17,000x, normalized for system size (30,000 atoms rather than 1,716). In addition, the errors in equilibrium water density with UMA and MACE models are ~15% relative to experiments (0.85 g/cm$^3$) and the melting point is over 40 °C.[102] Realistic simulations of aqueous interfaces are thus not feasible, even after $10^4$ times slower speed and high memory demands. Standard water models (flex. SPC, TIP3P) in IFF have <1% error in liquid density and in melting points (in Kelvin).

## 2.8. Opportunities and Limitations

The models can be extended to amorphous and porous alumina surfaces to investigate bulk properties as well as interactions with water and polymers.[103] γ-alumina, included here, provides an example for disordered surfaces, and thin amorphous alumina films on battery cathodes were recently studied with IFF MD.[13] Models of amorphous alumina glasses can be generated, for example, by melting models of α-alumina at temperatures of 2500 K to 3000 K, followed by instant supercooling to and relaxation at room temperature. Subsequent cleavage to generate surfaces should observe local electroneutrality and use annealing to allow reconstruction of the surfaces towards low energy.[104, 105] Implementing surface hydration and pH effects after cleavage and surface relaxation follows our protocol of stoichiometric hydration (Figure 3) and ionization of an appropriate percentage of surface $Al_2(OH)$ groups depending on pH value in solution (Figure 8). The degree of surface hydration can be informed by experimental data for specific use cases since thermal treatment (at high temperatures) modifies the area density of aluminol groups (Figures S12 and S13 in the Supplementary Material). In contact with electrolyte, we typically assume equilibrium hydration, which leads to gibbsite-like surfaces (Figure 8). The workflow to generate hydrated surfaces is the same for all types of alumina surfaces, and the nonbonded alumina models are more flexible to capture atomic reconstruction compared to bonded silica models.[63] Modeling amorphous and porous alumina interfaces can advance molecular understanding of catalysis, ceramics/polymer composites, batteries and electronics.

The experimentally informed models for pH-resolved alumina surfaces can be utilized for simulations with force fields and quantum methods. Protonation states are fixed, require stoichiometric additions of counterions and adsorbates for charge neutrality, and do not adapt in a reactive way. While "adaptive" and "reactive" changes are desirable, such implementations are challenging and hardly possible for the following reasons: (1) The concentration of $H_3O^+$ and $OH^-$ ions that function as physical carriers are too low at typical pH values. Even when using a model system with 100,000+ water molecules, addition of a single $H_3O^+$ ion (0.5 mM) drops the pH value from 7 to < 3. (2) The critical parameter that determines the structure of electrolytes and strength of molecular adsorption as a function of pH is the area density of surface charge, not its distribution (Figure 8c,d and Figure 9). Prior studies on pH-sensitive silica and phosphate surfaces have consistently shown that dynamic changes in distribution are second or third order in importance.[63, 68, 90, 106-108] While a statistical distribution of ionized groups is helpful, the influence of a less ideal distribution is lower than a changing the total area density equivalent to 1 pH unit. If desired, using two or more models with different statistical distributions of the ionized sites can provide specific quantitative insight at nominal computational cost. (3) Equilibration of charged, pH-resolved solid-electrolyte interfaces requires $>10^4$ atoms and time scales of ~100 ns, necessitating accurate and computationally efficient methods.

IFF was developed for standard conditions of 298 K and 1 atm pressure as a reference state. The alumina parameters are valid within a temperature range from 0 to 600 K with negligible loss in accuracy. As shown for similar oxides and metals, the parameters still perform well at temperatures above 1500 K[20] and beyond the melting point.[109] We anticipate shifts in surface energies by ~10% relative to experiment under these conditions and temperature-dependent adjustments of the LJ parameters can be made if necessary.

The models allow stepwise modeling of reactions, vacancy migration, and transition state probabilities.[20] Simulations of metal dissolution and deep surface reconstruction, involving reactions of water molecules with the alumina surface[110] are challenging, however, and could be modeled at discrete reaction stages corresponding to the long time scales. The reactive INTERFACE force field (IFF-R)[32] replaces remaining covalent bonds (e.g., in water, OH groups, or reactive organic molecules) with shifted Morse potentials, ready to simulate mechanical failure (including bond breaking) without further parameters. IFF-R can also be combined with DFT-level or CCSD(T) calculations at the local scale to incorporate changes in electronic structure and provides a path to model specific reactions upon hydration or cross-linking using reaction templates.

## 3. Conclusion

We present a unified and rigorously validated set of INTERFACE Force Field (IFF) parameters and surface models for aluminum oxides and hydroxides that enable quantitative predictions of bulk and interfacial properties across all major alumina phases. These primarily nonbonded models capture the lattice parameters, densities, surface energies, bulk moduli, and water contact angles of α-$Al_2O_3$, γ-$Al_2O_3$, diaspore, boehmite, and gibbsite with average deviations from experimental data of 0.8%, 2.1%, 4%, 7% and 0%, respectively. Simplicity and transferability eliminate the need for phase-specific reparameterization and ensure broad compatibility with CHARMM, AMBER, OPLS-AA, CVFF, and PCFF force fields to predict electrolyte interactions and organic assembly.

Beyond structural and energetic fidelity, we introduce a surface model database to represent equilibrium cleavage planes, pH-dependent surface chemistry, and charge distributions that are essential for realistic simulations of aqueous and organic alumina interfaces. The pH-resolved surface models bridge experimental titration and zeta potential data with atomistic resolution, enabling correct representation of protonated, neutral, and deprotonated alumina surfaces across a wide pH range. The surface charge can vary from approximately +1.5 e/nm$^2$ to -1.5 e/nm$^2$ and full equilibrium of the solid-electrolyte interfaces is reached upon a time scale of ~100 ns, about one to two orders of magnitude longer than for unconstrained diffusion. This framework is indispensable to study alumina-electrolyte interfaces and, beyond the corrosion-inhibitor adsorption examined here, offers opportunities to investigate adsorption of coatings, drugs, catalytic interfaces and interactions with cell membranes, with additional validation. Reactive models (IFF-R) and electronic structure representations can be embedded where required.

Compared to previous force fields and DFT/ML hybrid approaches (MLIPs), the IFF framework delivers higher accuracy, interpretability, and computational efficiency. It routinely supports model sizes of tens of nanometers and timescales beyond 100 ns, about $10^4$ to $10^5$ times faster than MLIPs and $10^{10}$ times faster than DFT methods, while maintaining atomic-level realism. The models are thus well suited to explore complex multiphase systems involving electrolytes, dopants, organic molecules, and polymers under experimentally relevant conditions.

In summary, the alumina models are a major advance in atomistic simulations and predictive design of alumina-containing materials across diverse areas from catalysis and corrosion protection to electrode coatings, implants, and biomedical applications. The inclusion of realistic, pH-resolved surface chemistry opens a pathway toward multiscale integration of chemical accuracy and materials functionality. The methodology can be readily extended to doped alumina, mixed-metal oxides, and emerging oxide–polymer hybrid systems.

# 4. Computational Methods

## 4.1. Building of Molecular Models

Models of the aluminum oxides and hydroxides with 1.5–3.0 nm side length were constructed from multiple supercells of the unit cells obtained by X-ray diffraction. The graphical user interface of Materials Studio was utilized to build and visualize all-atom models.[111] The models were larger than 12 Å in three directions for all simulations, and vacuum slabs of at least >50 Å in the vertical direction were utilized for the simulation of surface energies and contact angles to exclude interactions with periodic images in this direction.

## 4.2. Force Field Parameters

The IFF parameters were developed using the energy expressions of common class I and class II force fields. The IFF parameters are compatible with several class I force fields such as CVFF,[49] CHARMM,[64] Dreiding,[65] AMBER,[34] and OPLS-AA.[36] The energy terms include a 12-6 Lennard-Jones potential with Lorentz-Berthelot combination rules (arithmetic $r_{min,\, ij}$) or geometric combination rules (Equation (1)). IFF parameters are also compatible with class II force fields such as CFF,[37] PCFF,[38] and COMPASS.[39] Then, the energy terms include a 9-6 Lennard-Jones potential with Waldman-Hagler combination rules (Equation (2)). Class II force fields differ from class I force fields by additional cross-terms and higher order terms for bonded potentials. These additional terms are not necessary for or interfering with alumina phases and can be kept for originally included compounds.[37, 62] The workflow involved over 2000 independent energy minimizations and MD simulations to test the sufficiency of the α-$Al_2O_3$ parameters, derive the parameters for OH groups and associated pH-sensitive modifications, and validate the key properties (density/lattice parameters, surface energies, and bulk moduli) (Figure 2a). A summary of key parameterization targets and convergence criteria is shown below (Table 3).

**Table 3.** Overview of key parameterization targets, their roles, and convergence criteria.

| **Property class** | **Target property** | **Parameterization role** | **Target accuracy** | **Remarks** |
|---|---|---|---|---|
| Electrostatics | Atomic charge consistency, trends | Constrained by physics/chemistry | Within experimental/ literature range | XRD-derived trends, chemical consistency |
| Bulk structure | Lattice parameters, density | Primary fitting target | ≤1% | All phases; high priority for transferability |
| Surface properties | Surface energies | Primary validation target | ≤5% | Key (hkl) planes; critical for interfaces |
| Mechanical properties | Bulk modulus, Young's modulus | Secondary validation target | ≤5–10% | Ensures elastic modulus consistency |
| Interfacial properties | Contact angle, CTR-XRD, solvation energy | Prediction and further validation | Qualitative or ≤5% (where available) | Non-fitted, ready for materials design |
| Organic interactions (pH-resolved) | Binding energy, assembly, mobility | Prediction and further validation | ≤5–10% | Non-fitted, ready for materials design |

### 4.3. Molecular Dynamics Simulation of Crystal Structures and Lattice Parameters

For the simulations of crystal structures and bulk modulus, MD simulations were carried out in the isothermal-isobaric ensemble (NPT), and the canonical ensemble (NVT) was employed for the simulations of surface properties. The Discover program in Materials Studio and the Large-scale Atomic/Molecular Massively Parallel Simulator (LAMMPS) were applied when using IFF-OPLS/CVFF,[111] and IFF-PCFF,[112] and the Nanoscale Molecular Dynamics (NAMD) program was employed for IFF-CHARMM.[113] Regardless of the software used, a spherical cutoff of 12 Å was applied for the summation of pair-wise Lennard-Jones interactions, which is the standard IFF setting. Ewald methods such as Ewald summation in the Discover program, Particle Mesh Ewald (PME) in NAMD and particle-particle particle-mesh solver (PPPM) in LAMMPS in high accuracy of $10^{-6}$ were employed for the summation of electrostatic (Coulomb) interactions, which is also the standard IFF setting. Simulations were conducted for 2 ns and a timestep of 1 fs to achieve equilibrium, whereby all atoms were allowed to move freely (no fixed atoms). All simulations were repeated at least 3 times to obtain average results.

Though several settings are slightly different in these three simulation programs, they yield the same results (properties) within a typical difference of 0–2%, which is within the simulation errors. When using the Discover program, the temperature was controlled at 298 K using the velocity scaling thermostat with instantaneous fluctuations of ±10 K, and the pressure was maintained at 1 atm with the Parrinello–Rahman barostat. When using the NAMD program, the temperature was controlled at 298 K using the Langevin thermostat with a damping coefficient of 1 $ps^{-1}$. The average temperature remained within ±0.5 K of the target temperature with instantaneous fluctuations of ±20 K. In LAMMPS, the temperature was controlled at 298 K using the Nose-Hoover thermostats with temperature damping every 100 fs.

### 4.4. Simulation of Surface Energies

Surface free energies $\gamma_{SV}$, also called solid-vapor surface tensions, were computed using two NVT simulations, one with a 3D slab before and one with a 3D slab after cleavage. Before cleavage, a bulk slab with a length larger than 12 Å (typically > 20 Å) in all three directions was used. After cleavage, a vacuum slab of >50 Å length in the cleavage direction was created to expose the surface. For a given compound, these two models had the same number of atoms and the same equilibrium size in the *x* and *y* directions. The two model systems were subjected to NVT MD simulation at 300 K for 2 ns to reach equilibrium. The difference in average total energy between the non-cleaved structure $E_{NC}$ and the cleaved structure $E_C$ equaled the surface free energy per surface area $A$:

$$\gamma_{SV} = \frac{E_C - E_{NC}}{2A} - T\frac{S_C - S_{NC}}{2A} \approx \frac{E_C - E_{NC}}{2A}. \quad (3)$$

The entropy contribution $-T\frac{S_C - S_{NC}}{2A}$ is negligible compared to the cohesive energy contribution $\frac{E_C - E_{NC}}{2A}$ in the metal oxides, supported by minor oscillations of the metal atoms around their lattice points in the bulk and at the cleaved surfaces.[42] The entropy term was estimated to contribute less than 0.01 J $m^{-2}$ to the reported surface free energies, which is small compared to surface energies of metal oxides in the range around 0.3 J $m^{-2}$ to 2.0 J $m^{-2}$ and within the standard deviations of experimental measurements (≥0.01 J $m^{-2}$).

### 4.5. Simulation of Mechanical Properties

Mechanical properties of the alumina phases were investigated using equilibrium molecular dynamics (MD) simulations using LAMMPS. For example, a 5 x 5 x 2 supercell of α-$Al_2O_3$ was utilized. Starting with the equilibration structure as used in the calculation of lattice parameters, the structure was exposed to a controlled isotropic stress or anisotropic deformation and sampling of the resulting lattice parameters or pressure tensor, respectively. To obtain statistically meaningful averages, ten copies of each system were equilibrated for 10 ns in the NPT ensemble at $T$ = 298.15 K and $P$ = 1 atm using a Verlet integrator with a timestep of 1 fs. The deterministic Nose-Hoover thermostat and barostat with coupling constants of 100 and 1000 fs, respectively, were applied to maintain constant temperature and pressure. To ensure that the main elements of the stress tensor reached the desired pressure of 1 atm and that off-diagonal elements were close to zero, the simulation box was adjusted independently in all six directions. Following equilibration, the Enhanced Monte Carlo program was used to alter the dimensions of the simulation box in one direction, while keeping the others constant. Elongation and compression were performed up to 1% in the main directions and up to 3% in the off-diagonal directions. Each deformed system underwent further simulation in the NVT ensemble at $T$ = 298.15 K, during which the strain tensor was sampled every 1 ps. Using linear regression, the slope of the stress-strain curve was calculated and from it, the stiffness tensor C was obtained. The elastic compliance tensor S was then derived by inverting C, enabling computation of the elastic moduli $E_i = 1/S_{ii}$, $i = 1, 2, 3$, the Poisson ratio $v$, shear moduli $G_{ij} = E_i/(2+2v_{ij})$, i = 1, 2, 3, and bulk modulus $K = E/(3-6v)$ [7].

*Bulk Modulus.* The bulk modulus was calculated using the Discover program in Materials Studio with two independent NPT simulations at low pressure (0.0001 GPa) and high pressure (1 GPa) with the Parrinello–Rahman barostat, respectively. Other simulation settings remained the same as described in simulation protocols for lattice parameters. The bulk modulus was calculated using the following equation:

$$BM = -\frac{P_l - P_h}{V_l - V_h} \times V_l. \quad (3)$$

$P_l$ was the simulated value of low pressure, $P_h$ the simulated value of high pressure, $V_l$ the simulated cell volume at low pressure, and $V_h$ the simulated cell volume at high pressure.

*Elastic Constants.* LAMMPS was employed to simulate the elastic constants with IFF-OPLS/CVFF with reference to the LAMMPS examples.[114] The Ewald method (PPPM) was used for the summation of electrostatic interactions (Coulomb) with a high accuracy of $10^{-6}$ kcal $mol^{-1}$. Several variables were set for the simulations, e.g., the sampling interval was 20 timesteps, the number of samples was 20, and the frequency of one average sampling was 400 timesteps. The system was first equilibrated for 2 ns followed by simulations for 0.5 ns with a timestep of 1 fs at 298.15 K for each perturbation cycle. Positive and negative deformations (equal to 1% of the strain) were exerted on the lattice structure in six directions during the perturbation cycles, including *xx*, *yy*, *zz*, *xy*, *xz*, and *yz*.

### 4.6. Simulation of Water Contact Angles

The water contact angle was computed using MD simulations in the NVT ensemble as an average over 2 ns using the LAMMPS program (IFF-OPLS/CVFF, IFF-PCFF) and the NAMD program (IFF-CHARMM). For the hydrated α-$Al_2O_3$ (0001) surface, as an example, a typical cell size of 41.215×47.589×80.000 $Å^3$ including a >50 Å vacuum slab was built to allow enough space for the water droplet to wet the surface. A water cube consisting of 515 SPC water molecules was placed 5 Å above the hydrated α-$Al_2O_3$ (0001) surface. The settings in simulation protocols were the same as for the computation of lattice parameters and surface energies. During subsequent MD simulation, the water cube gradually spread out and reached a wetting equilibrium near 0° contact angles on the hydrated α-$Al_2O_3$ (0001) surface.

100 snapshots over the last 2 ns in equilibrium were collected for visual analysis using circular overlays as described in ref. [81]. While near-zero contact angles are easy to recognize, the formal method involved (1) drawing a circle of a size that best approximates the contour of the liquid, (2) reading out the angle between the diameter that ends at the solid–liquid interface and the diameter aligned with the surface normal (see Figure 4c in ref. [81]), and (3) repeating this protocol for all 100 snapshots to report an average value and the standard error.

Larger droplets >1000 SPC water molecules on larger surface areas, different initial drop shapes, and non-hydrated alumina surfaces were also tested and led to 0° contact angles.

### 4.7. Simulation of Solid-Liquid Interfaces

The radial distribution functions (Figure 6e) were calculated using the "measure gofr" routine in VMD[115] between O surface atoms and O atoms in water molecules over 100 frames during a 5 ns simulation at 298 K. The radial distribution function was normalized by the shell volume $4\pi r^2 \Delta r$ and the average number density of the selected atoms, following the standard implementation in VMD, such that $g(r) \to 1$ at large distances for a homogeneous system. Because the RDF was computed at a solid–liquid interface rather than in a homogeneous isotropic bulk phase, only approximately half of the spherical shell volume around the surface atoms was accessible within the liquid phase. As a result, the RDF approaches values near 2 at larger distances instead of converging to unity as in bulk systems.

Given the average shape of hexagonal gibbsite crystals of 120 nm wide and 8 nm thick in experiments (aspect ratio 8),[83] the immersion energy in water was computed by analyzing the two bounding facets. They include the (001) basal plane and the (400) equilibrium edge surfaces, and the total immersion energy, normalized by surface area, was: $\Delta E_{imm} = 0.224 \cdot \Delta E_{imm}^{(004)} + 0.776 \cdot \Delta E_{imm}^{(001)}$ (see details in Section S5 in the Supplementary Material). By convention, the negative sign of the exothermic immersion energy was omitted, and the results are reported as positive values. For each surface, two models were prepared: (1) a >2 nm thick 2D periodic gibbsite slab with the (hkl) surface in contact with an at least 4 nm thick slab of water, (2) and the same gibbsite slab with the (hkl) surface in contact with an at least 4 nm thick slab of vacuum. In addition, a 3D periodic box of water with the same number of water molecules as in (1) was prepared (Figure 6f). The 4 model systems with gibbsite surfaces were subjected to MD calculations in the NPT ensemble for 1 ns to determine equilibrium gibbsite lattice parameters normal to the (hkl) surface (e.g., $x$ and $y$), and the water box was run longer for 3 ns. Then, the 4 model systems with gibbsite surfaces were subjected to further MD simulations in the NVT ensemble for 3 ns with equilibrium lattice parameters and density in case of gibbsite-water systems, and maintaining the vacuum slab in gibbsite-vacuum systems. The average energies were used to compute the immersion energy for each facet as a difference $\Delta E_{imm}^{(hkl)} = E_{gibbs-water} - E_{gibbs-vacuum} - E_{bulk\ water}$, and then normalized per surface area ($2A$) to obtain values in $mJ/m^2$ for both the (400) and (001) planes. Subsequently, the immersion energy for the entire nanocrystal was computed with the 22.4% weight for the (400) surface and 77.6% for the (001) surface. Statistical uncertainties were on the order of $\pm 3$ $mJ/m^2$. Considering some variability in (400) surface reconstruction and physically justified charge assignments we estimate a conservative combined uncertainty of $\pm 10$ $mJ/m^2$.

The density profiles of the alumina surfaces in contact with water were simulated at pH values of 2, 5, 8, 10, and 12 using MD simulations in the NPT ensemble, allowing full equilibration of the liquid density. We built simulation boxes of ~3 nm x 3 nm lateral extension with the pH-specific surface models and a water slab of at least 5 nm height. Relaxation over 100 ns was necessary for equilibration of positions of $Cl^-$ or $Na^+$ ions on and off the surface (with ions initially placed away from the surface for equilibration), and the density profiles vertical to the alumina (0001) surfaces were sampled over the last 75 ns. The results (Figure 8d, e) represent 150 ns duration for each pH value using symmetry-based averaging of the two independent alumina-electrolyte interfaces in each 3D periodic system. The solution density was 1.00 g/cm$^3$ and MD simulations were carried using a time step of 1 fs at 298 K using the program LAMMPS, the PPPM method for high accuracy electrostatics ($10^{-4}$ sufficient), and a spherical cutoff for van-der-Waals interactions at 12 Å (IFF standard). Density profiles were created using custom python scripts.

### 4.8. Simulation of the Adsorption of Organic Molecules

The binding energies and conformations of p-hydroxybenzoate on the alumina (0001) surfaces were computed using the surface models for the respective pH values of 3, 5, 7, and 9 in contact with a water slab of 4 nm thickness (TIP3P model, 856 molecules) and the organic molecules (Figure 9b-g). As determined in the related experiments, the point of zero charge with full OH termination of the surface was set as pH 7 (rather than pH 8), with the other three models reflecting surface ionization 4 and 2 pH units below, and 2 pH units above the pzc at 3, 5 and 9 (rather than 4, 6 and 10). The molecular models of p-hydroxybenzoate and its deprotonated form at high pH were based on the IFF/CHARMM36 force fields, including IFF virtual π electrons.[93] The solution density was 1.000 g/cm$^3$ and multiple MD simulations in the NVT ensemble were carried out for each system for 200 ns. We used a time step of 1 fs, a temperature of 298 K, the PPPM method for high accuracy electrostatics ($10^{-4}$), a spherical cutoff for van-der-Waals interactions at 12 Å (IFF standard), and the program NAMD.

The two surfaces in each model were equivalent (same pH value and surface ionization) and the molecules reversibly adsorbed on the top and on the bottom of the slab or stayed detached in solution. Transitions between the two surfaces and dissolved states enhanced statistical sampling and averaging (Figure 9b, d, f). The time in contact, within 4 Å from superficial O atoms, was recorded to compute the percentage of time in contact and the free energy of binding $\Delta G$, along with the concentration in solution ($c_{sol}$ = 0.064 M) and the approximate concentration of available adsorption sites (~1 M, ~2 sites per $nm^2$ surface area in flat-on binding configuration):

$$\Delta G = -RTlnK \text{ with } K = \frac{t_{ads}}{t_{des} \cdot c_{sol}} \quad (4)$$

Upon carrying out multiple simulations, the residence times in percent were reproducible within 98.8±1% (pH 3), 83±4% (pH 5), 13±4% (pH 7), and 9±3% (pH 9), which translates into uncertainties in computed binding free energies of -4.2 ±0.5 kcal/mol (pH 3), -2.6 ±0.3 kcal/mol (pH 5), -0.5 ±0.2 kcal/mol (pH 7), -0.25 ±0.2 kcal/mol (pH 9).

**Supplementary Material Available**

Additional Figures, Tables, discussion, and Supplementary Files that contain the force fields in multiple formats (CVFF, PCFF, CHARMM), ready to use 3D models of the unit cells, the database of 3D surface models in vacuum, the database of 3D surface models in water across pH values from 2 to 12, simulation run scripts, examples to reproduce the reported data, and extensive documentation.

**Acknowledgements**

We acknowledge the helpful comments provided by Dr. Pieter J. in ’t Veld, BASF SE.

**Author Contributions: CRediT**

C.Z. contributed to conceptualization, data curation, methodology, analysis, writing the original and revised drafts. K. K. contributed to methodology, validation, and review. S. D., S. P. F., and K. S. contributed to data curation, visualization, and writing the original draft. P. K., N. S. K., and E. S. contributed to analysis and review. R. K. M. contributed to conceptualization, resources, supervision, investigation, writing and editing. H. H. contributed to conceptualization, resources, supervision, investigation, data curation, writing, editing, and validation.

**Funding Sources**

The authors acknowledge support by the National Science Foundation (OAC 1931587, CMMI 1940335, DMREF 2323546) and the University of Colorado Boulder. We utilized allocations of computational resources at the Argonne Leadership Computing Facility, supported by the Office of Science of the U.S. Department of Energy (DE-AC02-06CH11357), and at the Summit supercomputer, a joint effort of the University of Colorado Boulder and Colorado State University supported by the National Science Foundation (ACI 1532235 and ACI 1532236). We also utilized allocations of computational resources at the BASF supercomputer Quriosity.

**Supplementary Material**

for

# INTERFACE Force Field for Alumina with Validated Bulk Phases and a pH-Resolved Surface Model Database for Aqueous, Electrolyte, and Molecular Interfaces

Cheng Zhu,[1] Krishan Kanhaiya,[1] Samir Darouich,[2] Sean P. Florez,[1] Karnajit Sen,[2] Patrick Keil,[3] Nawel S. Khelfallah,[4] Eduard Schreiner,[2] Ratan K. Mishra,[2*] Hendrik Heinz[1*]

[1] Department of Chemical and Biological Engineering, University of Colorado Boulder, Boulder, CO 80303-0596, USA

[2] Group Research, BASF SE, Carl-Bosch-Strasse 38, 67056 Ludwigshafen am Rhein, Germany

[3] BASF Coatings GmbH, Glasuritstrasse 1, 48165 Münster, Germany

[4] Chemetall GmbH, Trakhener Strasse 3, 60487 Frankfurt am Main, Germany

* Correspondence to: ratan.mishra@basf.com, hendrik.heinz@colorado.edu

This PDF file includes:

Supplementary Figures S1 to S15

Supplementary Tables S1 to S8

Supplementary Discussion Sections S1 to S7

Supplementary References

## Supplementary Figures

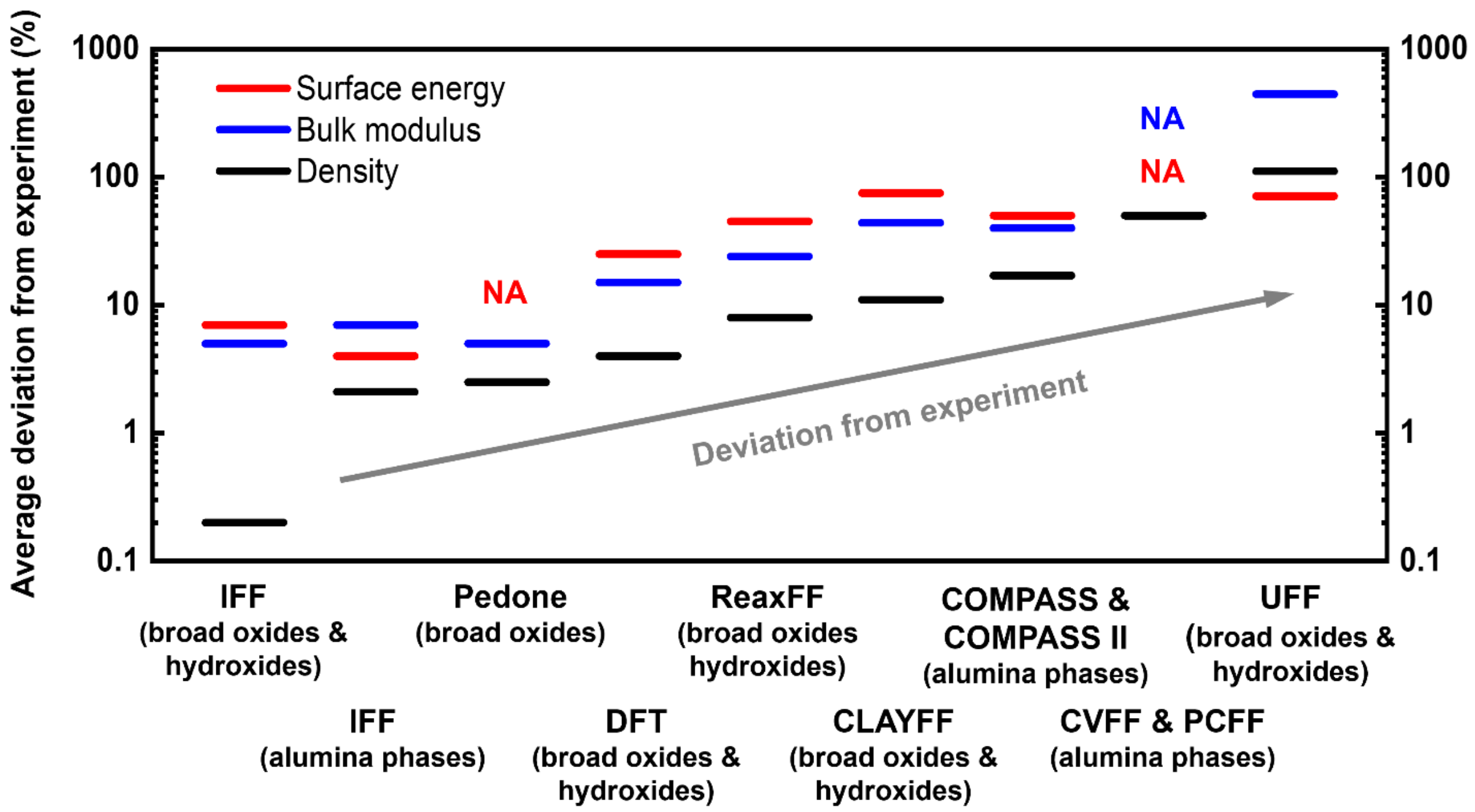

**Figure S1.** Comparison of the performance of IFF, other force fields, and DFT in key properties relative to experimental reference data for a broad set of oxides, hydroxides (see data in ref. [1]) and alumina phases. Average deviations in lattice parameters, surface energies, and bulk moduli are lowest in IFF models. Some force fields have limited coverage, e.g., Pedone does not include hydroxides and surfaces. Some models significantly distort or rearrange the structures of alumina phases (COMPASS and COMPASS II, CVFF, PCFF, and UFF). In these instances, surface energies and bulk moduli are only estimates or not meaningfully computable (NA). CVFF, PCFF, COMPASS, and COMPASS II have limited or no coverage of other oxides and hydroxides (e.g., Fe, Ni, Cr).

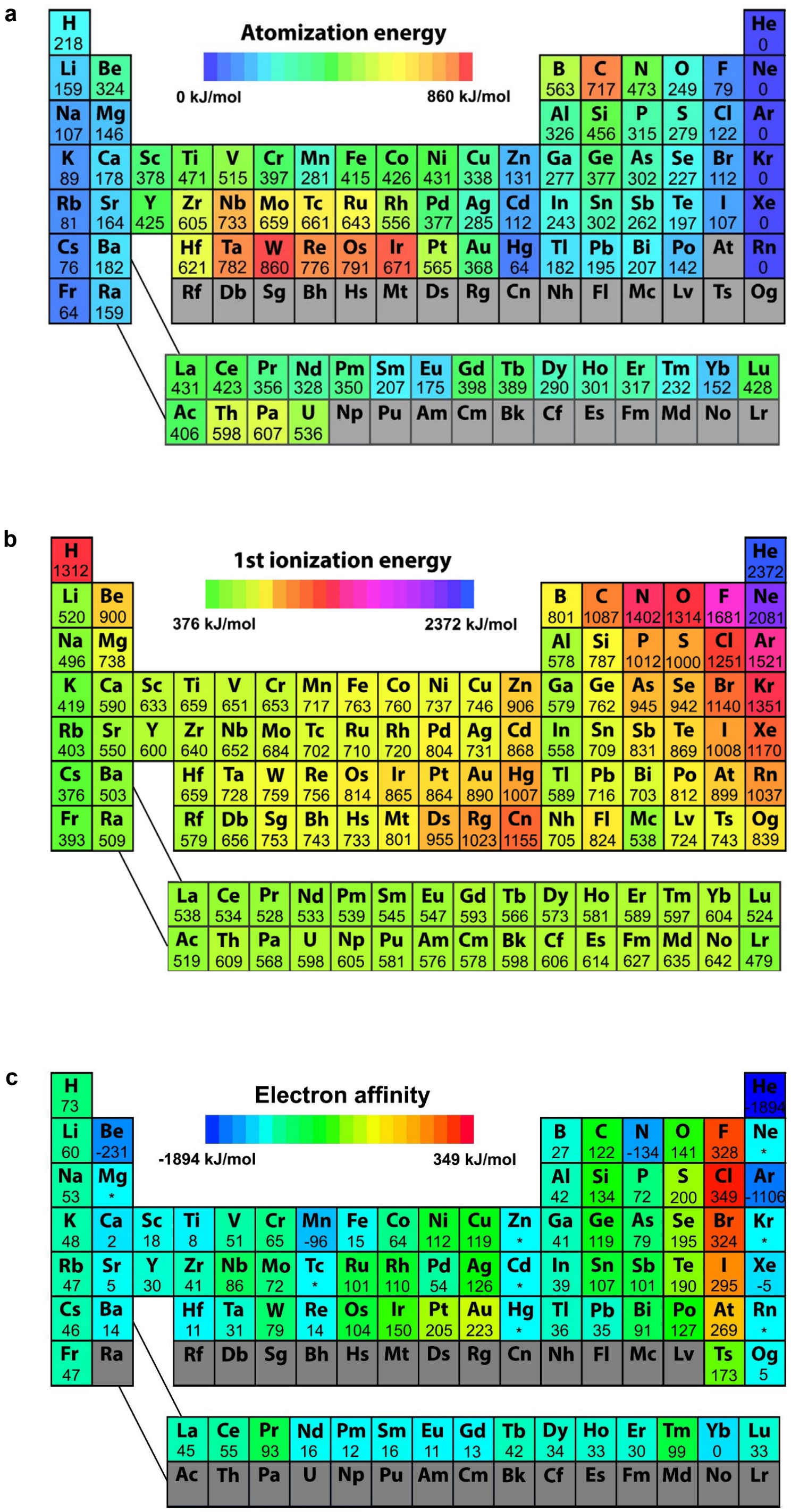

* Denotes elements that are expected to have negative or near-zero electron affinities on quantum mechanical grounds.

**Figure S2.** Periodic tables of atomization energy, 1$^{st}$ ionization energy, and electron affinity of the elements (all data from ref. [2], reproduced with permission from ref. [3]). The data aids in assigning and validating atomic charges for molecular simulations as described in the Extended Born Model.[4] (a) The magnitude of atomization energy represents the tendency of an element towards covalent versus ionic bonding, affecting the amount of charge transfer. (b) The magnitude of the 1$^{st}$ ionization energy represents the ease of ionization of an element and affects the amount of positive charge of a metal species in an oxide. (c) The magnitude of the electron affinity (note: sometimes represented with opposite sign) indicates the stabilization energy by attracting a single electron and reflects the amount of negative charge accepted by the species that is a valence electron acceptor. Maxima of the electron affinity may be found for fractional charges, such as for N (-0.5e) and O (-0.6e to -0.7e) (ref. [4]). The atomization energy and ionization energy of the metal, and the electron affinity of the oxygen (or other nonmetal) in an oxide correlate with the atomic charges and molar energies of formation.[4] For example, the atomization energy of Al of 326 kJ/mol is much lower than for Si or C, but higher than for Mg and Mn, indicating atomic charges higher than Si and C, but lower than for Mg and Mn in a comparable coordination environment. The 1$^{st}$ ionization energy of Al is 578 kJ/mol, e.g., clearly lower than C, somewhat lower than Si, Mg, and Mn, like Ca and higher than Na. This comparison corroborates a higher charge than C as well as Mg and Mn in similar environment, some similarity to Ca and a higher charge than $Na^+$ (as expected). Using experimental data from X-ray deformation electron densities,[4-7] multipole moments for similar compounds, these trends, and validated atomic charges for compounds containing the mentioned nearby elements in IFF, suggest an atomic charge of approximately +1.5e ±0.1e for Al and -1.0e ±0.07e for O in alumina phases (+1.62e ±0.1e for Al and -1.08e ±0.07e for O in $Al_2O_3$ phases in the nonbonded models).[1] The values are close to experimental data and consistent with the electronegativity difference of 1.83 relative to other compounds (Figure S3).

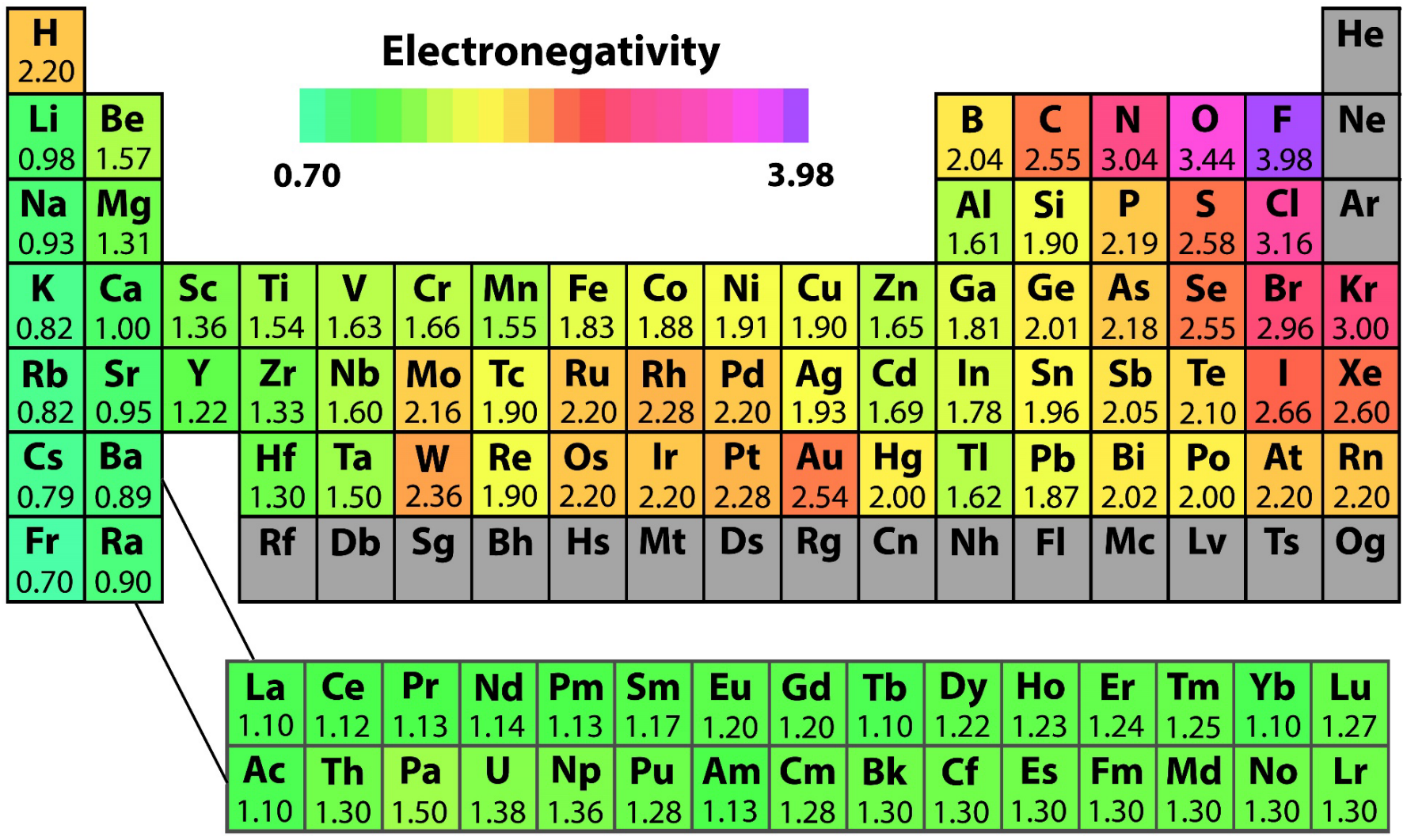

**Figure S3.** The periodic table of electronegativity according to Pauling (reproduced with permission from ref. [3]). Per definition, the square of the electronegativity difference between two elements quantifies the energy contribution to a pairwise bond due to electrostatic (non-covalent) interactions in eV (ref. [8]). The elemental electronegativity difference between O (3.44) and Al (1.61) is 1.83, which makes the Al–O bonds more ionic than covalent.

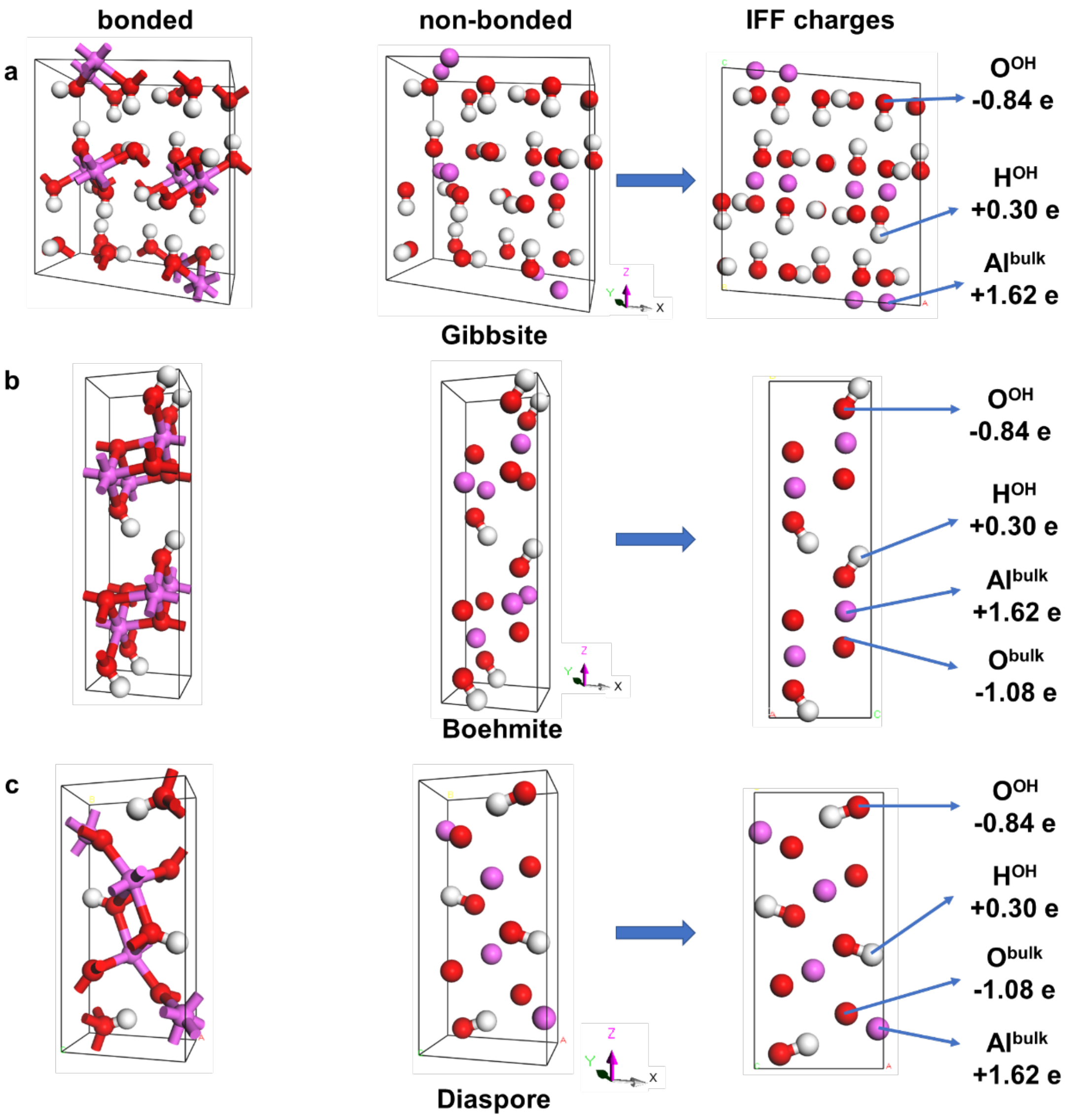

**Figure S4.** Representation of atomic charges on internal [Al O-H] groups in the minerals containing OH groups. One set of atom types is used for all minerals. (a) Gibbsite. (b) Boehmite. (c) Diaspore. When the [Al O-H] groups are exposed on external surfaces of these minerals, including hydrated surfaces of α-$Al_2O_3$ and γ-$Al_2O_3$, the same atom types and charges are used.

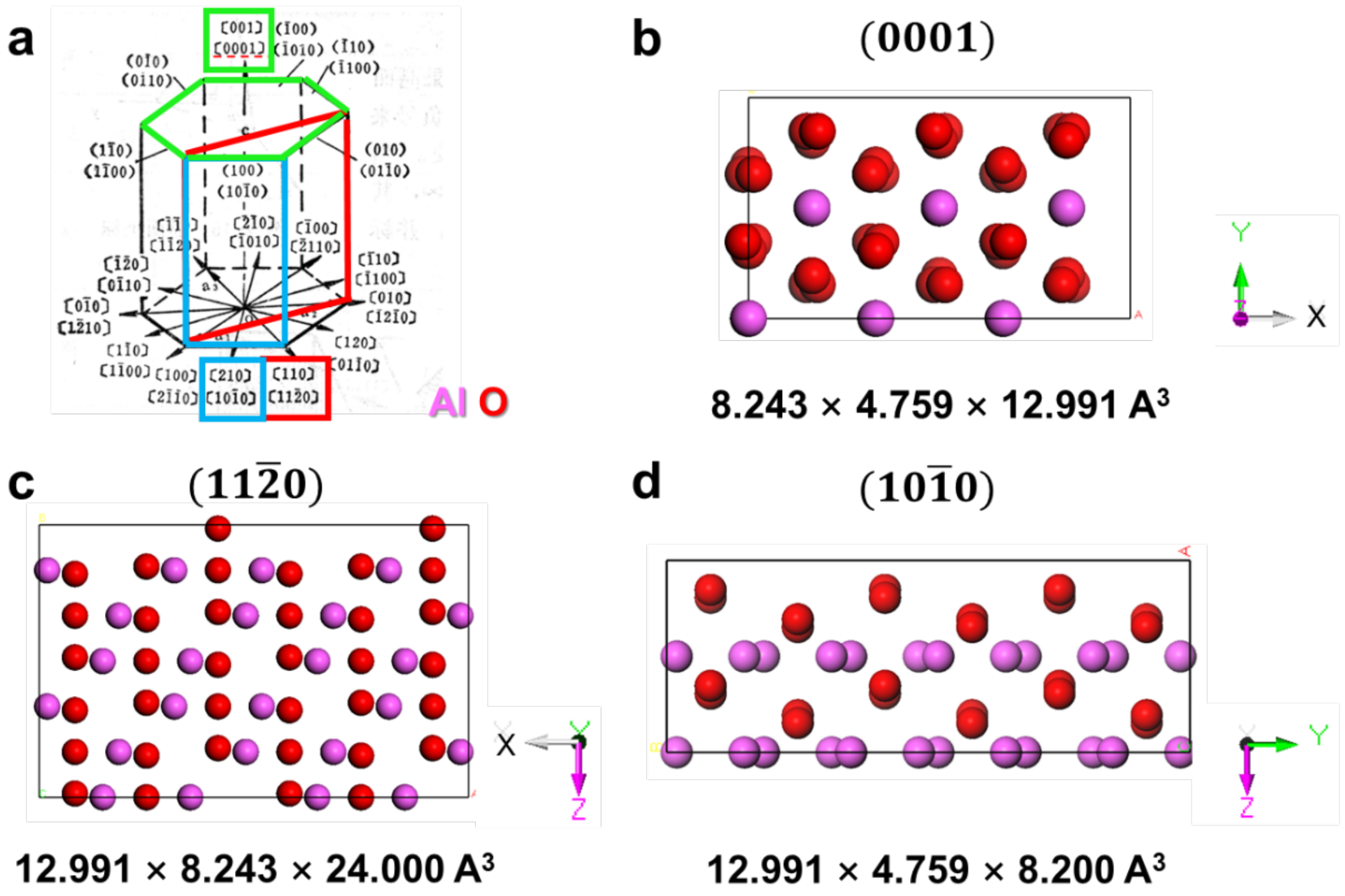

**Figure S5.** Cleavage planes in corundum (hexagonal α-$Al_2O_3$). (a) Schematic of multiple cleavage planes in the hexagonal crystal system. (b) Top view onto the $(0001)$ facet. (c) Top view onto the $(11\bar{2}0)$ facet. (d) Top view onto the $(10\bar{1}0)$ facet. The corresponding supercell sizes are listed below. The corresponding calculated surface energies are listed in Table S1.

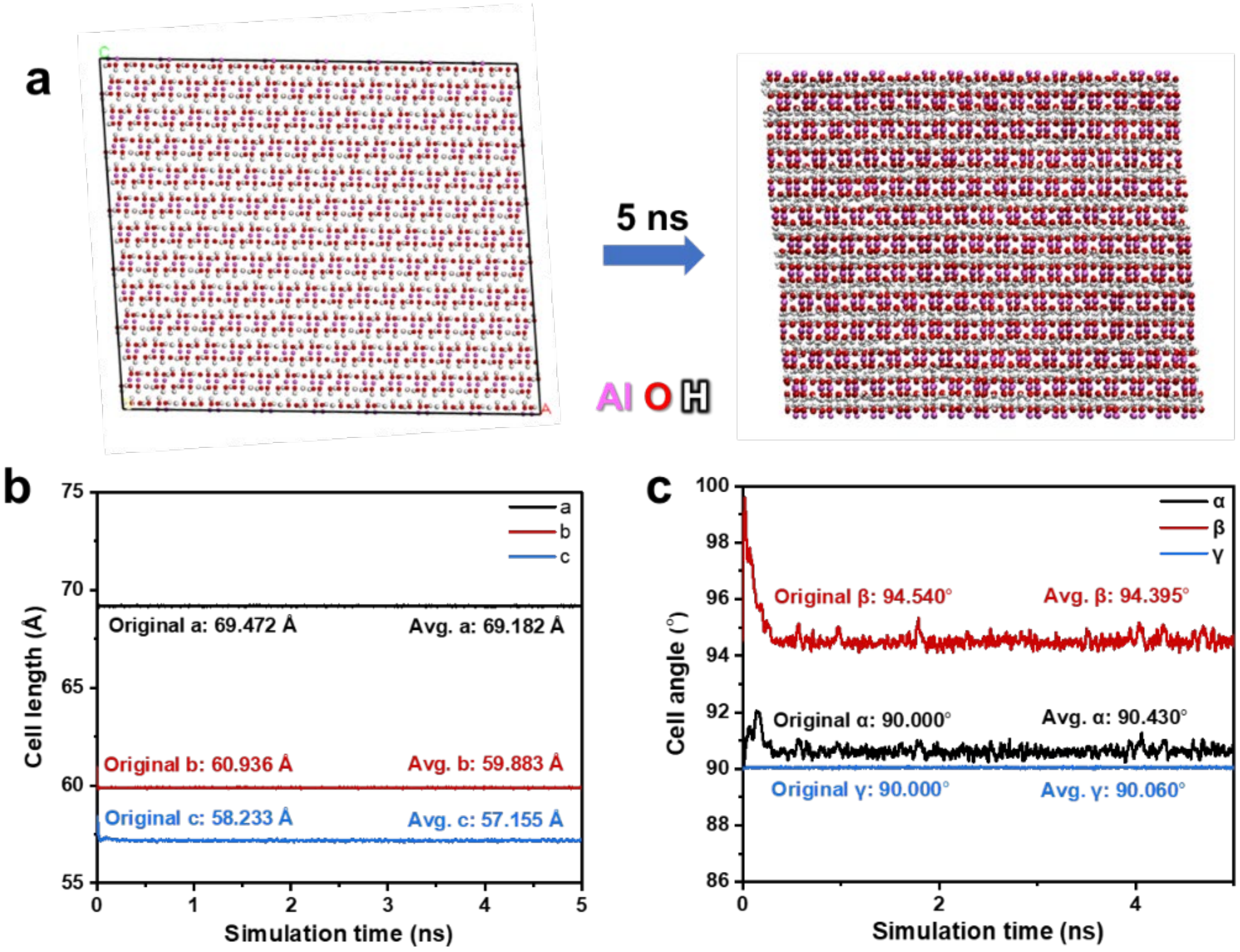

**Figure S6.** Stability of the gibbsite structures during IFF MD simulations (IFF-CVFF). (a) Starting and equilibrated structures after 5 ns for gibbsite (box size: 69.472 × 60.936 × 58.416 $Å^3$). (b) Values of the supercell lattice parameters *a*, *b*, *c* show high convergence as a function of simulation time. (c) Values of the cell angles as a function of simulation time, showing high convergence. We carried out MD simulations for 5 ns to ensure extensive relaxation of the lattice. Since there are only OH and Al layers inside the nonbonded models with relatively weak interactions, a large simulation cell is recommended (e.g., box size: 69.472 × 60.936 × 58.416 $Å^3$) to avoid sliding of atomic layers in the *x* and *y* directions. The simulated cell lengths and angles values are in very good agreement with XRD data with an average deviation <1%.[9] Notably, for the lattice angles,

there is a sharp change at the beginning of simulations resulting from reorientation of the OH groups, which eventually return to the initial values. In addition, the simulated density is 2.52 $g/cm^3$, only 4% higher than the experimental value 2.42 $g/cm^3$. Achieving this agreement by direct application of IFF parameters of $\alpha$-$Al_2O_3$ without additional adjustments provides strong proof of the high accuracy and compatibility of IFF parameters. If necessary, a higher accuracy of the density (<1% deviation) can be reached by slightly increasing the atomic charge on OH groups (e.g., +0.35e for H and -0.89e for O), presumably due to extensive internal hydrogen bonding.

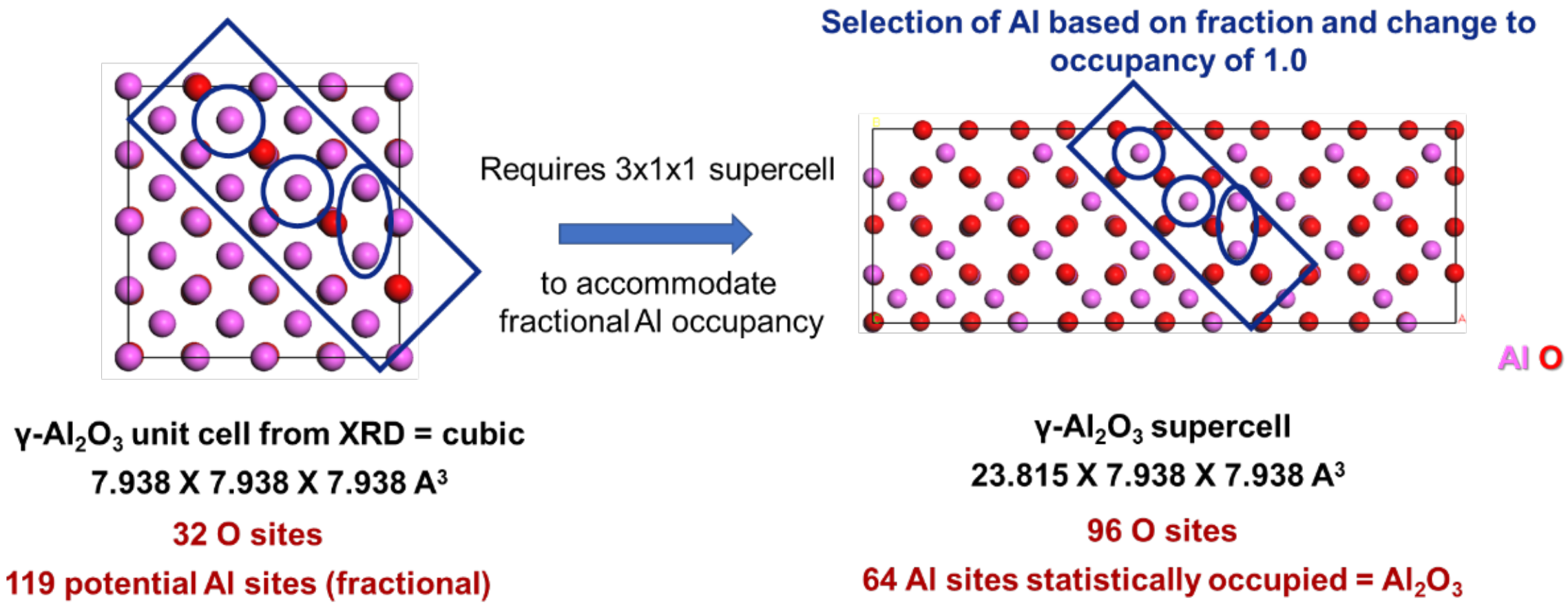

**Figure S7.** The construction of γ-$Al_2O_3$ models for simulations. The X-ray crystal structure contains well-defined O sites and many potential Al sites with fractional occupancy.[10] A triple size unit cell is required, at minimum, to build a stoichiometric supercell in which fractionally occupied Al sites are converted into fully occupied sites. We placed Al atoms with full occupancy in statistical distribution. Many alternative distributions are possible, especially for larger cells. The effect of the statistical distribution of Al on the computed properties is hardly noticeable when the distribution of Al atoms is spatially even. More quantitatively, γ-$Al_2O_3$ is a disordered cubic phase structure with random but stoichiometric occupancy of vacancies by Al in a well-defined oxygen lattice (32 O sites and 119 potential Al sites).[10] As the stoichiometric number of Al atoms among the fractionally occupied Al sites would be 21.33, the unit cell is not suitable for direct simulations. Larger cells and assumptions about a statistical full occupancy of Al sites are necessary to create a realistic simulation cell with Al and O in a stoichiometric ratio of 2:3. We built a γ-$Al_2O_3$ supercell from 3×1×1 unit cells that contain 96 O sites and 357 potential Al sites. Then, we refined the structure by randomly converting 64 fractionally occupied Al sites into fully occupied Al sites. Rectangles and circles in the figure mark the portion of the structure before and after refinement. The refined $\gamma$-$Al_2O_3$ cell has 64 Al atoms and 96 O atoms, matching the stoichiometric ratio of 2:3.

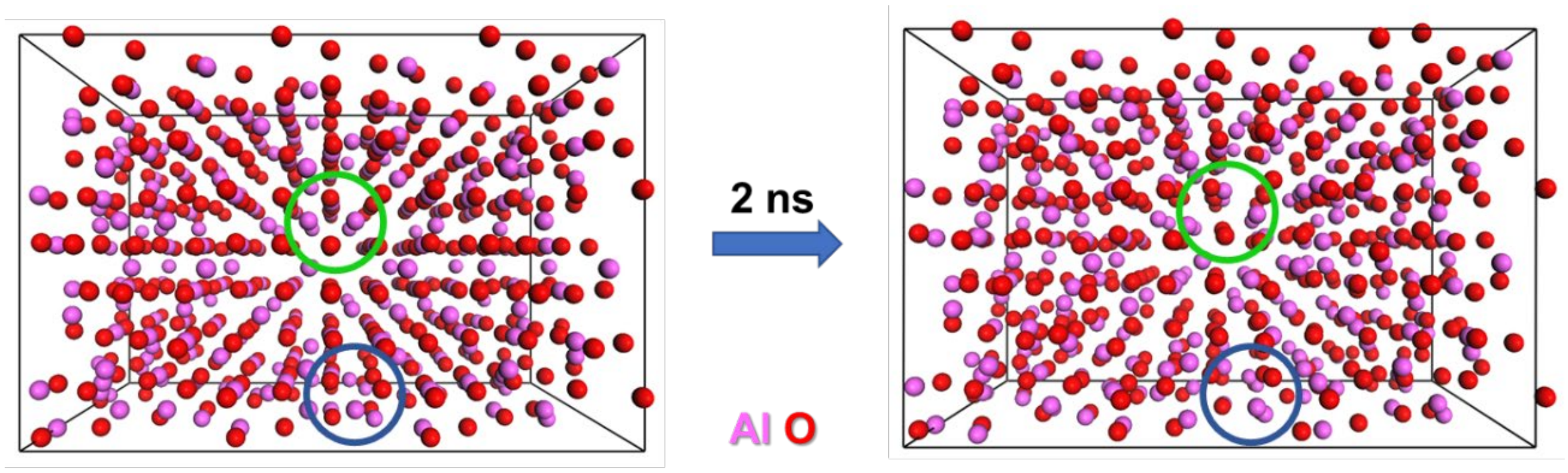

**Figure S8.** Perspective view of the starting and equilibrated structures of γ-$Al_2O_3$ after 2 ns MD simulations with IFF-CVFF (cell size: 23.815 × 15.876 × 15.876 $Å^3$). After 2 ns, the positions of several Al atoms rearranged related to the random distribution and opportunity of fractional occupancy of at least 5x more potential sites (green and blue circles mark the representative regions with atomic rearrangement). These observations are consistent with XRD measurements, which indicate 357 sites of fractional occupancy by 64 Al atoms inside the lattice. The lattice has a stable structure; the surface energy and bulk modulus can be accurately calculated (Table S5).

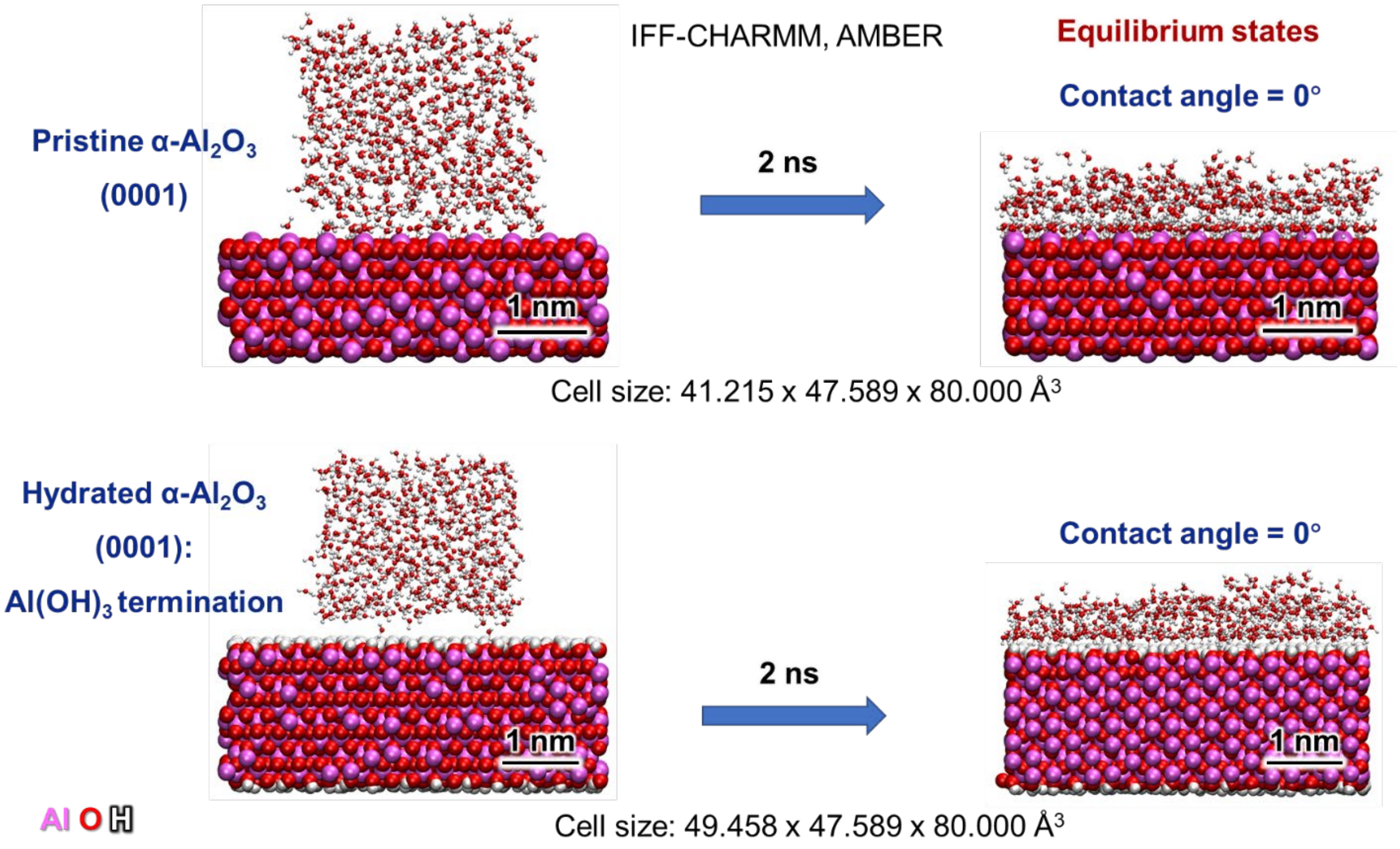

**Figure S9.** Simulated contact angle of water on pristine and hydrated α-$Al_2O_3$ (0001) surfaces at room temperature. The hydrated surface at pH 7 features an $Al(OH)_3$ layer on top, i.e., a disordered molecular layer of gibbsite. Computed contact angles on both surfaces are 0°, consistent with experimental data.[11] MD simulations were carried out using IFF-CHARMM, AMBER and the flexible SPC water model (standard in IFF).

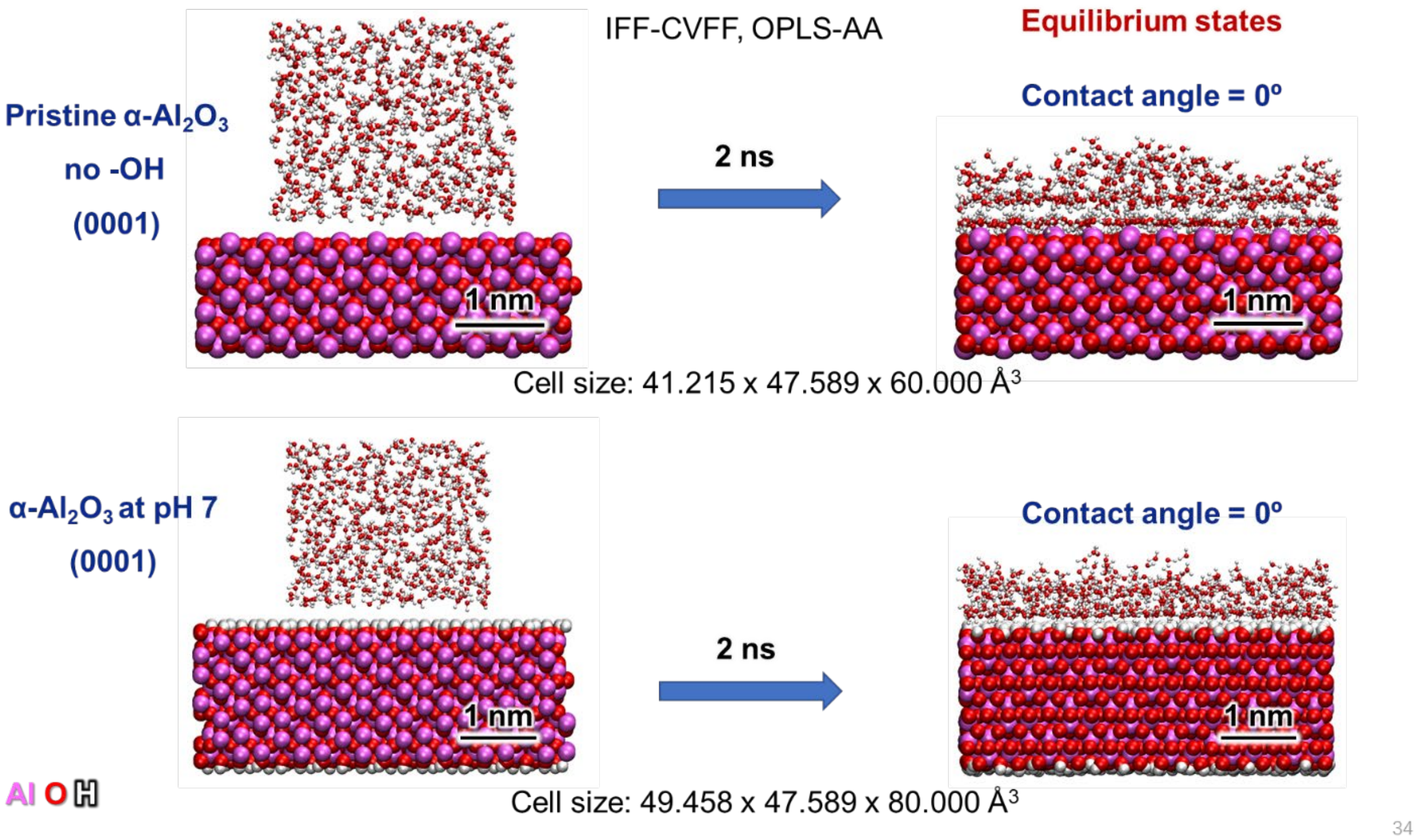

**Figure S10.** Simulated contact angle of water on pristine and hydrated α-$Al_2O_3$ (0001) surfaces at room temperature. The hydrated surface at pH 7 features an $Al(OH)_3$ layer on top, i.e., a disordered molecular layer of gibbsite. Computed contact angles on both surfaces are 0°, consistent with experimental data.[11] MD simulations were carried out using IFF-CVFF, OPLS-AA and the flexible SPC water model (standard in IFF).

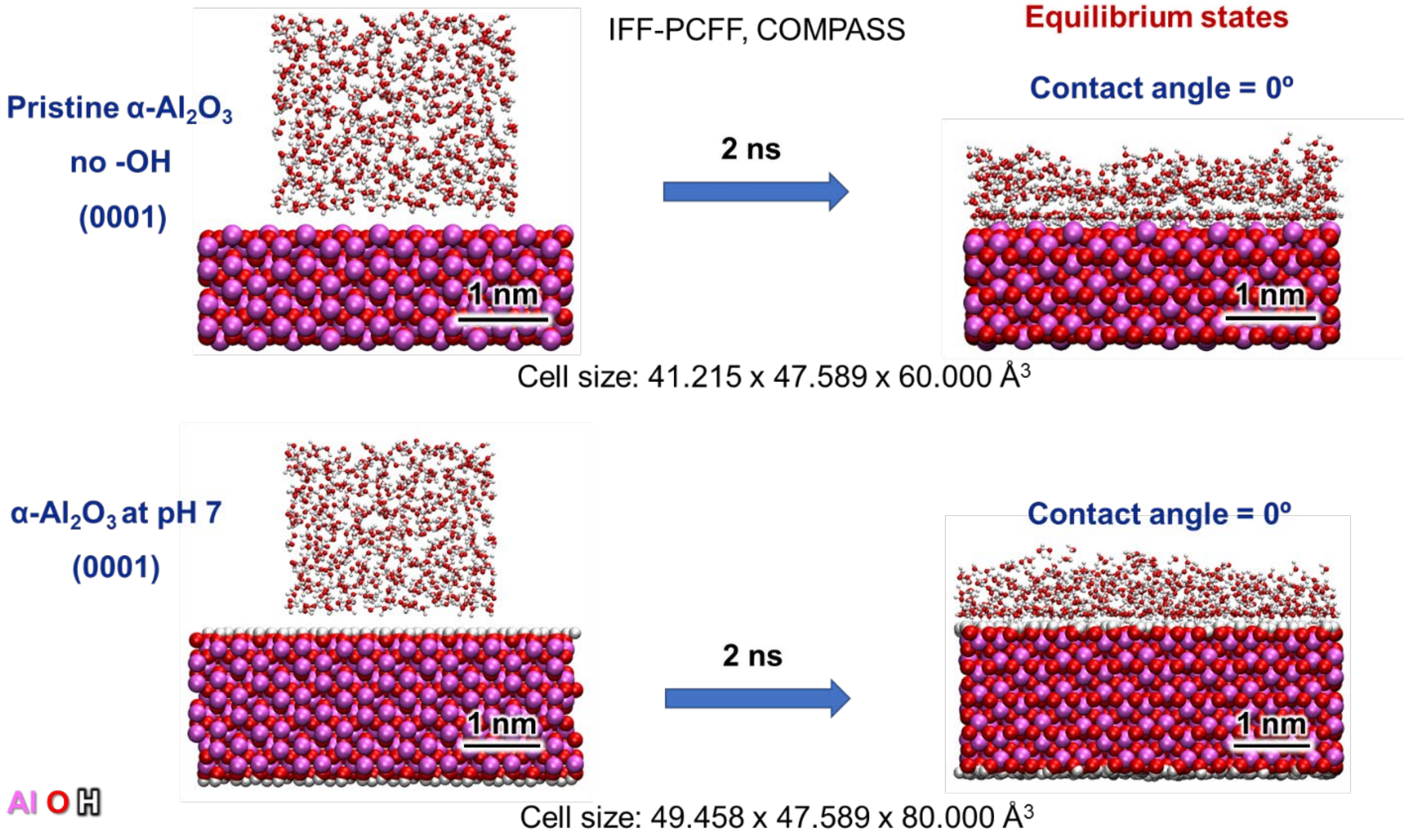

**Figure S11.** Simulated contact angle of water on pristine and hydrated α-$Al_2O_3$ (0001) surfaces at room temperature. The hydrated surface at pH 7 features an $Al(OH)_3$ layer on top, i.e., a disordered molecular layer of gibbsite. Computed contact angles on both surfaces are 0°, consistent with experimental data.[11] MD simulations were carried out using IFF-PCFF, COMPASS and the flexible PCFF water model.

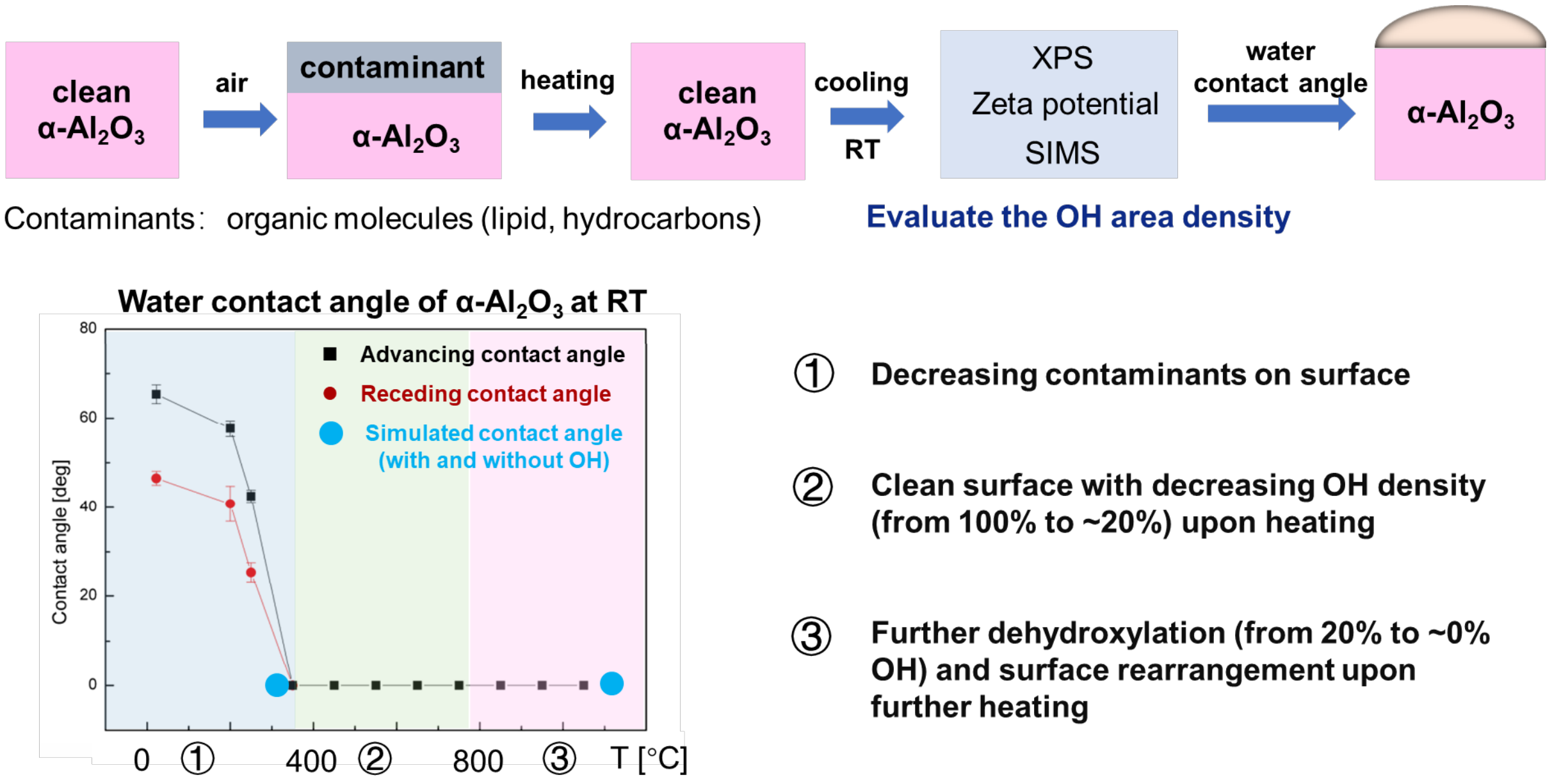

**Figure S12.** Experimental measurement of water contact angle on α-$Al_2O_3$ (0001) surfaces.[11]

The presence of contaminants often leads to contact angles between 50° and 60°, which can be removed by heating or cleaning. The Al-OH groups on the surface are stable up to about 500 °C. Contact angles reach 0° on clean hydrated surfaces. Above 500 °C, the area density of Al-OH decreases due to dehydration, and a certain fraction of Al-OH groups persists above 1050 °C. Surfaces instantly hydrate in contact in water, recreating a thin gibbsite layer.

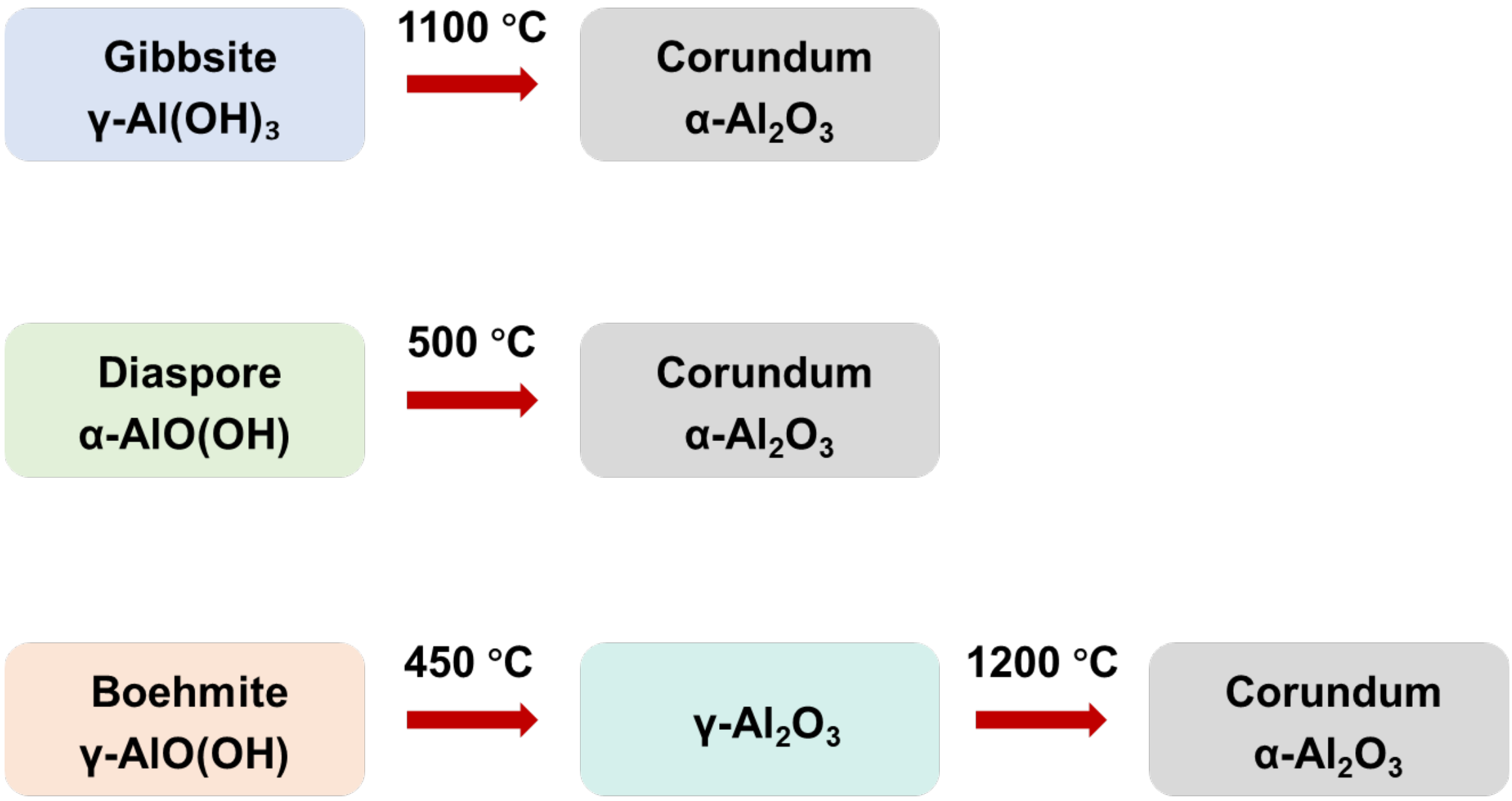

**Figure S13.** Common phase transitions of alumina phases.[12] α-$Al_2O_3$ is ultimately formed upon heating of any other oxyhydroxide phase.

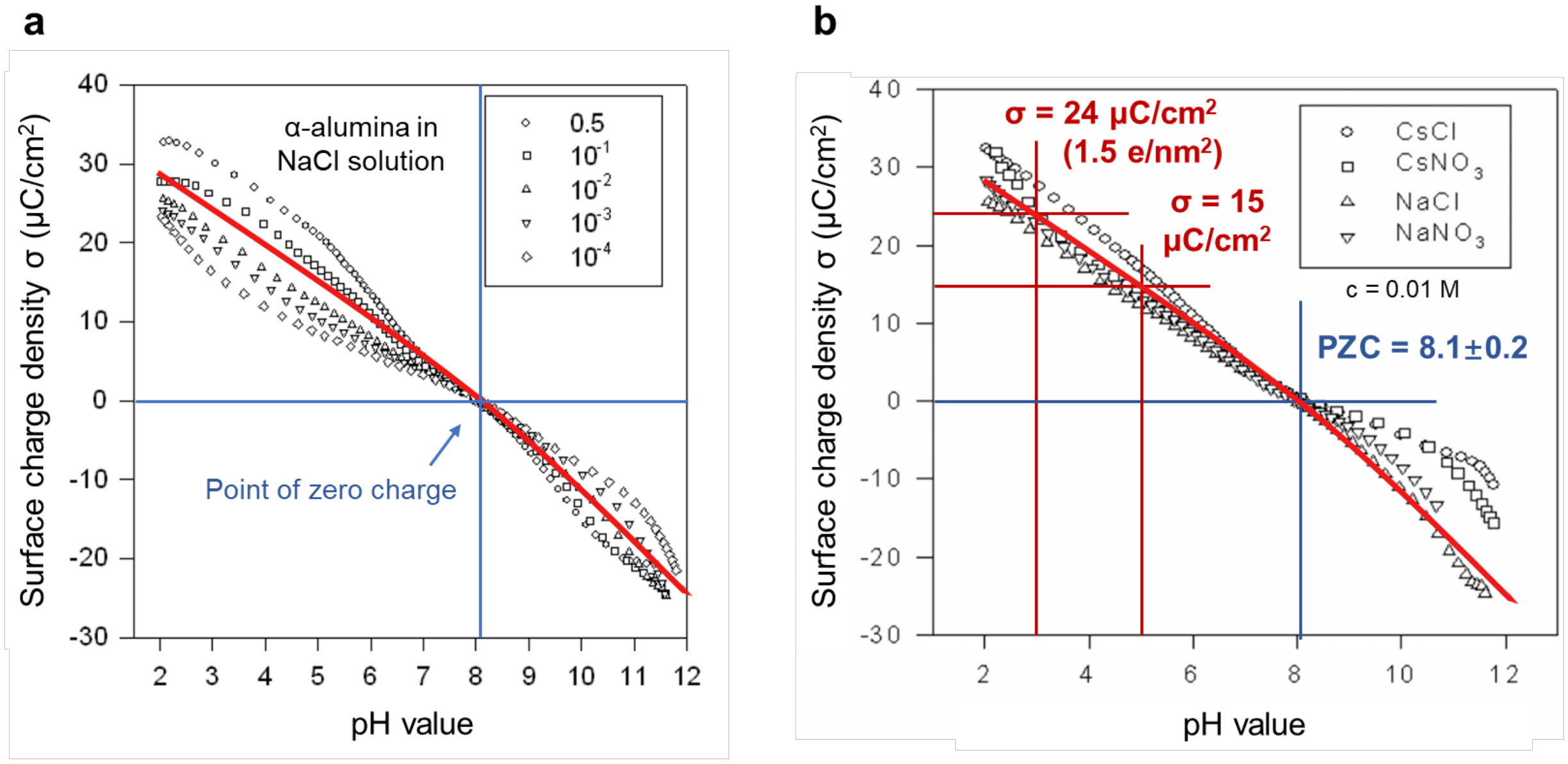

**Figure S14.** Experimental measurements of the surface charge of α-alumina as a function of pH in electrolytes at different concentrations (reproduced with permission from the Bulletin of the Chemical Society of Ethiopia).[13] The data were measured by surface titration. (a) The surface charge density of α-alumina as a function of pH value in aqueous NaCl solutions for multiple concentrations. Very dilute solutions of $10^{-4}$ M NaCl to concentrated solutions of 0.5 M NaCl were tested. The red solid line represents average values to inform molecular models. The point of zero charge at pH = 8.1 is identified by two blue solid lines. (b) Measurements of the surface charge of α-alumina as a function of pH value in different electrolytes at 0.01 M concentration.[13] $NaNO_3$ and NaCl lead to similar results whereas CsCl causes less ionization at high pH values due to more covalent bonding contributions of $Cs^+$ ions to the alumina surface than $Na^+$ ions. Two example readouts and the pzc are indicated. The data shown here is representative among literature reports[13-15] and includes the broadest pH ranges, which is a good reference to establish the IFF atomistic models. 16 $\mu C/cm^2$ = 1.0 $e/nm^2$. Deviations among different literature resources due to measurement methods, ion concentration, and electrolyte types are approximately within ±5 $\mu C/cm^2$, or ±0.3 $e/nm^2$.

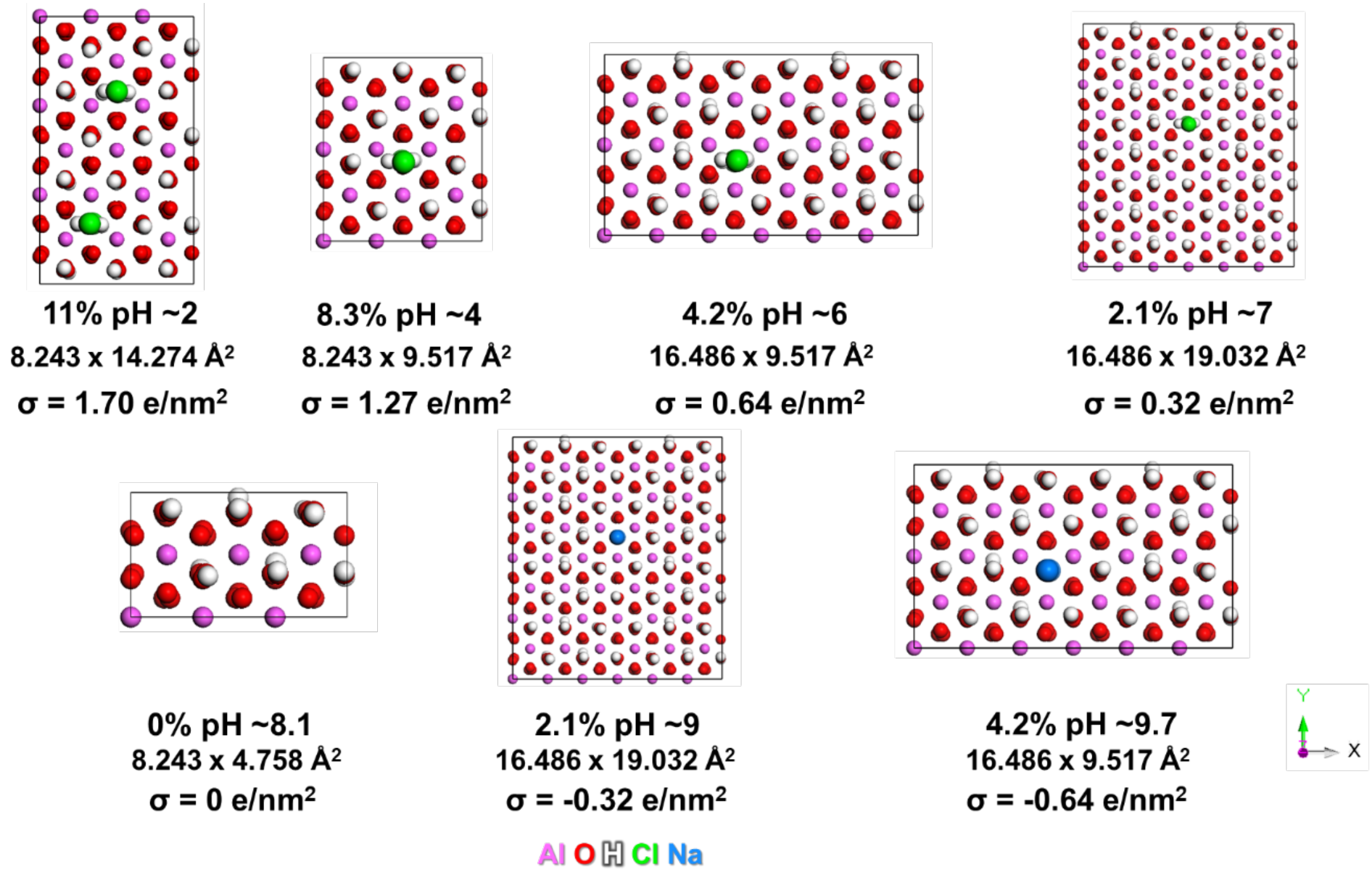

**Figure S15.** Representation of the surface chemistry of hydrated α-alumina (α-$Al_2O_3$ (0001)) surfaces using small repeat units. Annotations include the percentage of ionized OH groups, approximate pH value, the area of the surface, and the surface charge density (σ). We recommend using larger supercells, especially for high charge densities, to allow for a statistical or otherwise intentional surface charge distribution as well as sufficient space for electrolyte and other components.

**Supplementary Tables**

**Table S1.** Lattice constants, density, bulk modulus and interfacial properties of α-$Al_2O_3$ from experimental measurements and IFF MD simulations. Simulation results are displayed in IFF-CHARMM, IFF-CVFF, and IFF-PCFF, respectively.

| **Property** | **Exp.** | **IFF-CHARMM, AMBER** | **IFF-CVFF, OPLS** | **IFF-PCFF** | **Avg. Dev. (%)** |
|---|---|---|---|---|---|
| a (Å) | 14.276[16] | 14.279 | 14.217 | 14.322 | 0.25 |
| b (Å) | 14.276 | 14.282 | 14.217 | 14.315 | 0.24 |
| c (Å) | 12.991 | 12.990 | 13.070 | 12.932 | 0.36 |
| α (°) | 90.00[16] | 89.99 | 90.00 | 89.97 | 0.02 |
| β (°) | 90.00 | 90.00 | 90.00 | 90.00 | 0 |
| γ (°) | 120.00 | 119.98 | 119.99 | 119.96 | 0.02 |
| RMSD (Å) | N/A | 0.15 | 0.14 | 0.16 | N/A |
| Density (g/cm³) | 3.986[16] | 3.986 | 3.996 | 3.980 | 0.13 |
| Surface energy | 1.65±0.05[1, | 1.65 | 1.67 | 1.65 | 1.8 |
| (J/m$^2$) | 17] | | | | |
| (0001) | | | | | |
| $(11\bar{2}0)$ | N/A[a] | 1.88 | 1.85 | 1.86 | N/A |
| Bulk modulus (GPa) | 254±3[18, 19] | 265 | 273 | 207 | 10.1 |
| Water contact angle (°) | ~0[11], [b] | 0 | 0 | 0 | 0 |

[a] Some data on surface energies have been reported for both facets using DFT calculations, indicating the same trends $((0001)<(11\bar{2}0)<(10\bar{1}0))$.[20] [b] The contact angle is moderately hydrophilic in laboratory atmosphere due to surface contamination and reduces to 0° upon cleaning the surfaces or heating to about 350 °C. Thermal treatment up to 350 °C removes organic contaminants but does not remove Al-OH groups on the surfaces, which requires temperatures above 500 °C.[11] Pure oxide surfaces terminated by Al or O (without OH) groups are also hydrophilic with a contact angle of 0° and the presence of water rapidly leads to hydration reactions with $Al(OH)_3$ surface layers.

**Table S2.** The experimental and calculated lattice parameters and surface energy of boehmite, γ-AlO(OH), using molecular dynamics simulation with IFF parameters.

| Property | Exp. | IFF-CHARMM, AMBER | IFF-CVFF, OPLS | IFF-PCFF | Avg. Dev. (%) |
|---|---|---|---|---|---|
| **Lattice a (Å)** | 28.760[21] | 29.22 | 28.92 | 29.11 | 1.1 |
| **Lattice b (Å)** | 24.480 | 25.55 | 24.76 | 24.61 | 2.0 |
| **Lattice c (Å)** | 14.836 | 14.65 | 14.58 | 14.58 | 1.6 |
| **α (°)** | 90.000[21] | 89.90 | 89.85 | 89.92 | 0.12 |
| **β (°)** | 90.000 | 90.01 | 89.98 | 90.04 | 0.03 |
| **γ (°)** | 90.000 | 89.27 | 90.36 | 90.09 | 0.44 |
| **RMSD (A)** | N/A | 0.6 | 0.5 | 0.46 | N/A |
| **Density ($g/cm^3$)** | 3.03[21] | 2.90 | 3.03 | 3.04 | 1.5 |
| **Surface energy ($J/m^2$)** | (001) most likely:[a] ~0.9; DFT:[b] 0.96±0.12[22, 23] Experiment: 0.52 (facet average)[23] | 0.86 | 0.95 | 0.86 | ~5 |
| | (010) DFT:[b] 0.28±0.02[20, 24] | 0.17 | 0.24 | 0.25 | NA |
| **Bulk modulus (GPa)** | Likely: 100±10[c] Expt: 98[25][d] and 110±15[26][e], DFT: 82-97 [27] | 95 | 100 | 70 | ~10 |

[a] Given well-known surface energies for α-$Al_2O_3$ of 1.69 J/m$^2$, for MgO of 0.95 J/m$^2$ and surface energies much lower of hydroxides such as $Mg(OH)_2$ of 0.225 J/m$^2$ that are closely reproduced by IFF,[1] a surface energy of ~0.9 J/m$^2$ appears to be reasonable for γ-AlO(OH). [b] DFT calculations can show substantial deviations for interfacial and surface properties of oxides, including errors exceeding 50% in some cases.[1] [c] The experimental bulk modulus is somewhat uncertain to-date, see references and interpretations below. [d] Ref. [25] reports bulk moduli for boehmite nanotubes at 98 GPa. [d] Ref. [26] reports predominantly high-pressure data at 5-50 GPa. At low to medium pressure (0 to 3 GPa), the reported compressibility points to a bulk modulus between 95 and 120 GPa (Figure 5a in ref. [26]). As the bulk modulus would require more data near zero pressure, the error bars remain high at 110 ±15 GPa. Extrapolation from high pressure moduli from 0 to 50 GPa are 135 GPa.

**Table S3.** Experimental and calculated lattice parameters and surface energies of diaspore, α-AlO(OH), using IFF models and parameters. The lattice parameters *a* and *b* using IFF-CHARMM parameters showed large differences from experiments of +10% and +5%., which could be remedied by specific modifications of the OH parameters.

| **Property** | **Exp.** | **IFF-CHARMM, AMBER** | **IFF-CVFF, OPLS** | **IFF-PCFF** | **Avg. Dev. (%)** |
|---|---|---|---|---|---|
| **Lattice a (Å)** | 17.604[5] | 19.25 | 17.91 | 17.76 | 4.0 |
| **Lattice b (Å)** | 18.842 | 17.92 | 18.68 | 18.47 | 2.6 |
| **Lattice c (Å)** | 14.225 | 14.33 | 14.43 | 14.56 | 1.5 |
| **α (°)** | 90.000[5] | 90.02 | 90.35 | 90.06 | 0.16 |
| **β (°)** | 90.000 | 89.58 | 89.91 | 90.55 | 0.39 |
| **γ (°)** | 90.000 | 90.21 | 89.88 | 89.53 | 0.30 |
| **Density (g/cm³)** | 3.38[5] | 3.23 | 3.31 | 3.34 | 2.6 |
| **Surface energy ($J/m^2$)** | (001) N/A | 1.21 | 1.2 | 1.3 | NA |
| | (100) N/A | 1.10 | 1.22 | 1.32 | NA |
| **Bulk modulus (GPa)** | Likely: 120; 90±20,[28][a] 134,[29][b] 120[30],[c] high p:140-156[30, 31][d] | 107 | 115 | 86 | -15 |

[a] The high pressure study in ref. [28] shows a steep slope at low pressure below 1 GPa, suggesting a bulk modulus of 90 ±20 GPa (Figure 2 and Table 1 in ref. [28]). At higher pressures up to 20 GPa, the slope flattens considerably. [b] Most studies have difficulties accurately measuring the volume response below 0.5 GPa or 1 GPa stress. [c] Bulk moduli down to 120 GPa have been reported at low stress.[30] At higher stress, 140-153 GPa were observed. [d] At high pressures up to 50 GPa, extrapolations also suggest 140-153 GPa[30] and 146-156 GPa.[31]

**Table S4.** The experimental and calculated bulk properties of gibbsite, γ-$Al(OH)_3$, using IFF models and parameters. The lattice parameters *a* and *b* using IFF-PCFF parameters showed notable differences from experiment of +4% and +2.5%. The surface energy with IFF-OPLS/CVFF appears to be somewhat overestimated. Specific modifications of the OH parameters could remedy the differences.

| **Property** | **Exp.** | **IFF-CHARMM** | **IFF-OPLS/CVFF** | **IFF-PCFF** | **Avg. Dev. (%)** |
|---|---|---|---|---|---|
| **Lattice a (Å)** | 69.472[9] | 70.85 | 69.18 | 72.26 | 2.1 |
| **Lattice b (Å)** | 60.936 | 61.10 | 59.88 | 62.54 | 1.5 |
| **Lattice c (Å)** | 58.233 | 58.15 | 57.16 | 57.79 | 0.92 |
| **α (°)** | 90.000[9] | 90.00 | 90.43 | 90.35 | 0.29 |
| **β (°)** | 94.540 | 94.44 | 94.40 | 94.64 | 0.12 |
| **γ (°)** | 90.000 | 90.00 | 90.06 | 90.10 | 0.06 |
| **Density (g/cm³)** | 2.42[9] | 2.38 | 2.52 | 2.29 | 3.7 |
| **Surface energy ($J/m^2$)** | (001) DFT: 0.29[24][a] | 0.29 | 0.30 | 0.27 | ~5 |
| **Bulk modulus (GPa)** | Est. 30-35[b] Low pressure: 36[32] High pressure: 49[32] 70-85[33],[c] | 30 | 32 | 21 | ~15 |

[a] DFT calculations can have large deviations for interfacial and surface properties, even though good agreement is found here. [b] Gibbsite, γ-$Al(OH)_3$, is a soft mineral, easily scratched by a knife. It has perfect cleavage in the (001) plane, which consists of 2D planes of OH groups on both sides. A complex response to pressure has been found experimentally and the bulk modulus is likely the same or lower than for $Mg(OH)_2$ (45 GPa[1]) or $Ca(OH)_2$ (35 GPa[1]), which have a similar structure and lower density of (OH) groups per metal atom. The experimental data by Liu et al.[32] from 0 to 0.7 GPa compressive stress (425.2 $Å^3$ to 416.9 $Å^3$) suggest $B$ = 36 GPa. At lower pressure (0.1 GPa or less), the value is likely to decrease, leading to our estimate of 30-35 GPa. Measurements on high pressure phases yielded bulk moduli from 49 GPa[32] up to 85 GPa,[33] which relate to the response under a pressure of many GPa, which leads to different $Al(OH)_3$ phases.

**Table S5.** The experimental and calculated bulk properties of γ-$Al_2O_3$ using IFF parameters and refined structures. A 3×1×1 supercell was used to account for the statistical occupancy of Al atoms of many possible lattice sites. The positions of O atoms are uniquely determined.[10]

| Property | Exp. | IFF-CHARMM | IFF-OPLS/CVFF | IFF-PCFF | Avg. Dev. (%) |
|---|---|---|---|---|---|
| **Lattice a (Å)** | 23.815[10] | 23.97 | 24.08 | 23.91 | 0.72 |
| **Lattice b (Å)** | 15.876 | 15.99 | 15.88 | 16.00 | 0.50 |
| **Lattice c (Å)** | 15.876 | 16.01 | 15.83 | 15.93 | 0.51 |
| **α (°)** | 90.000[10] | 89.42 | 90.13 | 89.70 | 0.37 |
| **β (°)** | 90.000 | 89.67 | 89.88 | 89.50 | 0.35 |
| **γ (°)** | 90.000 | 90.12 | 89.93 | 89.88 | 0.11 |
| **RMSD (A)** | N/A | 0.17 | 0.18 | 0.16 | N/A |
| **Density (g/cm³)** | 3.65[10] | 3.53 | 3.58 | 3.55 | 2.65 |
| **Surface energy ($J/m^2$)** | (001) 1.66±0.1[23] | 1.69 | 1.74 | 1.57 | 4 |
| **Bulk modulus** | 162 | 162 | 206 | 143 | 13 |

**Table S6.** The experimental[18] and simulated elastic constants of α-$Al_2O_3$ using IFF-CVFF, OPLS-AA. The agreement is excellent for the shear modulus and Poisson ratio, and very good for the principal components. Values for less common shear-normal coupling such as $C_{14}$ diverge and could be uncertain in experiments.

| **α-$Al_2O_3$** | **Shear modulus (GPa)** | **Poisson ratio** | **Elastic constant (GPa)** | | | | | | |
|---|---|---|---|---|---|---|---|---|---|
| | | | **$C_{11}$** | **$C_{33}$** | **$C_{44}$** | **$C_{66}$** | **$C_{12}$** | **$C_{13}$** | **$C_{14}$** |
| Exp. | 165 | 0.24 | 497.0 | 501 | 147.0 | 167.5 | 163 | 116 | 22.0 |
| IFF-CVFF | 155.1 | 0.26 | 471.1 | 517.3 | 117.2 | 140.4 | 179.6 | 162.7 | 8.8 |
| Dev. (%) | -6.0 | 8.3 | -5.2 | 3.3 | -20.3 | -16.2 | 10.2 | 40.2 | -60.0 |

**Table S7.** The surface charge density of α-alumina as reported from acid-base titration in different units.[13] Measurements in μC/cm$^2$ are converted into the number of charged sites per nm$^2$, and the percentage of affected OH surface groups, assuming a (001) or (0001) plane with $Al(OH)_3$ groups. Surface charge densities of +0.5 e/nm$^2$ and +1.5 e/nm$^2$, for example, lead to triple the charge density and qualitatively different water interfacial structure and binding of molecules, especially when containing ionic groups. The surface is electroneutral near pH 8.1±0.2[13, 15] (other studies report a range from pH 7 to 9).[14] Negatively charged surfaces at pH values larger than 9 lead to attraction of oppositely charged molecules. These major differences in nanoscale behavior have been largely neglected in prior simulations and are critical for predictive simulations. Different phases of alumina and oxyhydroxides exhibit semi-quantitatively the same surface chemistry, point of zero charge, and surface terminations.

| **pH values** | **Exp. Surface charge density (μC/cm$^2$)[13, 14], [a, b]** | **Exp. Surface charge density (e/nm$^2$)[c]** | **Percentage of $Al(OH)_3$ surface groups ionized[b]** |
|---|---|---|---|
| **2 (diss.)** | 28 ±2 | 1.75 ±0.1 | ~12% |
| **3 (diss.)** | 24 ±2 | 1.5 ±0.1 | ~10% |
| **4 (diss.)** | 19 ±2 | 1.2 ±0.1 | ~8% |
| **5** | 15 ±2 | 0.9 ±0.1 | ~6% |
| **6** | 10 ±2 | 0.6 ±0.1 | ~4% |
| **7** | 5 ±2 | 0.3 ±0.1 | ~2% |
| **8.0-8.2** | 0 ±2 (pzc at 8.1±0.2) | 0 ±0.1 | 0 |
| **9** | -5 ±2 | -0.3 ±0.1 | ~2% |
| **10 (diss.)** | -12 ±2 | -0.75 ±0.1 | ~5% |
| **11 (diss.)** | -18 ±2 | -1.1 ±0.1 | ~8% |
| **12 (diss.)** | -24 ±2 | -1.5 ±0.1 | ~10% |

[a] Surface charge density of α-$Al_2O_3$ in NaCl solution at a concentration of 0.1 M NaCl from ref. [13]. The values remain similar in a broader concentration range <$10^{-4}$ M to >0.5 M in good approximation. $NaNO_3$ solutions show similar data within ±2 μC/cm$^2$. [b] CsCl solutions, also included in ref. [13], show slightly higher positive surface charge at low pH (ca. +31 μC/cm$^2$ at pH = 2) and a lower negative surface charge at high pH (ca. -15 μC/cm$^2$ at pH = 12). A lower negative surface charge density at high pH values likely results from covalent-type bonding of $Cs^+$ ions to $OH^-$ ions in solution ($Cs^+$ has a typical atomic charge of only +0.6e to +0.7e),[4] thereby reducing the amount of $OH^-$ ions available in solution to form $AlO^-$ groups from AlOH groups. In contrast, $Na^+$ ions are fully ionic in solution (+1.0e), form no complexes with $OH^-$ ions and enable the maximum activity of $OH^-$ ions to form $AlO^-$ sites. Consistent with this interpretation, the surface charge density of α-$Al_2O_3$ at low CsCl concentrations of $10^{-4}$ M is identical to that in $10^{-4}$ M NaCl solution (see Figure S14). [c] Unit conversion from μC/cm$^2$. [d] We assume an area density of 15.3 or 13.6 OH groups per nm$^2$ as found in hydrated (0001) α-alumina and (001) gibbsite, respectively.

**Table S8.** The percentage of ions dissociated from (0001) α-alumina surfaces at various pH values according to MD simulations, along with the density per area. Below pH 8, the groups include $Al(OH_2)^+$ $Cl^-$, at pH 8 there were no charged groups (pzc), and above pH 8, the groups include $AlO^-$ $Na^+$. The fraction of dissociated ions is larger at lower total surface charges (near the pzc) than at high surface charge, and chloride ions tend to dissociate further than sodium ions. The data are averaged over the last 75 ns of 100 ns simulation time and uncertainties are ±5% of the given values.

| pH | Ion species | Dissociated (> 5 Å) | Total area density of ions ($nm^{-2}$) | Area density of dissociated ions ($nm^{-2}$) |
|---|---|---|---|---|
| 2 | Cl- | 13 % | 1.87 | 0.25 |
| 5 | Cl- | 45 % | 0.85 | 0.38 |
| 8 | None | None | 0 | 0 |
| 10 | Na+ | 13 % | 0.85 | 0.11 |
| 12 | Na+ | 8 % | 1.53 | 0.12 |

## Supplementary Discussion

### S1. Prior Validation of the Interface Force Field (IFF) and Performance Comparison to Earlier Force Fields, DFT, and DFT-MLIPs

**S1.1. Prior Validation of the Interface Force Field (IFF).** IFF is a class I force field, that is, a combination of electrostatics, Lennard-Jones, and bonded terms, as is commonly used in MD simulations. It includes 12-6 and 9-6 options for Lennard-Jones parameters for most compounds. The concept of IFF has been briefly described in the main text and in detail in the introductory publication.[34] In comparison to other force fields, like OPLS-AA, CHARMM, and others, IFF uses a more quantitative, consistent approach for a wider range of chemistries that leads to order-of-magnitude higher accuracy, i.e., inorganic, organic compounds, and biomaterials. The simulations of mixtures of compounds, including compatibility with other force fields, succeed because the parameters for each material on their own reproduce internal multipole moments, molecular/lattice geometry, as well as the surface energy and interactions with other molecules. IFF builds on a self-consistent classical Hamiltonian for each validated compound, which enables a high degree of compatibility and transferability. In addition, IFF uses only one interpretable set of parameters, which lowers the uncertainty in model assumptions (also called epistemic uncertainty).[35] This approach differs from grey-box and black-box approaches with multiple parameter sets for the same physical systems as, for example, in density functionals with a long menu of dispersion corrections and ReaxFF with multiple parameters sets.

The protocol of IFF development builds on step by step validation of the atomic charge, bonded terms (bond, angle, and dihedral values and constants), Lennard-Jones parameters, structural and surface properties, leading to estimations of bulk and interfacial properties in quantitative agreement with experimental measurements (Figure 2a).[34] Rigorous validations and consistent protocols for development facilitate high accuracy (few percent average deviation), interpretability, compatibility and transferability. IFF has been extensively validated and applied for MD simulations of metals, oxides, minerals, 2D materials, polymers, ionic liquids, gases, and inorganic-organic interfaces, including specific molecular recognition, self-assembly, dynamics, competitive adsorption, electrolyte interfaces, crystal growth, and reactivity at interfaces.[36-43]

**S1.2. Performance Comparison of IFF to Earlier Force Fields and DFT.** We have previously compared the performance of IFF for a broader set of oxides and hydroxides (including our parent model of α-alumina for the alumina phases) to Pedone, DFT, ClayFF, ReaxFF, and UFF in depth (Figure S1).[1] These comparisons are representative of the accuracy of each of these models for alumina phases. Here, we added specific tests for alumina phases for CVFF, PCFF, COMPASS, and COMPASS II for alumina phases (also included in Figure S1). CVFF, PCFF, COMPASS, and COMPASS II do not cover the broader set of oxides and hydroxides discussed in ref. [1]. CHARMM, AMBER, and OPLS-AA include no parameters for alumina phases or other oxides, but are compatible with IFF. Density functionals have considerable deviations in key alumina properties (see ref. [1] for details). Common density functionals such as PBE or rPBE yield equilibrium water densities of 0.89 and 0.75 g/cm$^3$, which tend to be too far from experimental data (1.00 g/cm$^3$) for predictive simulations. Dispersion corrections (D2, D3, SCAN, etc) can mitigate some challenges although still face issues with unphysical bond dissociation curves in case of modeling reactions.[43, 44]

Machine-learned interatomic potentials from DFT such as MACE and Allegro-FM have higher deviations and less transparency than the baseline DFT data. For example, we expect major deviations in surface energies (up to 100%) and elastic properties (up to 30%) for the bulk phases. The methods cannot reproduce the water density (~0.85 to ~0.9 g/cm$^3$) and lead to ice phases at room temperature with melting points above 40 °C, which prohibits realistic simulations of aqueous, pH-sensitive, or organic interfaces. A major drawback of MLIPs is the lack of rigorous benchmarking against experimental data, which has been standard for IFF, CHARMM, AMBER, OPLS-AA, PCFF, and other force fields for decades (e.g., including in this contribution). Therefore, MACE and other DFT-learned MLIPs are insufficiently validated and can often be misleading. The computational speed is approximately 100,000 times slower than IFF MD for systems with 100,000 atoms so that simulations of realistic nanomaterials and inorganic-organic interfaces are highly ineffective, if feasible at all (see main text). We recommend CCSD(T) levels of theory or higher for local additions of electronic structure with reliability. Carefully vetted density functionals with clear monitoring of limitations and errors can also be used at a local scale. In the remainder of this subsection, we describe specific test results for the prior force fields CVFF, PCFF, COMPASS, and COMPASS II for the five alumina phases, which were not discussed in ref. [1].

Tests with CVFF indicate missing charge assignments. If using the charge equilibration Qeq

(recommended in Materials Studio, but misleading for inorganic materials[4]) and primarily nonbonded representations as in IFF, lattice parameters of all phases expanded by approximately 50%, leading to density errors of -70% and higher relative to experiments. When using bonded models, lattice parameters contracted between 13% and 17% and the density deviated +50% relative to experiments. The strong lattice distortions and density errors preclude physically meaningful results for other properties, and we carried out no further tests of surface energies and bulk moduli.

Tests with PCFF led to similar performance as CVFF, including missing charge assignments. If using Qeq to assign atomic charges and primarily nonbonded models as in IFF, the lattice parameters of all phases expanded unevenly between 20% and 60% (average density error of -60%). Using bonded models for the same calculations was not feasible due to missing bond parameters. We carried out no further tests of surface energies and bulk moduli due to major structural distortions.

Tests with COMPASS in nonbonded mode (after typing and charge assignment in bonded mode) reproduce lattice parameters of corundum with less than 0.5% error. Surface energies of corundum (0001), however, are ~45% too low relative to experiment, and the computed bulk modulus is too low by a similar amount. γ-alumina showed larger deviations than corundum with an expansion of all lattice parameters by ~11% and a density decrease of -30% relative to experiment (which appears to involve atom typing issues). For diaspore, COMPASS yields lattice parameters that are 5% to 12% too large, resulting in -20% error in density. Boehmite expands

+19% in the *a* direction and features ~5% error in *b* and *c* direction, also leading to -20% error in density relative to experiments. Gibbsite experiences +45% elongation in the *a* direction, and a -34% reduction in overall density. The extent of lattice distortions of the hydroxide phases allows only approximate simulations of surface energies and bulk moduli. Surface energies and moduli have larger errors than corundum (>50%), compounded by structural distortions.

Tests with COMPASS II showed the same high structural definition of corundum as COMPASS with less than 0.5% error in lattice parameters. Surface energies of corundum are, however, ~45% too low relative to experiments, and the modulus is too low by a similar amount. γ-alumina also showed an expansion of all lattice parameters by ~11% and a density decrease of -30% relative to experiment. Lattice parameters of diaspore expand by ~6%, leading to a density error of -18% relative to experiment. Boehmite expands +19% in the *a* direction and has ~5% error in *b* and *c* directions, leading to -20% error in density. COMPASS II shows a strong distortion of the gibbsite lattice and its internal geometry, including expansion in the *b* direction by over 50% and contraction in *c* by ~20%, resulting in an unphysical structure with -23% density. These distortions are more severe than in COMPASS. Computed surface energies and moduli of all phases have large deviations between 30% and 80% relative to experiment (typically, underestimates) and are estimates due to major deformations in equilibrium.

In summary, CVFF and PCFF models are not suited for alumina phases due to density errors >50%. COMPASS and COMPASS II models mitigate some issues with lattice parameters (e.g., accurate for corundum) but include significant internal structural distortions and density deviations of ~25% for the other four alumina phases. All models, including for corundum, feature deviations in surface energies and bulk moduli of ~50% and higher relative to experimental data, which substantially limits predictive use in synthesis, process-level decision making, and materials design.

The primary limitation in COMPASS and COMPASS II are less physically consistent atomic charges of +1.20e for Al in combination with LJ parameters that do not adequately reproduce the balance of structural and energetic properties. These shortcomings are remedied in the IFF parameter set, which is based on accurate parameters for the α-alumina parent compound[1] and reproduces the key properties for all 5 alumina phases (Figure S1).

## S2. Tradeoffs Between Bonded Models and Nonbonded Models

**S2.1. Limitations of Bonded Models.** The major limitation of bonded models lies in the complex bond connections inside the atomic models, especially compounds with more than 6 bond connections for an atom, which require an extensive and complete set of bonded terms in the force field file to carry out a successful molecular dynamics simulation. The definition and validation of the bonded terms, e.g., settings for bond, angle, and dihedral equilibrium values and force constants, even when they are given by crystal structures, vibration spectra, and require no actual parameter search, is very time consuming. In addition, the atomic charges and Lennard-Jones parameters require thorough analysis to reproduce key bulk and interfacial properties for validation, including lattice constants, density, surface energy, water contact angle, adsorption energy, and mechanical properties.

When modifications are needed in the system, such as a change from oxide ions to hydroxide ions, elemental doping, and surface hydration to Al-OH groups, it is even more challenging to use bonded models due to the introduction of new atoms, known or unknown new bond topology. Any new bond, element, or force field type of an existing element requires a new set of bond, angle, dihedral, and possibly out of plane (improper) parameters and force constants. The known structure and chemical analogy often help in finding the added parameters quickly, however, the large number of necessary terms can become burdensome.

The definition of bonded terms can thus be a challenge for new compounds. We recognize, however, some key observations: (1) Predominantly covalent compounds such as organic and biological matter composed of carbon, nitrogen, oxygen, and hydrogen usually have low coordination numbers (1 to 4), which limit the number of bonded terms. Also, analogy considerations to derive bonded terms, which for the most part have clear starting points in experimental data and involve narrow physical ranges for optimization, can be taught to AI-based agents for automatic presentation, reducing the workload of researchers. (2) Predominantly ionic compounds, or about half ionic compounds, such as alumina, salts, and several other oxides have higher coordination numbers. Higher coordination numbers lead to an exponential increase in the number of angle and torsion terms if the bonded interactions were to be included.[45, 46] From a physical viewpoint, a large number of these terms becomes increasingly redundant as interactions are mostly ionic and would be more adequately represented by nonbonded terms. Therefore, nonbonded models that can eliminate a long batch of bonded parameters introduce much greater flexibility. If needed to reproduce the equilibrium structure in borderline cases, bonded terms can still be utilized. However, when bonding is clearly more covalent than ionic, e.g., in silica, zirconia, and other oxides, the use of nonbonded parameters without bonded terms is unphysical. Charges higher than chemically realistic values are often necessary to compensate for some missing covalent bonding contributions. (3) Using nonbonded models, the parameterization for different alumina phases is much faster (e.g., 1 week with prior experience) than with bonded models (e.g., 10 weeks with prior experience). Overall, the advantages of the nonbonded models surpass minor limitations such as sensitivity to both the type of LJ potential and combination rules.

AI can be used to guide elements of the workflow for parameter derivation. However, the thought processes to incorporate chemical knowledge, the identification, retrieval, and evaluation of experimental reference data, and diverse sets of test simulations for validation are highly complex challenges towards automation.

**S2.2. Utility of Nonbonded Models.** The nature of chemical bonding in oxides is governed by the chemical environment and intrinsic atomic properties, such as atomization energy, ionization energy, and electronegativity.[47, 48] Based on the extent of valence electron transfer, metal-oxygen bonds in oxides can be broadly categorized as covalent, ionic, or intermediate (covalent/ionic).[49, 50] Typical covalent-dominated oxides include $SiO_2$, $CeO_2$, $MnO_2$, $TiO_2$, $ZrO_2$, and $MoO_3$. Ionic-dominated oxides include CaO, MgO, CoO, and $Na_2O$, while mixed-bonding compounds such as $Al_2O_3$, $V_2O_3$, and $Fe_2O_3$ exhibit both covalent and ionic character. In general, increased ionic character correlates with higher chemical reactivity.

In this context, it is chemically justified to represent ionic and mixed ionic/covalent oxides using nonbonded models in molecular dynamics (MD) simulations. These models eliminate explicit bond connections among atoms and instead rely on electrostatic interactions between positively and negatively charged species to reflect valence electron transfer. This approach is conceptually similar to modeling metallic bonding in FCC metals, where atoms interact without net charge transfer.[51] In nonbonded models, all bonded terms are omitted—except for surface hydroxyl (OH) groups where needed—greatly simplifying the parameterization process.

Validated nonbonded models offer accurate predictions for key bulk and interfacial properties, including lattice constants, density, surface energy, bulk modulus, water contact angle, and adsorption energy with average deviations of 0.8%, 2.1%, 4%, 7%, 0%, and <1 kcal/mol relative to experiments, respectively. Notably, surface energies—which are often difficult to capture by bonded models due to bond constraints—can be readily calculated using nonbonded models. Elemental doping is easily implemented by substituting existing atoms with desired anions or cations while maintaining stoichiometry. Surface hydroxylation can also be customized to match experimental surface chemistry. Since bonded terms are absent, the force field parameters can often be transferred to other polymorphs of the same compound without requiring reparameterization. Nonbonded force fields can also perform better in the computation of vibrational spectra (IR, Raman).

**S2.3. Limitations of Nonbonded Models.** Despite these advantages, nonbonded models have reduced compatibility across different force fields (e.g., CHARMM, CVFF, PCFF).[1] The reduced compatibility is primarily due to differences in LJ parameters and in combination (mixing) rules. Among potentials with 12-6 Lennard-Jones potentials, therefore, we distinguish (1) force fields with an arithmetic mean for $r_{min,\ ij}$, including CHARMM, AMBER, and DREIDING, (2) force fields with a geometric mean for $r_{min,\ ij}$, including CVFF and OPLS-AA. Accordingly, modifications to both IFF LJ parameters $r_{min,\ ii}$ and $\varepsilon_{ii}$ are necessary to ensure compatibility within the chosen group of force fields (Table 1). When using force fields with a 9-6 LJ potential and Waldman-Hagler combination rules[52, 53] such as CFF, PCFF, and COMPASS, $r_{min,\ ii}$ and $\varepsilon_{ii}$ also change due to both the different mathematical form and combination rules. Bonded models, if used, are less reliant on LJ parameters and the same set of 12-6 LJ parameters for different combination rules is often sufficiently accurate.

## S3. Using IFF Parameter Sets Tailored to the Type of Energy Expression

IFF has been using parameter sets tailored to the energy expression since 2005, including (1) the LJ type, (2) the default combination rule for $r_{min,\mathrm{ij}}$ and $\varepsilon_{\mathrm{ij}}$, and (3) the scaling convention of nonbond interactions between 1,4-bonded atoms.[34, 45, 54] The parameters are developed for broad compatibility, yet cater for specific energy expressions, which include their default combination rules and 1,4 scaling conventions. The origin of these differences has been previously described[45, 54-57] and is reviewed here to illustrate the benefits as well as losses in case of simplification.

The differentiation of LJ parameters to the level of combination rules matters specifically when relying on nonbonded models, which include no further bonded terms to stabilize interatomic distances and geometries. Nonbonded models are thus more impacted by the type of LJ potential and the combination rules. In contrast, bonded models stabilize largely via bond stretching, angle bending, and torsion terms, resulting in less sensitivity to the details of the LJ potential. Differences in 12-6 LJ parameters due to different combination rules (e.g., CHARMM/AMBER vs CVFF/OPLS-AA) are often small enough to neglect in bonded models, which broadens transferability.

Bonded models, however, usually require differentiation of LJ parameters (and/or bonded parameters and charges) to the level of scaling conventions of nonbonded interactions between 1,4 bonded atoms (e.g., OPLS-AA vs AMBER vs. CHARMM). These interactions are typically significant and impact molecular geometries. Nonbonded models such as for alumina phases have no 1,4 bonded topologies, however, they need differentiation of combination rules.

In reference [1], we introduced nonbonded IFF parameter sets for a series of oxides and hydroxides tailored to the type of energy expression and combination rules, e.g., IFF-CHARMM, IFF-CVFF, and demonstrated that arithmetic versus geometric combination rules for $r_{min,\, ij}$ (e.g., CHARMM vs OPLS) require distinct 12-6 LJ parameters to reproduce key properties for validation with minimal errors.

These differentiations can be neglected, if desired, and IFF parameters for CHARMM but used with OPLS-AA still yield a better model than, e.g., COMPASS or ClayFF (Figure S1).

However, the best practice is to utilize parameter sets tailored to the specific energy expression because accuracy and transferability within the distinct framework can be impressively high. For example, staying within the specific energy expression versus intermixing can make the difference between a model with 5% error versus 20% error in organic binding energies on alumina relative to experiments. The typically high accuracy can greatly enhance active feedback loops with experimental collaborators and platforms towards autonomous materials discovery.

## S4. Interpretation of Force Field Parameters for the Alumina Phases

The ability to describe a structurally diverse set of minerals using a single parameter set highlights the robustness of the chemical information embedded in the IFF parameters (Table 1). It also demonstrates that minor variations in coordination environments do not necessitate complex polarizable models or additional empirical fit parameters, as previously suggested.[58, 59] The key lies in accurately representing bonding through experimentally and theoretically backed atomic charges, along with validating structures and energies—a unique principle of IFF and IFF-R.[34]

The Lennard-Jones (LJ) and bonded parameters for O-H groups remain the same for all hydroxide bulk phases and for OH surface groups of all minerals, and are similar to those of other oxides.[1] This assumption is practical and sufficiently accurate, however, the atomic charges require more detailed consideration. The assigned values are essential in representing interactions in the bulk minerals and at surfaces, including ionization as a function of pH value (Figure 3 and Table 1). Across pH values from 0 to 14, the presence of $AlOH_2^+\cdots Cl^-$, AlOH, and $AlO^-\cdots Na^+$ groups on the surface modifies zeta potentials and charge density on the surface across orders of magnitude, changing water interfaces, types of attracted charged and functional groups in molecules, and polyelectrolyte interactions.

To represent these effects, we differentiate between two atomic charge settings for OH groups based on their chemical environment, "bulk" and "surface" (Table 1). In gibbsite, boehmite, and diaspore, bulk AlOH groups inside the crystal lattice carry a charge of +0.30e on H atoms and -0.84e on the O atoms (Figure 3a). This charge distribution reflects the bond polarity and correlates with a trend in acidity (when also considering the stability of the resulting oxide ion). For example, the OH bond in AlOH groups is less polar compared to SiOH groups. As a result, the H charge in AlOH groups of +0.3e is lower than in SiOH groups with a H charge of +0.4e and an O charge of -0.675e.[60] The acidity of AlOH groups in water is weaker than in SiOH groups (pzc of ~8.4 for alumina phases versus ~3 for silica).

Additionally, the Al-O bond in alumina is more polar than the Si-O bond in silica, leading to a higher oxygen charge in AlOH groups. This increased polarity necessitates a lower H charge of +0.30e to prevent artificially high charges and multipole moments on the O atoms in AlOH groups. Bulk OH groups thus experience moderate local polarity while surrounded by highly charged oxygen and aluminum atoms with -1.08e and +1.62e in the vicinity, respectively, which introduce very strong attraction as well as strong repulsion. Surface OH groups, which are exposed to water with lower charges of -0.82e (O) and +0.41e (H), encounter an environment of similar polarity and retain atomic charges of -0.84e (O) and +0.30e (H) (Figure 3b and Table 1). In vacuum, the charge maybe slightly lower at -0.83e (O) and +0.29e (H), corresponding to a small amount of polarization (analogous but smaller differences than $Ca^{2+}$ ions in anhydrite of ca. +1.6e charge versus $Ca^{2+}$ ions in gypsum of ca. +1.7e charge).[61] The charge assignments for bulk OH were validated by their agreement with experimental bulk moduli, and the surface OH groups by their agreement with the immersion energy of gibbsite.

Second, we distinguish two charge settings for protonated and deprotonated surface AlOH groups, namely, $AlOH_2^+ \cdots Cl^-$ groups at low pH values and $AlO^- \cdots Na^+$ groups at high pH values in solution (Figure 3b and Table 1). The Al atoms in $AlOH_2^+ \cdots Cl^-$ groups retain +1.62e charge, the oxygen charge also retains -0.84e, and the H charge is increased to +0.65e, leading to an overall positive charge of +1.0e that is compensated by an anion such as $Cl^-$ (Figure 3b). The two nearest Al atoms in $AlO^- \cdots Na^+$ groups absorb some of the incoming negative charge and have +1.40e charge, and the oxygen atom carries a higher charge of -1.10e, leading to an overall negative charge of -1.0e that is compensated by a cation such as $Na^+$ (Table 1). Note that there are two Al atoms per such deprotonated OH group on the surface of alumina phases that are affected by these adjustments.

We tested different charge assignments, such as reduced negative charges in O atoms in $AlOH_2^+ \cdots Cl^-$ down to -0.5e and reduced H charges down to +0.48e, which would be justified if also using bonded terms for Al-O bonds. However, we represent covalent bonding here through nonbonded interactions, which requires upkeep of high charges for structural stability, such as - 0.84e for O. If charges are lowered, some $OH_2^+$ groups would dissociate from the surface. Keeping the O charge in $AlOH_2^+$ groups at the same level as in neutral Al-OH groups prevents this failure mode, represents the interfacial properties well, and simplifies modifications for protonated groups. In $AlO^- \cdots Na^+$ groups at high pH values, similar considerations apply. Lowering the nearby Al charges more significantly below +1.40e, which would be realistic in bonded models (e.g., Al down to +1.30e and O to -0.90e), can lead to instabilities in the nonbonded structure. Therefore, the Al charges of the two nearest neighbors are only reduced from +1.62e to +1.40e, and the oxygen charge remains at -1.10e.

The origin of these charges increments lies in partitioning the bulk Al charge of 1.62e into 6 equal contributions of +0.27e from the 6 coordinative Al-O bonds per Al atom. For example, bulk O atoms are bound to 4 Al atoms and then carry the opposite charge of $-4 \cdot 0.27e = -1.08e$. In case of an AlOH group, the H atom has a charge of +0.3e in bulk and on the surface, which is closely the same contribution to the O charge as another Al atoms. The oxygen atoms in AlOH groups are bound to two Al atoms and one hydrogen atom, resulting in a charge of $-2*0.27e + (-0.3e) = -0.84e$ (bulk), for example. Finally, when AlOH groups are protonated to $AlOH_2^+$ groups, the incoming charge of +1.0e is distributed over the two H atoms, increasing the H charge to 2x +0.65e and (which is +1.0e more than the original H charge of 0.30e), and the oxygen charge remains at -0.84e (Figure 3b and Table 1). When AlOH groups are deprotonated, $AlO^-$ groups form on the surface, and the negative charge of -1.0e is distributed over the oxygen atoms and the two nearby Al atoms. Hereby, removal of the H atom leaves -0.70e for distribution, of which a larger share would remain on the oxygen atom (-0.26e) and slightly smaller parts be distributed over the two attached Al atoms (-0.22e each), reducing their charge from +1.62e to +1.40e and increasing the oxygen charge from -0.84e to -1.10e. These charges are relatively high but suitable for a nonbonded model, which needs to cover some contributions by covalent through stronger ionic bonding.[1]

The Lennard-Jones parameters for the bulk alumina phases reflect the relative nonbonded atomic sizes of Al and O, and have been thoroughly validated for $\alpha$-$Al_2O_3$.[1] The LJ parameters for the Al-OH groups in bulk and surface environments are similar to the ones used for silanol groups on silica, which were suitable to accurately describe hydration energies and contact angles. In alumina, the oxygen atom has a similar environment and charge. The hydrogen atoms in OH groups have zero LJ parameters (or nearly zero in IFF-PCFF) and are only characterized by an atomic charge, which is lower than in silanol groups due to higher charge of Al and reduced acidity compared to silanol groups as known from experiments.

IFF parameters compatible with CVFF and OPLS-AA parameters are identical because the combination rules for $r_{min,\ ij}$ are the same (geometric) (Equation (1)). A separate set of 12-6 LJ parameters is used for compatibility with CHARMM, AMBER, and DREIDING, which use arithmetic combination rules for $r_{min,\ ij}$ (Equation (1)) and typically require slightly smaller $r_{\min}$ values. Finally, a set of LJ parameters with 9-6 LJ options is included for use of IFF with PCFF, CFF, and COMPASS (Equation (2)). The origin of these differences was previously described.[45, 54-57] Different scaling of nonbond parameters between 1, 4-bonded atoms can be neglected due to the lack of angle and dihedral parameters (Equation (1)).

If covalent constraints for some bonds are desired, such as Al-O in protonated ($Al_2OH_2^+$) or dissociated ($Al_2O^-$) surface groups, they can be added along with appropriate harmonic bond and angle parameters, including adjustments in atomic charges and maintaining overall charge neutrality.

**S5. Measurement and Computation of the Immersion Energy of Gibbsite Nanocrystals**

The experimental measurement of the immersion energy of a hexagonal nanocrystal with an aspect ratio of 8 (120 nm wide, 8 nm thick) involved the (001) basal planes and a significant edge contribution from (400) planes.[62] Specifically, if the width of the hexagonal platelets is 2*a* (*a* = 60 nm) and the thickness *t* (*t* = 15 nm), then the total surface area has a contribution by two basal planes of $A_{basal} = 3\sqrt{3}a^2$ ($\frac{\sqrt{3}}{4a^2}$ per every triangle) and a contribution by 6 edge planes of $A_{edge} = 6at = \frac{3}{4a^2}$. Therefore, the basal area is 3.464 times the edge area and the total area has fractional contributions of:

$$A_{total} = 0.776 A_{basal} + 0.224 A_{edge}. \qquad \text{(S1)}$$

The energy of immersion $\Delta E_{imm}$ per unit area is therefore:

$$\Delta E_{imm} = 0.776 \Delta E_{basal} + 0.224\, \Delta E_{edge}. \qquad \text{(S2)}$$

We carried out computations of the immersion energy for the (001) basal and (400) edge surfaces and used equation (S2) to process these facet-specific immersion energies into an immersion energy for the entire crystal for comparison with experimental data. We utilized fully relaxed equilibrium model surfaces. In particular, the (400) edge surfaces reconstructed via outward mobility of internal OH groups near the surface to coordinate exposed Al sites on the edges.

Computed immersion energies on both facets are sensitive to the H charge on the surface OH groups. Using the same charges as in internal OH groups (-0.84e on O and +0.30e on H) leads to quantitative agreement with experimental measurements of -223 ±15 $mJ/m^2$ if we assume that the charge on the OH groups in vacuum is slightly lower (-0.83e on O and +0.29e on H). A slightly lower surface charge in vacuum, equal to a decrease in H charge from 0.30e to 0.29e in OH surface groups, appears to be justifiable due to the absence of external hydrogen bonds and contributes ca. -88 $mJ/m^2$ to the overall immersion energy of the nanocrystal. Alternatively, if we require identical charges to be present on surface OH groups in the hydrated state and in the vacuum state, higher charges of -0.89e on O and +0.35e on H in surface OH groups (both edge and basal surface) are necessary to reproduce experimental measurements of the immersion energy of -223 ±15 $mJ/m^2$. This setup is equally supported and requires different charge assignments on surface OH groups versus internal OH groups. Hereby, an increase by +0.01e in the H charge in the surface OH groups lead to approximately -17 $mJ/m^2$ larger immersion energy (e.g., -240 $mJ/m^2$ at -0.90e/+0.36e).

Overall, we recommend using the same charges for OH surface groups in aqueous interfaces and internal OH groups (and slightly lower OH charges in vacuum if needed), rather than separate charges for OH surface groups. This choice is physically justifiable and simplifies model preparation and simulation workflows.

**S6. Point of Zero Charge of Gibbsite, Hydrated Alumina, and Other Phases**

Data on the point of zero charge vary somewhat by source. An authoritative, up-to-date compilation is given by Kosmulski.[63] The most universal suggestion for the point of zero charge (neutral, OH-terminated surface) is a pH value of 9 ±0.3, and a wider reported range of 9 ±1. Some alumina samples, diaspore, boehmite, amorphus alumina, alumina nanoparticles, or other phases may have lower pzc if not given enough time to hydrate on the surface. Therefore, also reports of pzc as low as 7 are common.[64] On balance, therefore, we suggest a pzc of 8 to 8.5, which reasonably covers common situations.

The proposed models can be shifted to a specific point of zero charge that corresponds to the experimental data for the system of interest. Hereby, we can assume that the master surface titration curve (Figure 8a and Figure S14) will only be shifted on the horizontal pH axis, e.g., to a pzc to 7, 8 (default), or 9, and otherwise look closely the same. I.e., only the intersection point with the horizontal axis would shift to the new pzc. For example, if current models assume full OH termination at pH 8 and an ionization of +0.3 e/nm$^2$ at pH 7 ($OH_2^+$ $Cl^-$), the same models can be used to represent pH 9 and pH 8 if it is known that the point of zero charge is at pH = 9 instead of pH = 8. This approximation is up to an order of magnitude better than not making it. Experimental reference data and “master curves” for surface titration are known for a few similar OH-terminated oxides (silica, titania, ceria, alumina) share common features, even when the pzc is different, indicating a similar amount of ionization per surface area for a similar pH distance from the pzc, as well as some dependence on the electrolyte composition (Figure S14). In the future, it may be feasible to utilize machine learning to determine the expected ionization properties, however, currently the total amount of reference data remains limited.

## S7. Computation of Binding Free Energy from Langmuir Adsorption Isotherms: Theory and Limitations

The experimental adsorption isotherms of p-hydroxybenzoic acid show a clear trend as a function of pH (Figure 9a and Figure 5 in ref. [64]). The decrease in adsorbed amount for higher pH values is consistent with alumina surface chemistry and suggests a smaller negative binding free energy as seen in MD simulations for single molecules. Calculations of the associated binding free energies from the experimental Langmuir isotherms, however, showed a less intuitive correlation.[64] In the Langmuir model, [64, 65] the adsorbed amount $\Gamma$ is given as:

$$\Gamma = \Gamma_{max} C_e / (\frac{1}{K_S} + C_e), \tag{S3}$$

with a maximum adsorbed amount $\Gamma_{max}$, the equilibrium concentration of p-hydroxybenzoate in solution $C_e$, and the adsorption coefficient $K_S$. Furthermore, the adsorption coefficient $K_S$ is defined as:

$$K_S = \frac{N_{ads}}{(N_S - N_{ads}) C_e}, \tag{S4}$$

with $N_{ads}$ equal to the number of molecules adsorbed, $N_S$ the number of available surface sites, and $C_e$ the equilibrium concentration as in equation (S3).

Using the measured data (Figure 5 and Table 2 in ref. [64]), the values of $K_s$ at pH values of 5, 6, 7, 8, and 9 were reported as 0.42, 0.32, 0.11, 0.18, and 0.38 l/mmol, respectively, with uncertainties between ±0.03 and ±0.12 l/mmol. The corresponding values of $K_S$ are 420, 320, 110, 180, and 380 $M^{-1}$. This trend appears at odds with the measured values of $\Gamma_{max}$, which decreased as shown in Figure 9a (equal to Figure 5 in the original paper ref. [64]) from 3.82, 2.11, 1.09, 0.49, to 0.41 $\mu mol/m^2$ from pH 5 to 9 and is well explicable based on surface chemistry.

The corresponding free energies of adsorption, calculated as:

$$\Delta G = -RTln\, K_S \tag{S5}$$

are -3.64, -3.47, -2.84, -3.13, and -3.58 kcal/mol at pH 5, 6, 7, 8, and 9, respectively. Like the trend in $K_S$, the increase in negative binding free energy towards pH 9 does not follow the trend in adsorption capacity in the isotherms (Figure 9a). Given the original measurements of adsorption isotherms including $\Gamma_{max}$, we expect that strong adsorption at pH 5 would correspond to the larger negative free energies of binding (-3.6 kcal/mol), while adsorption is expected to be much less favorable at pH 9 as seen in the isotherms and in MD simulation (Figure 9).

In conclusion, the calculated binding constants from the Langmuir models in equations (S3) and (S4) may not represent the experimental data well. Possible reasons lie in the definition of available surfaces sites $N_S$ and occupied surface sites $N_{ads}$ on these chemically diverse surfaces (Figures 7 to 9). Simulations show high mobility across the surface paired with a certain level of attraction and average distance, driven by the amount electrostatic attraction and ion pairing. The original Langmuir models from the 1910s, as still used today, focused on mica, glass, and clean metal surfaces with rare gas adsorption in vacuum.[65] The Langmuir models were not intended for a ternary mixture including solvent, chemically adaptable surfaces with pH-dependent charge regulation, absorbents with pH-sensitive chemistry in electrolytes, ion pairing, and the associated pH-dependent restructuring of the interfacial environment (Figures 7 to 9). These effects clearly exceed the assumptions of classical Langmuir adsorption theory and may explain the gaps between fitted adsorption constants and experimentally observed adsorption capacities. Future studies may therefore venture further into thermodynamic descriptions of reactive surfaces and their adsorption isotherms, for example, including our methods of computation of $K$ and $\Delta G$ from MD simulations. In addition, infrared spectroscopy can provide further insight into binding modes and mechanisms as shown in ref. [64].

**Supplementary References**